\documentclass[prb,twocolumn,showpacs,floatfix,amsmath,amssymb,superscriptaddress]{revtex4}

\usepackage{graphicx}

\usepackage{dcolumn}

\usepackage{bm}

\usepackage{subfigure}

\usepackage{amssymb}

\usepackage{float}
\usepackage{array}
\usepackage{amsthm}
\usepackage{amsmath}
\usepackage{epsfig}
\usepackage{graphicx}
\usepackage{wasysym}
\usepackage{bbm}
\usepackage[all]{xy}

\newcommand{\SG}{SG}
\newcommand{\gsg}{u}
\newcommand{\GG}{GG}
\newcommand{\gGG}{g}
\newcommand{\G}{G}
\newcommand{\gG}{s}
\newcommand{\hG}{h}

\newcommand{\ccy}{\omega}
\newcommand{\ccygroup}{U(1)}
\newcommand{\manif}{M}

\newcommand{\colo}{\varphi}
\newcommand{\ccysgn}{\epsilon}
\newcommand{\area}{A}
\newcommand{\loopc}{P}
\newcommand{\pathg}{\Gamma}
\newcommand{\bstring}{L}
\newcommand{\wstring}{w}
\newcommand{\phase}{\Phi}
\newcommand{\phasec}{\Theta}
\newcommand{\phasew}{\Psi}
\newcommand{\upt}{\bigtriangleup}
\newcommand{\downt}{\bigtriangledown}
\newcommand{\ccyc}{c}
\newcommand{\ccye}{\varepsilon}

\newcommand{\transl}{\mathcal{T}}
\newcommand{\gs}{gs}
\newcommand{\state}{\psi}
\newcommand{\vis}{h_v}
\newcommand{\gvis}{g}
\newcommand{\gvish}{h_g}
\newcommand{\gvishone}{h_{g1}}
\newcommand{\gvishtwo}{h_{g2}}
\newcommand{\gvisg}{\tilde{g}}
\newcommand{\gd}{g}
\newcommand{\gdh}{h}
\newcommand{\gdg}{\tilde{g}}
\newcommand{\symg}{{\tilde{g}}}
\newcommand{\rep}{M}
\newcommand{\gcop}{\hat{e}}
\newcommand{\pathgA}{\Gamma}
\newcommand{\pathgB}{\bar{\Gamma}}
\newcommand{\ket}[1]{\left|#1\right\rangle}
\newcommand{\bra}[1]{\left\langle#1\right|}

\begin{document}

\title{A classification of symmetry enriched topological phases with exactly solvable models}

\author{Andrej Mesaros}
\affiliation{Department of Physics,
Boston College, Chestnut Hill, MA 02467, USA}

\author{Ying Ran}
\affiliation{Department of Physics,
Boston College, Chestnut Hill, MA 02467, USA}

\begin{abstract}
Recently a new class of quantum phases of matter: symmetry protected topological states, such as topological insulators, attracted much attention. In presence of interactions, group cohomology provides a classification of these [X. Chen \textit{et al.}, arXiv:1106.4772v5 (2011)]. These phases have short-ranged entanglement, and no topological order in the bulk. However, when long-range entangled topological order is present, it is much less understood how to classify quantum phases of matter in presence of global symmetries. Here we present a classification of bosonic gapped quantum phases with or without long-range entanglement, in the presence or absence of on-site global symmetries. In 2+1 dimensions, the quantum phases in the presence of a global symmetry group $SG$, and with topological order described by a finite gauge group $GG$, are classified by the cohomology group $H^3(\SG \times\GG, U(1))$. Generally in d+1 dimensions, such quantum phases are classified by $H^{d+1}(\SG\times\GG, U(1))$. Although we only partially understand to what extent our classification is complete, we present an exactly solvable local bosonic model, in which the topological order is emergent, for each given class in our classification. When the global symmetry is absent, the topological order in our models is described by the general Dijkgraaf-Witten discrete gauge theories. When the topological order is absent, our models become the exactly solvable models for symmetry protected topological phases [X. Chen \textit{et al.}, arXiv:1106.4772v5 (2011)]. When both the global symmetry and the topological order are present, our models describe symmetry enriched topological phases. Our classification includes, but goes beyond the previously discussed projective symmetry group classification. Measurable signatures of these symmetry enriched topological phases, and generalizations of our classification are discussed.
\end{abstract}

\maketitle

\tableofcontents

\section{Introduction}

Recently there has been significant interest in topological phases of matter, which are quantum phases of matter beyond the Ginzburg-Landau symmetry-breaking description\cite{VLGinzburg:1950p7433}. After the discovery of fractional quantum Hall states, the notion of topological order was proposed\cite{Wen:1990p5870}. Topologically ordered phases of matter feature ground state degeneracies on torus\cite{Wen:1991p6377}, and anyonic quasiparticle excitations in the bulk in 2+1 dimensions\footnote{The phrase ``topological order'' has been widely used in literature, sometimes with different meanings. In this paper, we particularly reserve the phrase ``topological order'' for those quantum phases that feature nontrivial topological ground state degeneracies, and in 2+1 dimensions have quasiparticles with anyonic statistics.}. These features are robust against arbitrary local perturbations. Therefore, the global symmetry is NOT a requirement for topologically ordered quantum phases.

More recently, symmetry protected topological (SPT) phases have been discovered. SPT phases are defined to have no topological order in the bulk (and thus no anyons in the bulk nor ground state degeneracies on torus); nevertheless their distinctions are protected by the global symmetry\cite{Turner:2012p7511,Mong:2010p3871,Fu:2011p6126,Hughes:2011p7526,Liu:2012p7513,Liu:2012p7514,Gu:2012p6669,Wen:2012p7523,Slager:2012p7518}. Examples of SPT phases include topological insulators and superconductors\cite{RevModPhys.82.3045,Qi:2011p7509,Hasan:2011p7517,PhysRevB.78.195125,2009AIPC.1134...22K}. One experimental signature of the SPT phases are the symmetry protected gapless boundary states, which can be obtained from a Chern-Simons based classification of phases\cite{Lu:2012p7468}. Another classification of interacting bosonic SPT phases for on-site global symmetries is provided in the original work by Chen et al.\cite{Chen:2011p6670} using the group cohomology method. At the superficial level, there is no relation between the SPT phases and the topologically ordered phases. For example, in 1+1d, it can be shown that in the presence of interactions, there is no topological order but there are nontrivial SPT phases\cite{FDMHaldane:1983p7434,Kitaev:2012p7445,Fidkowski:2010p7507,Turner:2011p7498,Tang:2012p7515,Chen:2011p6670}, such as the AKLT\cite{Affleck:1987p7421,Pollmann:2012p7510} integer spin chain. A recent beautiful work\cite{Levin:2012p7190} shows that SPT phases and topologically ordered phases are related via certain duality in spatial dimensions higher than one.

How to understand/classify gapped quantum phases when both the topological order and the global symmetry are present? This is an important question for both fundamental and practical purposes, and this paper is an attempt to answer it, at least partially. To illustrate the importance of this question, as an example, we can consider one famous topologically ordered phase: the Laughlin's $\nu=1/3$ fractional quantum Hall liquid\cite{LAUGHLIN:1983p4122} (FQHL), which has three-fold ground state degeneracy on torus\cite{Wen:1990p5870} and anyonic quasiparticle excitations in the bulk. In the physical realization of the Laughlin FQHL in 2DEG, there is also a global symmetry: the $U(1)$ charge conservation for electrons. One can imagine what would happen if the $U(1)$ charge conservation was absent, for instance, if a small electronic pairing was introduced via proximity effect. Because the topological order is robust towards arbitrary perturbation, the three-fold ground state degeneracy and the anyonic statistics of quasiparticles would still be present.

Is the $U(1)$ global symmetry unimportant for the FQHL physics then? Obviously, this is not the case. In fact, this $U(1)$ symmetry allows one to find two striking experimental signatures of Laughlin's state: the quantized Hall conductance $\sigma_{xy}=e^2/3h$, and the $e^*=e/3$ fractional charge carried by quasiparticles. The second signature is very interesting: the quasiparticles of a topologically ordered phase can carry a fraction of the quantum number of the fundamental degrees of freedom (electrons here) in the quantum system. Such phenomena are often referred to as ``symmetry fractionalization''. This phenomena only occur when the system has topological order. The $e^*=e/3$ charge of quasiparticles is a remarkable demonstration of how the global symmetry can ``act'' on the topological order in a non-trivial fashion\cite{Yao:2010p7521,Kou:2008p7524,Kou:2009p7527}.

Another collection of fascinating quantum phases is the quantum spin liquid (QSL). Quantum spin liquids are often defined to be featureless Mott insulator phases, namely phases that respect full lattice symmetry as well as the $SU(2)$ spin rotational symmetry, with a half-integer spin per unit cell. Based on the Hastings' generalization\cite{Hastings:2004p7424} of Lieb-Schultz-Mattis theorem\cite{Lieb:1961p7423} in higher dimensions, we know that gapped quantum spin liquids in two and higher spatial dimensions must host non-trivial ground state degeneracies on torus. But because there is no symmetry-breaking-induced ground state degeneracy, this indicates that the gapped QSLs are topologically ordered.

How can one classify/understand QSL phases? For instance, recent numerical simulations\cite{Yan:2011p7425} point out that the spin-1/2 Heisenberg model on a Kagome lattice hosts a gapped quantum spin liquid phase. It is then an important issue to understand the nature of this QSL phase. As a matter of fact, numerical evidence for topological order described by a $Z_2$ gauge theory has been found\cite{Jiang_Kagome_entropy,PhysRevLett.109.067201}. Is this topological order enough to determine the nature of this QSL phase? The answer is negative. It turns out that there are more than one QSL phase on the Kagome lattice even for a given $Z_2$ topological order\cite{Lu:2011p7442,Wang:2006p6704,Misguich:2002p7441,Sachdev:1992p7443,Huh:2011p7439}. Their distinctions are protected by the global symmetries. Roughly speaking, the way that the global symmetries act on the topological order are different for different phases. These phenomena have been called ``symmetry enriched topological phases'' or ``symmetry enriched topological order''\cite{Maciejko:2010p7431,Swingle:2011p7432,Levin:2012p7429,Cho:2012p7426}. When the global symmetries are absent, all these phases are no longer distinguishable and are adiabatically connected to one another. But when the global symmetries are present, one necessarily encounters phase transitions while going from one phase to another. Therefore, for the Kagome lattice gapped spin liquid example, it remains an unresolved issue to understand which among all the symmetry enriched topological phases is the one found in the numerical simulations.

The above physical examples motivate us to consider the following questions: How are symmetry enriched topological (SET) phases generally classified? Or, how can one classify different ways in which the global symmetry ``acts'' on the topological order? What are the experimental/numerical signatures of different SET phases? The last question is quite urgent for the above Kagome QSL example: although there are nice numerical methods (e.g., the topological entanglement entropy\cite{Kitaev:2006p7455,Levin:2006p7456}) to detect the $Z_2$ topological order\cite{Furukawa:2007p7454,PhysRevLett.109.067201,Zhang:2011p7452,Isakov:2011p7453}, due to the lack of theoretical understanding it is still unknown how to numerically distinguish different SET phases.

This paper attempts to address these questions to a certain level. We consider on-site global symmetries only; namely the global symmetry transformation is a direct product of unitary transformations, and each transformation only acts in the local Hilbert space. In addition, we focus on bosonic systems with finite unitary symmetry groups $\SG$ and topological orders that can be described by finite gauge groups $\GG$. Generalizations of these conditions will be discussed at the end of the paper. Under these assumptions, we propose that gapped bosonic quantum phases with $SG$ and $GG$ are classified by group coholomogy $H^3(SG\times GG,U(1))$ in 2+1 dimensions, and generally $H^{d+1}(SG\times GG,U(1))$ in $d$+1 dimensions ($d\ge 2$). Here ``$\times$'' is the direct product (or the cross product) of two groups, and we will explain the notion of group cohomology shortly.

Let's consider some special limits of our classification. When the system does not have topological order, $GG=Z_1$, our classification becomes $H^{d+1}(SG,U(1))$. This in fact goes back to the group cohomology classification of SPT phases\cite{Chen:2011p6670}. When the system does not have global symmetry, $SG=Z_1$, our classification becomes $H^{d+1}(GG,U(1))$. In 2+1 dimensions, this coincides with the Dijkgraaf-Witten classification\cite{Dijkgraaf:1990p7194} of topological quantum field theories with discrete gauge groups.

When both the $SG$ and the $GG$ are non-trivial, we will show that the indices of the classification $H^{d+1}(SG\times GG,U(1))$ can be expanded as:
\begin{align}
 H^{d+1}&(SG\times GG,U(1))=H^{d+1}(SG,U(1))\notag\\
&\times H^{d+1}(GG,U(1))\times SET(SG,GG),
\end{align}
where $SET(SG,GG)$ will be introduced later. $SET(SG,GG)$ describes the non-trivial interplay between the topological order and the global symmetry, and classifies the symmetry enriched topological phases.

Some detectable signatures of SET phases, for example, the symmetry protected degeneracy of excited states, are studied in this paper. We leave the general numerical/experimental signatures of SET phases as a subject for future investigation. Nevertheless, we provide exactly solvable local bosonic models for every phase in our classification, in which the topological order is emergent. These models would be useful tools to further study the properties of these phases, including detectable signatures.

The plan of this paper is as follows. In Sec.\ref{sec:classification}, we provide the mathematical background of our classification. We review a previously known partial classification of SET phases: the projective symmetry group (PSG), and comment on the general notion of ``symmetry fractionalization''. In particular, we show that our classification includes the mathematical structure underlying the PSG classification in 2+1 dimensions, and goes beyond it. Namely, our classification contains phases that are \emph{not} described by the PSG. In 3+1 dimensions, our classification becomes very different from the PSG classification and we will explain the reason in Sec.\ref{sec:conclusion}. In Sec.\ref{sec:model}, we focus on 2+1 dimensions and present the geometric interpretation of group cohomology, leading to a class of exactly solvable models. Each model corresponds to a phase in our classification. Generalizations to higher dimensions will be briefly discussed. Staying in 2+1 dimensions, in Sec.~\ref{sec:elem-excit} we study the elementary excitations of these models, namely gauge fluxes and charges, by introducing string-like operators. In Sec.\ref{sec:example}, we will solve these models in 2+1 dimensions for some illuminating examples. One particularly important case is the simplest example that is NOT described by the PSG classification nor ``symmetry fractionalization''. In that example the global symmetry transformation interchanges the quasiparticle species. Detectable signatures of these examples will be studied. In Sec.\ref{sec:conclusion} we consider generalizations of our study, comment on relations with previous work, and conclude.

\section{The Classification}\label{sec:classification}

\subsection{Mathematical preparation}\label{sec:math}
\subsubsection{Definition of the cohomology group}

We begin with a brief introduction to group cohomology. A detailed introduction can be found in Ref.\cite{Chen:2011p6670}; in this paper, we will not present the most general definition of group cohomology.

For a finite group $G$, and an abelian group $M$ ($M$ does not need to be finite or discrete), one can consider an arbitrary function that maps n elements of $G$ to an element in $M$; $\omega: G^n\rightarrow M$ or equivalently $\omega(g_1,g_2,...,g_n)\in M$, $\forall g_1,g_2,...g_n\in G$. Such a group function is called an n-cochain. The set of all n-cochains, which is denoted as $C^n(G,M)$, forms an abelian group in the usual sense: $(\omega_1\cdot\omega_2)(g_1,g_2,...,g_n)=\omega_1(g_1,g_2,...,g_n)\cdot \omega_2(g_1,g_2,...,g_n)$, in which the identity n-cochain is a group function whose value is always the identity in $M$.

One can define a mapping $\delta$ from $C^n(G,M)$ to $C^{n+1}(G,M)$: $\forall \omega\in C^n(G,M)$, define $\delta \omega \in C^{n+1}(G,M)$ as
\begin{align}\label{eq:n-cocycle}
\delta &\omega (g_1,...,g_{n+1})=\omega(g_2,...,g_{n+1})\cdot\omega^{(-1)^{n+1}}(g_1,...,g_n)\notag\\
&\times \prod_{i=1}^n \omega^{(-1)^i}(g_1,..,g_{i-1},g_i\cdot g_{i+1},g_{i+1},..,g_{n+1}).
\end{align}
It is easy to show that the mapping $\delta$ is nilpotent: $\delta^2 \omega=1$ (here $1$ denotes the identity (n+2)-cochain). In addition, for two n-cochains $\omega_1,\omega_2$, obviously $\delta$ satisifies $\delta(\omega_1\cdot\omega_2)=(\delta\omega_1)\cdot(\delta\omega_2)$.

An n-cochain $\omega(g_1,...g_n)$ is called an n-cocyle if and only if it satisfies the condition: $\delta\omega =1$, where $1$ is the identity element in $C^{n+1}(G,M)$. When this condition is satisfied, we also say that $\omega(g_1,...g_n)$ is an n-cocycle of group $G$ with coefficients in $M$. The set of all n-cocycles, denoted by $Z^n(G,M)$, forms a subgroup of $C^{n}(G,M)$.

Not all different cocyles are inequivalent. Below we define an equivalence relation in $Z^n(G,M)$. Because $\delta$ is nilpotent, for any (n-1)-cochain $c(g_1,...,g_{n-1})$, we can find the n-cocyle $\delta c$. And if an n-cocyle $b$ can be represented as $b=\delta c$, for some $c\in C^{n-1}(G,M)$, $b$ is called an n-coboundary. The set of all n-coboundaries, denoted by $B^n(G,M)$, forms a subgroup of $Z^n(G,M)$. Two n-cocycles $\omega_1,\omega_2$ are equivalent (denoted by $\omega_1\sim\omega_2$) if and only if they differ by an n-coboundary: $\omega_1=\omega_2\cdot b$, where $b\in B^n(G,M)$.

The n-th cohomology group of group $G$ with coefficients in $M$, $H^{n}(G,M)$, is formed by the equivalence classes in $Z^n(B,M)$. More precisely: $H^{n}(G,M)=Z^n(G,M)/B^n(G,M)$.

In this paper we will make a lot of use of 3-cocycles $\omega$. We will always choose them to be in ``canonical'' form, which means that $\omega(g_1,g_2,g_3)=1$ if any of $g_1,g_2,g_3$ is equal to $\openone$ (the identity element of group $G$). For any of the inequivalent cocycles mentioned above, it is always possible to choose a gauge such that $\omega$ becomes canonical~\cite{Chen:2011p6670}. Specifically, the explicit elementary cocycles that we will use in studying examples of our models in Section~\ref{sec:example} are going to be canonical.

So far the notions of cocycle and cohomology group are quite formal. But it turns out that they have clear geometric/topological meanings, which we will describe in Sec.\ref{sec:model}.

\subsubsection{Examples}\label{sec:math_example}
$\mathbf{H^1(G,U(1))}$ \textbf{and one-dimensional representations of groups}: Let's consider the first cohomology group of a finite group $G$ with coefficients in $U(1)$: $H^1(G,U(1))$. In this case, the cocycle condition becomes:
\begin{align}
 \omega(g_1)\cdot\omega(g_2)/\omega(g_1\cdot g_2)=1
\end{align}
This means the 1-cocycle $\omega(g)$ is a one-dimensional unitary representation of the group $G$. And clearly different 1-cocycles are different representations. A 0-cochain is defined to be a constant $c_0\in U(1)$. Consequently, a 1-cocycle is a 1-coboundary if and only if it is identity: $\omega(g)=c_0/c_0=1$. We conclude that the $H^1(G,U(1))$ is formed by inequivalent one-dimensional unitary representations of $G$.

For instance: 
\begin{align}
 H^1(Z_n,U(1))=Z_n,\;\;H^1(Z_n^k,U(1))=Z_n^k.
\end{align}
More generally, for any finite abelian group $G$, due to a fundamental theorem, we know that $G$ can be decomposed as $G=Z_{n_1}^{k_1}\times Z_{n_2}^{k_2}...\times Z_{n_q}^{k_q}$. Because we know the one-dimensional representations of all the components, clearly, \begin{align}
H^1(G,U(1))=G, \mbox{ $\forall$ finite abelian $G$.} \label{eq:abelianH1}
\end{align}

$\mathbf{H^1(G,Z)=Z_1}$, \textbf{ $\mathbf{\forall}$ finite $\mathbf{G}$}: Following the above discussion, $H^1(G,Z)$ is formed by the group of all group homomorphisms from $G$ to $Z$. It is straightforward to show that the only group homomorphism between a finite $G$ and $Z$ is the trivial one.

$\mathbf{H^2(G,U(1))}$ \textbf{and projective representations of groups}: The condition for 2-cocycles is:
\begin{align}
 \omega(g_1,g_2)\cdot \omega(g_1\cdot g_2,g_3)=\omega(g_2,g_3)\cdot \omega(g_1,g_2\cdot g_3)\label{eq:2-cocycle}
\end{align}
In fact, 2-cocycle is related to the so-called projective representations of groups. In usual unitary group representations, each group element $g$ in $G$ is represented by a unitary matrix $D(g)$, which satisfies: $D(g_1)\cdot D(g_2)=D(g_1\cdot g_2)$. But for projective representations, this relation can be modified by a phase factor $\omega(g_1,g_2)\in U(1)$: $D(g_1)\cdot D(g_2)=\omega(g_1,g_2)D(g_1\cdot g_2)$. And the phase factor $\omega(g_1,g_2)$, which is a function of $g_1,g_2$, is called a factor system. A factor system cannot be arbitrary. In order to satisfy the associativity condition: $[D(g_1)\cdot D(g_2)]\cdot D(g_3)=D(g_1)\cdot [D(g_2)\cdot D(g_3)]$, the factor system must satisfy Eq.(\ref{eq:2-cocycle}) --- the same condition as for 2-cocycles.

What is a 2-coboundary? A 2-coboundary $\omega(g_1,g_2)$ can be written as $\omega(g_1,g_2)=c(g_1)\cdot c(g_2)/c(g_1\cdot g_2)$ for a certain 1-cochain $c(g)$. If two 2-cocyles, $\omega_1,\omega_2$ , differ by a 2-coboundary:
\begin{equation}
  \label{eq:gauge2coc}
  \omega_1(g_1,g_2)=\omega_2(g_1,g_2)\cdot\frac{ c(g_1)\cdot c(g_2)}{c(g_1\cdot g_2)},
\end{equation}
it is obvious that they correspond to equivalent projective representations, because one can absorb the 1-cochain into $D(g)$ by redefining $\tilde D(g)=c(g)\cdot D(g)$, after which the two factor systems becomes the same (this is actually the definition of equivalent projective representations). We conclude that $H^2(G,U(1))$ classifies all inequivalent factor systems of projective representations.

The calculation of $H^2(G,U(1))$ is non-trivial. We list some useful results:
\begin{align}
 &H^2(Z_n,U(1))=Z_1, \;\; H^2(Z_n^k,U(1))=Z_n^{k(k-1)/2}, \notag\\
&H^2(Z_n\times Z_m,U(1))=Z_{gcd(n,m)},
\end{align}
where $Z_1$ is the trivial group.

$\mathbf{H^3(G,U(1))}$. The 3-cocycle condition is:
\begin{align}
 &\omega(g_1,g_2,g_3)\cdot\omega(g_2,g_3,g_4)\cdot \omega(g_1,g_2\cdot g_3,g_4)\notag\\
=&\omega(g_1\cdot g_2, g_3,g_4)\cdot \omega(g_1,g_2, g_3\cdot g_4),\label{eq:3-cocycle}
\end{align}
and a 3-cocyle $\omega$ is a 3-coboundary iff it can be represented as
\begin{align}  \label{eq:gauge3coc}
 \omega(g_1,g_2,g_3)=\frac{c(g_2,g_3)\cdot c(g_1, g_2\cdot g_3)}{c(g_1,g_2)\cdot c(g_1\cdot g_2,g_3)}.
\end{align}

These equations may look strange. But after we introduce a geometric interpretation in Sec.\ref{sec:model}, their meanings will become clear.

We list some useful results for $H^3(G,U(1))$:
\begin{align}
 &H^3(Z_n,U(1))=Z_n,\notag\\
 &H^3(Z_n^k,U(1))=Z_n^{k+k(k-1)/2!+k(k-1)(k-2)/3!}.
\end{align}
For instance, $H^3(Z_2^2,U(1))=Z_2^3$ and $H^3(Z_2^3,U(1))=Z_2^7$.  

\subsubsection{Some useful theorems}
First, it is known that for any finite group $G$, its every n-th cohomology group with $n>0$ is a finite Abelian group. Below we list a couple of theorems on group cohomology that will be used in the following.

\textbf{Universal coefficients theorem:} This theorem relates cohomology groups with different coefficients:
\begin{align}
 H^n(G,\mathbf{B})=[H^n(G,Z)\otimes \mathbf{B}]\times \mbox{Tor}(H^{n+1}(G,Z),\mathbf{B}).
\end{align}
This formula allows one to compute cohomology groups with coefficients in some Abelian group $\mathbf{B}$ by using the cohomology groups with coefficients in the group of integers $Z$.

Here ``$\times$'' is the usual direct product of groups, and we need to define the two new operations: ``$\otimes$'' and ``$\mbox{Tor}$''. ``$\otimes$'' stands for the ``symmetric tensor product'' (over $Z$) between two abelian groups; while ``$\mbox{Tor}$'' stands for the ``torsion product''. 

Instead of explaining the rigorous mathematical definitions of these products, we simply list some useful results. $A\otimes B$ always equals $B\otimes A$ (up to isomorphism), and
\begin{align}
 Z_n\otimes Z_m&=Z_{gcd(n,m)},\notag\\
Z_n\otimes Z&=Z_n,\notag\\
Z_n\otimes U(1)&=Z_1,\notag\\
Z\otimes U(1)&=U(1),\notag\\
Z\otimes Z&=Z,\notag\\
(A\times B)\otimes C&=(A\otimes C)\times(B\otimes C)
\end{align}
The last relation means that $\otimes$ is distributive.

Concerning the torsion product, one also has $\mbox{Tor}(A,B)=\mbox{Tor}(B,A)$. In addition,
\begin{align}
\mbox{Tor}(Z_n,Z_m)&=Z_{gcd(n,m)},\notag\\
\mbox{Tor}(Z_n,U(1))&=Z_{n},\notag\\
\mbox{Tor}(A,Z)&=Z_1, \;\;\;\forall A,\notag\\
\mbox{Tor}(A\times B, C) & =\mbox{Tor}(A,C)\times \mbox{Tor}(B,C).
\end{align}
The torsion product is also distributive (the last relation above).

Using the universal coefficients theorem, one can compute the cohomology groups with coefficients in $Z$ from those with coefficients in $U(1)$, and vice versa:
\begin{align}
 &H^n(G,U(1))\notag\\
=&[H^n(G,Z)\otimes U(1)]\times \mbox{Tor}(H^{n+1}(G,Z),U(1))\notag\\
=&H^{n+1}(G,Z), \;\;\mbox{for $n>0$},\label{eq:Z_U1}
\end{align}
where we used the fact that $H^n(G,Z)$ is a finite abelian group for $n>0$ so that $[H^0(G,Z)\otimes U(1)]=Z_1$. Note that the above equation is invalid if $n=0$. At this moment, let's define the 0th cohomology group. In this paper, $H^0(G,M)=M$. Therefore $[H^0(G,Z)\otimes U(1)]=U(1)\neq Z_1$.

Using Eq.(\ref{eq:Z_U1}) and Eq.(\ref{eq:abelianH1}), we have:
\begin{align}
 H^2(G,Z)=G, \mbox{ $\forall$ finite abelian $G$.}
\end{align}

\textbf{The K\"{u}nneth formula}: This theorem allows one to compute the cohomology group of a direct product of groups, using the cohomology groups of its components:
\begin{align}
 H^n(A\times &B, Z)=\prod_{i=0}^{n}[H^i(A,Z)\otimes H^{n-i}(B,Z)]\notag\\
&\times \prod_{i=0}^{n+1} \mbox{Tor}(H^i(A,Z),H^{n+1-i}(B,Z)),
\end{align}
where ``$\prod$'' is the usual direct product. For example, this formula and the following basic results:
\begin{align}\label{eq:HpZnZ}
 H^p(Z_n,Z)= \left\{ \begin{array}{ll}
         Z_n & \mbox{if $p$ is even};\\
         0 & \mbox{if $p$ is odd};\\
Z &\mbox{if $p=0$},
\end{array} \right.
\end{align}
together with Eq.(\ref{eq:Z_U1}), allow one to obtain the results for $H^p(Z_n^k,U(1))$ listed previously.

\subsection{The notion of symmetry fractionalization and the Projective Symmetry Group}\label{sec:psg}

The main goal of this paper is to address the non-trivial interplay between global symmetry and topological order. What is already known about that interplay? One phenomenon famously connected to such an interplay is the so-called ``symmetry fractionalization''.

Topologically ordered phases feature anyonic quasiparticle excitations in the bulk. In fact, in some sense these quasiparticles are non-local, because one cannot create a single quasiparticle excitation in a system with periodic boundary conditions (PBC). One must at least create a pair: a quasiparticle and its anti-quasiparticle. Therefore, a single anyonic quasiparticle state is not in the excitation spectrum of a system with PBC.

The fact that a single anyonic quasiparticle excitation is not a physical excitation in a system with PBC has an important physical consequence. \footnote{We consider PBC systems for simplicity. For open boundary systems, one can put quasiparticles on the edge and there can be only one quasiparticle in the bulk. But this does not modify the argument below.} When the system has a certain global symmetry group SG, based on quantum mechanics, we know that all excitated states of the quantum system can be labeled by irreducible representations (irreps) of SG. The irreps characterize how the ground state and excited states transform under the global symmetry. Then it is natural to imagine that each anyonic quasiparticle also has to transform as a certain irrep of SG. But this does not need to be true, exactly because a single quasiparticle is not a physical excitated state!

Let's consider a famous example, the $\nu=1/3$ Laughlin state. There the physical system has a $U(1)$ charge conservation symmetry, and therefore any physical state should be an irrep of this $U(1)$: $|\psi\rangle\rightarrow e^{im\theta}|\psi\rangle$, for $\forall e^{i\theta}\in U(1)$. Here the integer $m\in Z$ is nothing but the total electric charge of the state. But we were also told that the anyonic quasiparticle carries $1/3$, a non-integer, electric charge. A fractional charge is NOT an irrep of the global symmetry.

Clearly, the fractional charge of an anyonic quasiparticle can be realized exactly because the single anyonic quasiparticle is not a physical state. Only when there are three (generally multiple of three) quasiparticles in the bulk can it be a physical state, which carries one more electric charge compared to the ground state. This phenomenon is called symmetry fractionalization.

In this example, we can ask a further question: why the quasiparticles have to carry 1/3, not 1/5, or some other fraction of electric charge? Or, what is the guiding principle that dictates this fractional charge?

One obvious guiding principle is the fusion rule. We know that three quasiparticles become an electron after fusion, which must carry electric charge one. Consequently each quasiparticle must carry 1/3 charge. In fact, this point of view is conceptually very general. For example, one may even be able to consider topologically ordered phases with non-abelian quasiparticles. However, the mathematical framework behind this point of view, for the most general topologically ordered phases, is technically highly nontrivial\footnote{The general topological orders are mathematically described by tensor category theories. See, for instance, Ref.\onlinecite{Kitaev:2006p6266,Levin:2005p3468} for detailed discussions.}, and is beyond the scope of this paper.

In this paper, we choose a different point of view, which involves a simpler mathematical framework --- projective representations of symmetry group. The trade-off will be that we can only use this point of view to understand symmetry fractionalization in certain subclasses of topologically ordered phases\footnote{More precisely, we discuss projective representations determined by a gauge group, so the topological order is described by a discrete gauge group. However, symmetry fractionalization can be studied in topologically ordered phases beyond this subclass.}. However, this is enough for the purpose of this paper.

The point of view that we choose is the following. Because only multiples of three quasiparticles correspond to physical states, we can define a so-called Invariant Gauge Group (IGG): $IGG=Z_3=\{1,e^{i2\pi/3},e^{i4\pi/3}\}$. We can multiply each quasiparticle in the system by a fixed element in IGG. Clearly the total phase becomes unity and the physical wavefunction is not modified.

This IGG tells us that when we implement the global $U(1)$ transformation on each quasiparticle, it is perfectly fine to have a phase ambiguity, if and only if (iff) this phase ambiguity is an element in IGG --- because this ambiguity does not modify the physical state at all. Therefore, a single quasiparticle does not have to form an irrep of SG, but it can form a so-called projective representation of SG with coefficients in IGG. Formally, this means that a single quasiparticle can transform under the global $U(1)$ as: $\psi_{qp}\rightarrow D(e^{i\theta})\psi_{qp}$, $\forall e^{i\theta}\in U(1)$, where $D(e^{i\theta})$ only needs to be a projective representation:
\begin{align}
 D(e^{i\theta_1})\cdot D(e^{i\theta_2})=\omega(e^{i\theta_1},e^{i\theta_2})D(e^{i(\theta_1+\theta_2)}),
\end{align}
where $\omega(e^{i\theta_1},e^{i\theta_2})\in IGG$ (this is why we say that the projective representation has coefficients in IGG). In addition, as we learned in Sec.\ref{sec:math_example}, the associativity condition is satisfied iff $\omega(e^{i\theta_1},e^{i\theta_2})\in H^2(U(1),IGG)$ --- a 2-cocycle of $U(1)$ group with coefficients in IGG. 

In fact, the 1/3 electric charge (in general $n/3$ fractional charge with $n$ being an integer) is exactly a projective representation of $U(1)$ with coefficients in IGG. One can check it explicitly: Let's define the transformation law of a single quasiparticle under $U(1)$ as $D(e^{i\theta})=e^{i\theta/3}, \forall\theta\in[0,2\pi)$. Clearly $\omega(e^{i\theta_1},e^{i\theta_2})=e^{i2\pi/3}$, iff $\theta_1+\theta_2\ge 2\pi$, $\forall \theta_{1,2}\in[0,2\pi)$, and $\omega(e^{i\theta_1},e^{i\theta_2})=1$ otherwise. This $\omega(e^{i\theta_1},e^{i\theta_2})$ is a 2-cocycle $\in H^2(U(1),IGG)$, because the associativity condition is obviously satisified by $D(e^{i\theta})$.

In this example, we learned that a quasiparticle of a topologically ordered phase can transform under the global symmmetry group $SG$ as a projective representation of $SG$ with coefficients in a certain abelian group $IGG$. This point of view is also quite general and is enough to characterize symmetry fractionalization in this paper. Actually, we will use this point of view to classify symmetry fractionalization. Firstly, we comment on our choice of notation.

In this paper, ``symmetry fractionalization'' is a phrase reserved to characterize how the global symmetry is implemented \emph{locally} on a single quasiparticle. Here ``locally'' is the key word. It basically means that when we claim symmetry fractionalization, we already made a basic assumption --- that the global symmetry transformations can indeed be implemented by local transformations of each quasiparticle.

\begin{figure}
\includegraphics[width=0.35\textwidth]{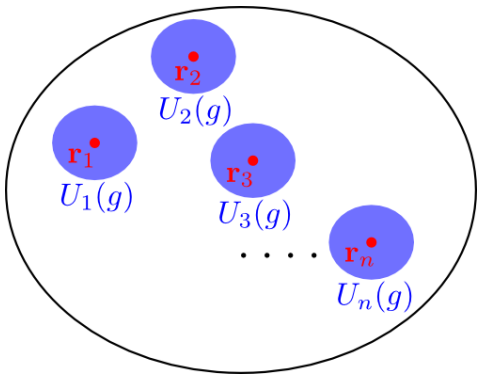}
\caption{Illustration of the basic assumption of symmetry fractionalization Eq.(\ref{eq:s_f_assumption}): Under a symmetry transformation $U(g)$ with $\forall g\in SG$, an excitated state is transformed by the product of local transformation operators $U_i(g)$, with each operator only acting on one quasiparticle locally.}
\label{fig:s_f_assumption}
\end{figure}

\textbf{The Basic Assumption of Symmetry Fractionalization:} Consider an excitated state of a topologically ordered phase with a global symmetry group $SG$, having $n$-quasiparticles (which do not have to be of the same type) spatially located at positions $\mathbf{r}_1,\mathbf{r}_2...\mathbf{r}_n$, far apart from one another. Let's denote this state by $|\psi(\mathbf{r}_1,\mathbf{r}_2,...\mathbf{r}_n)\rangle$. For any symmetry transformation $U(g)$ by a group element $g\in SG$, clearly $U(g)$ will generally  transform this state into another state: $U(g):|\psi(\mathbf{r}_1,\mathbf{r}_2,...\mathbf{r}_n)\rangle\rightarrow |\tilde\psi(\mathbf{r}_1,\mathbf{r}_2,...\mathbf{r}_n)\rangle$. The basic assumption of symmetry fractionalization is that there exist local operators $U_1(g),U_2(g),...,U_n(g)$, such that $U_i(g)$ is a local operator acting only in a finite region around the spatial position $\mathbf{r}_i$, and does not touch the other quasiparticles; in addition, $U_1(g),U_2(g),...,U_n(g)$ satisfy:
\begin{align}
 &U_1(g)\cdot U_2(g)\cdots U_n(g)|\psi(\mathbf{r}_1,\mathbf{r}_2,...\mathbf{r}_n)\rangle\notag\\
=&U(g)|\psi(\mathbf{r}_1,\mathbf{r}_2,...\mathbf{r}_n)\rangle=|\tilde\psi(\mathbf{r}_1,\mathbf{r}_2,..,\mathbf{r}_n)\rangle\label{eq:s_f_assumption}
\end{align}
Pictorially, this assumption is shown in Fig.\ref{fig:s_f_assumption}.

Wen\cite{Wen:2002p162} first attempted to classify symmetry fractionalization while investigating the parton mean-field states of quantum spin liquids in presence of global symmetries such as lattice space group symmetries, time-reversal symmetry, and spin-rotation symmetry. In Ref.\onlinecite{Wen:2002p162}, Wen introduced the notion of Projective Symmetry Group (PSG), a very useful tool to classify parton states, as well as the low energy gauge fluctuations, in the presence of these global symmetries. We will not provide a detailed review the PSG classification here, however, we introduce the core mathematical structure underlying the PSG, and comment on the connection between PSG and the current work, in Sec. \ref{sec:connection}. Also, a brief introduction to the parton construction and PSG can be found in Appendix \ref{app:PSG}.

One way to understand PSG is via the low energy effective theory of a state with topological order. Let's consider the following situation: there is a quantum state whose topological order is described by a gauge group $GG$. Therefore, there are gauge charge excitations of the gauge group $GG$ in the system, which are anyonic quasiparticles. In order to write down an effective theory in terms of these gauge charge quasiparticles, it is crucial to understand how they transform under the global symmetry $SG$. Similarly to the usual Ginzburg-Landau theory, any local term that is allowed by global symmetry will appear in the effective theory. Naively, one may expect that the gauge charges must form representations of the symmetry group $SG$.

However, Wen pointed out\cite{Wen:2002p162} that these gauge charges do not have to form representations of the SG. Instead, gauge charges transform under a larger group: the PSG. The relation between PSG, SG and GG is given by:
\begin{align}
 PSG/GG=SG.\label{eq:PSG}
\end{align}
Mathematically, $PSG$ is a group extension of $SG$ by the group $GG$. From here on to the end of this section, we assume for simplicity that $GG$ is a finite abelian group. In this case, $PSG$ can be shown to be the central extension (i.e., $GG$ is in the center of $PSG$) of $SG$ by $GG$ (see Appendix \ref{app:PSG}). The classification of how gauge charge quasiparticles transform under the SG becomes the classification of all inequivalent central extensions of the group. There is a nice mathematical theorem (see, for example, Ref\cite{robinson1996coursetheorygroups}) stating that all inequivalent central extensions of the group $SG$ by $GG$ are classified by $H^2(SG,GG)$.

The mathematical structure $H^2(SG,GG)$ underlying the PSG classification, which has been independently observed by several people\footnote{A. Kitaev in Ref.\onlinecite{Kitaev:2006p7455}, Ying Ran and Xiao-Gang Wen, unpublished (2002), Michael Hermele, private communication (2012) and in Ref.\onlinecite{Hermele_2012}}, is somewhat mysterious at this moment. But in fact its physical meaning can be easily understood. To proceed, again for simplicity, let's assume GG has the form $GG=Z_n\times Z_m$, and one can easily generalize the following discussion to any finite abelian gauge group. In this case, in order to understand the symmetry transformations of the gauge charges, we only need to consider two fundamental gauge charge excitations: $\psi_1$ and $\psi_2$, which carry gauge charge $(1,0)$ and $(0,1)$, respectively. (Note that we adopt the notation $(a,b)$ to label gauge charge here, $0\leqslant a<n,0\leqslant b<m$.) This is sufficient because one can build any gauge charge quasiparticle by fusing $\psi_1$ and $\psi_2$.

What are the most general possible ways in which $\psi_1$ and $\psi_2$ can transform under SG? This is a big question and we will attempt to provide an answer later in this paper. In this section, however, let's consider a smaller question: Under the assumption of symmetry fractionalization, what are the most general possible ways in which $\psi_1$ and $\psi_2$ transform under SG?

Under this assumption, symmetry transformations of quasiparticles are realized by local operators, which cannot change the quasiparticle's species (or more precisely, the superselection sector of a quasiparticle). Therefore, the gauge charge will be invariant under SG transformation: $\psi_1$ only transforms into $\psi_1$ while $\psi_2$ only transforms into $\psi_2$. However, similarly to the situation with fractional charge in fractional quantum Hall states discussed above, $\psi_1$ (or $\psi_2$) does not need to form a representation of SG. This is because any excited states with PBC must contain a multiple of $n$ number of $Z_n$ gauge charges $\psi_1$, and a multiple of $m$ number of $Z_m$ gauge charges $\psi_2$. Consequently, when we define symmetry transformations of $\psi_1$ ($\psi_2$), it is perfectly fine to have a phase $e^{i2\pi\frac{k_1}{n}}$ ($e^{i2\pi\frac{k_2}{m}}$) ambiguity. The state $\psi_1$ ($\psi_2$) only needs to form a projective representation of SG with coefficients in the $Z_n$ ($Z_m$) subgroup of $U(1)$, which is exactly classified by $H^2(SG,Z_n)$ ($H^2(SG,Z_m)$). Finally, because we can pair up any two transformation laws of $\psi_1$ and $\psi_2$, the symmetry transformations of gauge charge quasiparticles with $GG=Z_n\times Z_m$ are classified by $H^2(SG,Z_n)\times H^2(SG,Z_m)$.

Based on the universal coefficients theorem, it is straightforward to show that:
\begin{align}
 H^2(SG,Z_n \times Z_m)=H^2(SG,Z_n)\times H^2(SG,Z_m),
\end{align}
which is exactly the mathematical structure $H^2(SG,GG)$ underlying the PSG classification. 

Through this example, we learned that $H^2(SG,GG)$ is a classification of different ways in which anyonic quasiparticles transform under the global symmetry $SG$, under the assumption of symmetry fractionalization. Therefore, in this paper we will refer to $H^2(SG,GG)$ as the symmetry fractionalization classification, and the classes contained in $H^2(SG,GG)$ as the symmetry fractionalization classes.

There is one important point that we have not mentioned. We have shown that $H^2(SG,GG)$ classifies how gauge charges transform under SG. But we also know that there are other quasiparticle excitations in the system, such as gauge flux excitations. For instance, in a $GG=Z_2$ topologically ordered phase, there are three species of non-trivial quasiparticles: $Z_2$ gauge charge $e$, $Z_2$ gauge flux $m$, and their bound state $em$. In a usual topologically ordered state described by an abelian gauge group $GG$, the gauge charges and gauge fluxes are dual to each other in 2+1 dimensions. For instance, it does not matter if one labels $e$ or $m$ as the gauge charge in the usual $Z_2$ gauge theory.

In fact, the above discussion indicates that $H^2(SG,GG)$ is only a classification of symmetry fractionalization for $GG$ gauge charges (or gauge fluxes) only, but not for both gauge charges and gauge fluxes. That means that the full classification of symmetry fractionalization should go beyond $H^2(SG,GG)$. However, we will see shortly in Sec.\ref{sec:connection} that our classification of symmetry enriched topological phases only contains $H^2(SG,GG)$. In addition, in our exactly solvable models, we will show that this $H^2(SG,GG)$ only corresponds to the symmetry fractionalization of the gauge fluxes. It turns out that in these exactly solvable models, the gauge charges always have trivial symmetry fractionalization.\footnote{One may wonder how we fix our convention --- because for a usual finite abelian gauge theory $GG$ in 2+1 dimensions, the gauge charges and gauge fluxes are self-dual. In fact, for a general gauge theory with finite gauge group $GG$, the gauge fluxes and gauge charges are physically distinct. For example, gauge fluxes are labeled by conjugacy classes of $GG$, while gauge charges are labeled by the irreducible representations of the centralizer group of a conjugacy class. We will show that in our exactly solvable models, it is clear that the $H^2(SG,GG)$ classification corresponds to the symmetry fractionalization classes of the gauge fluxes.} We will comment on this issue in Sec.\ref{sec:connection}, and in Sec.\ref{sec:conclusion}.

Now let's consider some simple examples to see the power of $H^2(SG,GG)$. For the reason mentioned in the previous paragraph, in the following examples we describe $H^2(SG,GG)$ as the symmetry fractionalization classifcation of the \emph{gauge fluxes}.
\begin{enumerate}
 \item{$GG=Z_2$, and $SG=Z_2$.}

Let's denote the generator of SG as $\sigma$, and denote by $D_m(\sigma)$ the transformation of the $Z_2$ gauge flux $m$ by $\sigma$. Because $SG=Z_2$, we have:
\begin{align}
 \sigma^2=1. 
\end{align}
The universal coefficients theorem allows us to compute:
\begin{align}
 H^2(SG,GG)=H^2(Z_2,Z_2)=Z_2.
\end{align}
It means that there are two symmetry fractionalization classes of the $Z_2$ gauge flux. They correspond to:
\begin{align}
 D_m(\sigma)^2=\pm 1.\label{eq:sgZ2_ggZ2}
\end{align}
These two possible signs are exactly the two inequivalent cocycles in $H^2(Z_2,Z_2)$. The positive sign is the trivial symmetry fractionalization class, while the negative sign is the non-trivial class.

\item{$GG=Z_2^2$, and $SG=Z_2$}

Let's denote the generator of SG by $\sigma$. Now there are two fundamental $Z_2$ gauge fluxes: $m_1$, the $\pi$-flux in the first $Z_2$ gauge group, and $m_2$, the $\pi$-flux in the second $Z_2$ gauge group.
Straightforward computation gives:
\begin{align}
 H^2(SG,GG)=H^2(Z_2,Z_2^2)=Z_2^2.
\end{align}
There are 4 classes. The corresponding transformations of the $Z_2$ gauge fluxes $m_1,m_2$, denoted by $D_{m_1}(\sigma),D_{m_2}(\sigma)$, satisfy:
\begin{align}
 D_{m_1}(\sigma)^2=\pm 1,\;\;\;D_{m_2}(\sigma)^2=\pm 1.\label{eq:sgZ2_ggZ22}
\end{align}

\item{$GG=Z_2$, and $SG=Z_2^2$}

Let's denote the two generators of SG by $\sigma,\tau$. Because $SG=Z_2^2$, we have:
\begin{align}
 \sigma^2&=1,&\tau^2&=1,&\sigma\tau=\tau\sigma.
\end{align}
Straightforward computation gives:
\begin{align}
 H^2(SG,GG)=H^2(Z_2^2,Z_2)=Z_2^3.
\end{align}
There are 8 classes. The corresponding transformations of the $Z_2$ gauge flux $m$, denoted by $D_m(\sigma)$ and $ D_m(\tau)$, satisfy:
\begin{align}
 &D_m(\sigma)^2=\pm 1,\;\;\;\;D_m(\tau)^2=\pm 1,\notag\\
&D_m(\sigma)D_m(\tau)=\pm D_m(\tau)D_m(\sigma).\label{eq:sgZ22_ggZ2}
\end{align}

\end{enumerate}

\subsection{The classification and connection to previous work}\label{sec:connection}
Quite some time ago, Dijkgraaf and Witten pointed out that the topological orders in 2+1 dimensions, described by discrete gauge theories with a gauge group $GG$ are classified by its third cohomology group: $H^3(GG,U(1))$.\cite{Dijkgraaf:1990p7194} Different topological orders labeled by $H^3(GG,U(1))$ can be viewed as different discrete versions of the Chern-Simons terms\cite{Propitius:1995p6856,Propitius:1993CS_ccy,Propitius:1997ccy_nA}. For example, because $H^3(Z_2,U(1))=Z_2$, there are two distinct topological orders described by a $Z_2$ gauge group. In the language of the $K$-matrix, the two topological orders are described by $K=\begin{pmatrix} 0& 2\\ 2& 0 \end{pmatrix}$ and $K=\begin{pmatrix} 2& 0\\ 0& -2 \end{pmatrix}$, respectively. The first one is the usual $Z_2$ gauge theory while the second one is the so-called double-semion theory. The quasiparticle anyonic statistics in the two theories are different.

Recently, an original work by Chen et al.\cite{Chen:2011p6670} showed that bosonic SPT phases protected by a global (unitary) on-site symmetry group $SG$ in 2+1 dimensions are also classified by $H^3(SG,U(1))$. Here different phases labeled by $H^3(SG,U(1))$ can be viewed as different topological $\theta$-terms on a discrete space-time. For instance, because $H^3(Z_2,U(1))=Z_2$, there are two distinct Ising paramagnetic (namely disordered) phases (without topological order) in 2+1 dimensions. One is the usual Ising paramagnet, while the other one is the non-trivial Ising SPT phase which features symmetry protected gapless edge states.

It appears that the mathematical object $H^3(G,U(1))$ shows up in these two completely different physical contexts, and one may wonder if there is a certain underlying relation between them. A recent beautiful work by Levin and Gu\cite{Levin:2012p7190} demonstrated such an underlying relation explicitly. It was known that the deconfined phase of a usual $Z_2$ gauge theory is dual to the usual Ising paramagnetic phase. What was shown in Ref.\onlinecite{Levin:2012p7190} is that following the same duality, using exactly solvable models, the double-semion gauge theory is dual to the nontrivial SPT phase. And it was proposed that such dualities between the Dijkgraaf-Witten theories and the SPT phases are general\footnote{In Ref.\onlinecite{Levin:2012p7190}, string-net models are used to construct the exactly solvable models for Dijkgraaf-Witten theories. It appears to us that this construction may not be general; namely not all Dijkgraaf-Witten theories can be described by string-net models using the construction in Ref. \onlinecite{Levin:2005p3468}. In the present work, we use a different way to construct exactly solvable models for Dijkgraaf-Witten discrete gauge theories, which is general.}.

The observation made by Levin and Gu is illuminating and motivated us to consider the cases where both a global on-site symmetry group $SG$ and a topological order described by a gauge group $GG$ are present. Let's consider such a gapped quantum phase. On one hand, one can imagine following the route of duality transformation to transform $SG$ into a gauge group, and eventually having a quantum phase with topological order described by the gauge group $SG\times GG$. One the other hand, one can follow the backward duality transformation to transform $GG$ into a global on-site symmetry, which eventually gives a quantum phase with a global on-site symmety $SG\times GG$. If the initial phases are distinct, it is natural to expect that the phases after duality are also distinct, and vice versa.

Therefore it is reasonable to expect that, in 2+1 dimensions, bosonic phases with both a global on-site symmetry group $SG$ and a topological order described by a gauge group $GG$ are classified by $H^3(SG\times GG, U(1))$. We will construct exactly solvable models for these phases shortly, and we will solve these models in some examples and discuss the measurable differences between different phases.

\begin{figure}
\includegraphics[width=0.42\textwidth]{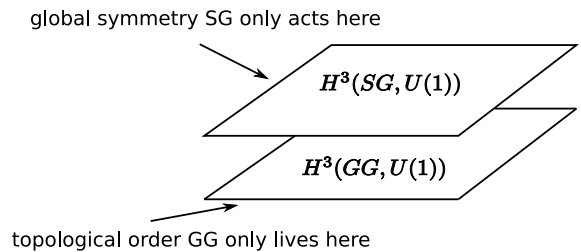}
\caption{A bilayer system in which the topological order and the global symmetry decouple.}
\label{fig:decouple}
\end{figure}

Intuitively, a classification of phases having both $SG$ and $GG$ should at least include $H^3(SG,U(1))$ and $H^3(GG,U(1))$. This is because one can always consider a system where the degrees of freedom which give rise to the topological order and the degrees of freedom on which the symmetry group acts completely decouple from each other. For instance, we can consider a bilayer system, in which the global symmetry $SG$ only acts non-trivially on the first layer, while the topological order described by $GG$ only lives on the second layer (see Fig.\ref{fig:decouple}). In this case, the possible phases living on the first (second) layer would be classified by $H^3(SG,U(1))$ ($H^3(GG,U(1))$). Because one can tune these phases separately, a classification of phases with both $SG$ and $GG$ should actually at least include the cross product: $H^3(SG,U(1))\times H^3(GG,U(1))$. These indices are labeling the phases with no interplay between the global symmetry and the topological order.

This intuitive argument also indicates that, if a classification contains more phases than $H^3(SG,U(1))\times H^3(GG,U(1))$, the extra phases must have non-trivial interplay between the global symmetry and the topological order.

Using the K\"{u}nneth formula and the universal coefficients theorem, we can immediately examine whether or not our classification is consistent with the above physical intuition:
\begin{align}
&H^3(SG\times GG,U(1))=H^4(SG\times GG,Z)\notag\\
=&H^4(SG,Z)\times H^4(GG,Z) \times SET(SG,GG)\notag\\
=&H^3(SG,U(1))\times H^3(GG,U(1)) \times SET(SG,GG).\label{eq:full_exp}
\end{align}
Note that to obtain the first two terms, we have used $H^0(G,Z)=Z$ and $K \otimes Z=K$, $\forall$ abelian finite group $K$. Here we define the abelian group $SET(SG,GG)$ as all the other terms in the K\"{u}nneth expansion formula. For reasons that will become clear shortly, we further decompose $SET(SG,GG)$ into two parts:
\begin{align}
 SET(SG,GG)\equiv SFC(SG,GG)\times EXTRA(SG,GG),\label{eq:SET_exp}
\end{align}
where
\begin{align}
 SFC(SG,GG)&\equiv[H^2(SG,Z)\otimes H^2(GG,Z)]\notag\\
&\times \mbox{Tor}(H^3(SG,Z),H^2(GG,Z)),\label{eq:PSG1}
\end{align}
and
\begin{align}
 EXTRA(SG,GG)\equiv\mbox{Tor}(H^2(SG,Z),H^3(GG,Z)),
\end{align}
where we used the fact that $H^1(G,Z)=Z_1,\;\forall \mbox{ finite }G$. In Eq.(\ref{eq:full_exp}), we see that indeed our classification contains $H^3(SG,U(1))\times H^3(GG,U(1))$, which is what one expects. When we choose the indices in $SET(SG,GG)$ to be trivial, i.e., the identity group element in $SET(SG,GG)$, these terms label the phases in which the topological order and the global symmetry are decoupled. Clearly, the indices in $SET(SG,GG)$ are characterizing the non-trivial interplay between the topological order and global symmetry; namely global symmetry and topological order together enrich the classification. The notation $SET(SG,GG)$ follows from ``symmetry enriched topological order''.

The potential physical meaning of $SFC(SG,GG)$ becomes clear if $GG$ is an abelian group. In this case we can consider the symmetry fractionalization classes, which are given by $H^2(SG,GG)$ as discussed in Sec.\ref{sec:psg}. Using the universal coefficients theorem:
\begin{align}
 &H^2(SG,GG)\notag\\
=&[H^2(SG,Z)\otimes GG]\times \mbox{Tor}(H^3(SG,Z),GG)\notag\\
=&SFC(SG,GG), \mbox{ if $GG$ is abelian,}
\end{align}
where we used the fact that $H^2(GG,Z)=H^1(GG,U(1))=GG,$ if $GG$ is a finite abelian group. Indeed, in this case, $SFC(SG,GG)$ has exactly the same mathematical structure as the symmetry fractionalization classification, leading to the notation ``SFC''.

When $GG$ is non-abelian, the projective symmetry group is no longer related to the central extensions of the SG by GG, and $H^2(SG,GG)$ is not even well-defined. In this case, the mathematical structure underlying PSG, for symmetry fractionalization classes, was unknown. However, $SFC(SG,GG)$ in Eq.(\ref{eq:PSG1}) is still well-defined. We propose that $SFC(SG,GG)$ is the correct counterpart of $H^2(SG,GG)$ when $GG$ is non-abelian.

At this moment, the expansion formula Eq.(\ref{eq:full_exp}) is completely mathematical. It appears that the above discussion is attaching physical meaning to the terms in this formula, such as symmetry fractionalization for $SFC(SG,GG)$, without justification. In fact, we will not mathematically prove our physical interpretation of the formula Eq.(\ref{eq:full_exp}) generally, although we believe it. However, because we have exactly solvable models for every phase in the classication $H^3(SG\times GG,U(1))$, we can at least justify our physical interpretation in some examples by solving these models. We will show in Sec.\ref{sec:example} that, in all the examples that we study, our physical interpretation is correct.

As mentioned earlier, a full classification of symmetry fractionalization classes should go beyond $H^2(SG,GG)$ even when $GG$ is abelian, because one should at least consider the symmetry fractionalization classes for both gauge charges and gauge fluxes. However, in the expansion Eq.(\ref{eq:full_exp}), only $SFC(SG,GG)$ appears. We will show that in the exactly solvable models, this $SFC(SG,GG)$ is characterizing all the symmetry fractionalization classes for gauge fluxes only. It turns out that gauge charges in these models always have trivial symmetry fractionalization. Intuitively, this means that our classification for the symmetry fractionalization is incomplete. This may be due to the fact that we only consider quantum phases with exactly solvable model realizations, which puts constraints on our classification.

The extra indices $EXTRA(SG,GG)$ in the expansion Eq.(\ref{eq:full_exp}) have a completely different mathematical structure than symmetry fractionalization classes, and intuitively this term must be related to the non-trivial interplay between the global symmetry and the topological order, but should not be associated with symmetry fractionalization. Indeed, we will show that $EXTRA(SG,GG)$ is related to the phenomena in which global symmetry transformations interchange the quasiparticle species (or more precisely, the superselection sectors). For instance, in the example mentioned in Sec.\ref{sec:psg}, in which $GG=Z_n\times Z_m$, $EXTRA(SG,GG)$ characterizes the phenomena where the global symmetry could transform a $\psi_1$ gauge flux into a $\psi_2$ gauge flux under certain conditions. Such a non-trivial interplay between the global symmetry and the topological order is beyond symmetry fractionalization, because it violates the basic assumption of symmetry fractionalization: it is impossible to change quasiparticle species by operators acting on the quasiparticles only locally.

Before we move to the exactly solvable models, let's present the examples that we will solve in Sec.\ref{sec:example}. We consider three simple cases:

\begin{enumerate}
 \item $SG=Z_2,GG=Z_2$.
\begin{align}
H^3(SG\times GG,U(1))=Z_2^3,
\end{align}
and among these:
\begin{align}
  \label{eq:H3Z2Z2}
 H^3(SG,U(1))&=Z_2,& H^3(GG,U(1))&=Z_2\notag\\
SFC(SG,GG)&=Z_2,&EXTRA(SG,GG)&=Z_1.
\end{align}
This means that among $Z_2^3$ indices, one $Z_2$ is labeling the two SPT phases, one $Z_2$ is labeling the two Dijkgraaf-Witten topological orders. And the remaining $Z_2$ is labeling the symmetry fractionalization classes, whose physical meaning is presented in Eq.(\ref{eq:sgZ2_ggZ2}). In this case there is no SET indices beyond the symmetry fractionalization classification.

\item $SG=Z_2^2, GG=Z_2$.
\begin{align}
 H^3(SG\times GG,U(1))=Z_2^7,
\end{align}
and among them:
\begin{align}
 H^3(SG,U(1))&=Z_2^3,& H^3(GG,U(1))&=Z_2\notag\\
SFC(SG,GG)&=Z_2^3,&EXTRA(SG,GG)&=Z_1.
\end{align}
This means that among $Z_2^7$ indices, one $Z_2^3$ is labeling the 8 SPT phases, one $Z_2$ is labeling the two Dijkgraaf-Witten topological orders. The remaining $Z_2^3$ is labeling the symmetry fractionalization classes, whose physical meaning is presented in Eq.(\ref{eq:sgZ22_ggZ2}). In this case there is also no SET indices beyond the symmetry fractionalization classification.

\item $SG=Z_2,\;\;\;GG=Z_2^2$.
\begin{align}
H^3(SG\times GG,U(1))=Z_2^7,
\end{align}
and among them:
\begin{align}
 H^3(SG,U(1))&=Z_2,& H^3(GG,U(1))&=Z_2^3\notag\\
SFC(SG,GG)&=Z_2^2,&EXTRA(SG,GG)&=Z_2.
\end{align}
This means that among $Z_2^7$ indices, one $Z_2$ is labeling the two SPT phases, one $Z_2^3$ is labeling the 8 Dijkgraaf-Witten topological orders. One $Z_2^2$ is labeling the symmetry fractionalization classes, whose physical meaning is presented in Eq.(\ref{eq:sgZ2_ggZ22}). Finally, the remaining $Z_2$ in $EXTRA(SG,GG)$ labels the phases beyond the symmetry fractionalization. This is the simplest example in which SET phases beyond symmetry fractionalization are realized.
\end{enumerate}

\section{Exactly Solvable Models}\label{sec:model}

In this Section we introduce the exactly solvable models which exhibit all the phases from the general classification introduced above. First we recall the Dijkgraaf-Witten topological invariant, and then introduce the general form of our exactly solvable models.

\subsection{The geometric interpretation of group cohomology and the Dijkgraaf-Witten topological invariants}
\label{sec:DW}

\subsubsection{The geometric interpretation of group cohomology}

In Sec.\ref{sec:math} we introduced group cohomology, which appears to be a group theoretical concept. However, group cohomology is actually about topology. In this section, we introduce the geometric interpretation of group cohomology, which is the mathematical foundation of our exactly solvable models.

An n-cocyle $\omega\in H^n(G,U(1))$ of a group $G$ allows one to construct a topological invariant for n-dimensional manifolds. Generally, different elements of $H^n(G,U(1))$ correspond to different topological invariants of n-manifolds. Below we will illustrate the construction of such topological invariants.

Let's consider a 3-dimensional manifold as an example. We know that tetrahedra can be viewed as building blocks for arbitrary 3-manifolds. To begin with, we show that a 3-cocycle $\omega\in H^3(G,U(1))$ allows one to assign a complex number to a tetrahedron following a simple procedure.
\begin{figure}
  \centering
\includegraphics[width=0.49\textwidth]{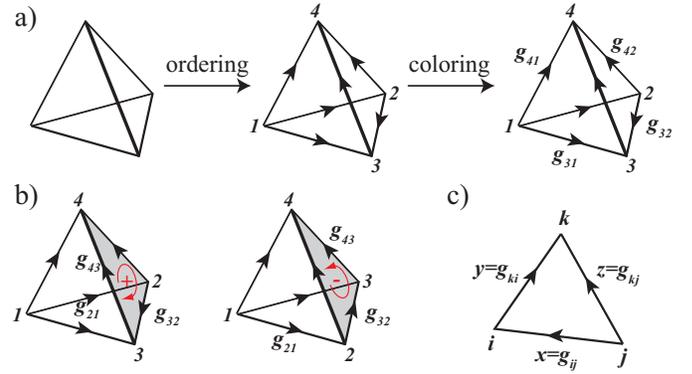}
\caption{The 3-cocycle $\omega(h_1,h_2,h_3)$ assigns a $U(1)$ complex number (i.e. a phase) to a 3-simplex (tetrahedron). (a) Left to Center: the ``ordering'' of tetrahedron's four vertices, we choose here $1\rightarrow 2\rightarrow 3\rightarrow 4$. An edge is oriented from lower to higher vertex, so no triangle forms an oriented loop. (Alternatively, one orients all edges without forming oriented triangle loops, and a consistent underlying vertex ordering is guaranteed.) Center to Right: ``Coloring'' assigns group element $g_{ij}$ to $j\rightarrow i$ edge, with $g_{ji}=g_{ij}^{-1}$. (Four out of six elements are shown explicitly.) The shown tetrahedron $1\rightarrow 2\rightarrow 3\rightarrow 4$ is assigned the phase $\ccy(g_{43},g_{32},g_{21})^\epsilon=\ccy(g_{43},g_{32},g_{21})$. The exponent $\epsilon=\pm1$ is determined by: (b) Chirality. For tetrahedron $1\rightarrow 2\rightarrow 3\rightarrow 4$, looking from vertex 1, (counter-)clockwise loop 234 means $\epsilon=-1$ ($+1$), which is realized in the right (left) tetrahedron. (c) The zero-flux rule applies to all tetrahedron faces, i.e. triangles. Generally, $g_{ki}\cdot g_{ij}\cdot g_{jk}=\openone$, the group identity element. Recall that $g_{jk}=g_{kj}^{-1}$. Choosing an ordering and assigning elements to ordered bonds, like shown, leads to the constraint$z=y\cdot x$.}
\label{fig:order_color}
\end{figure}

The procedure contains two steps (see Fig.\ref{fig:order_color}). The first step is called \textit{ordering}, in which one chooses an ordering of the 4 vertices of the tetrahedron. We can represent this ordering by assigning arrows going from lower to higher ordered vertices on the edges of the tetrahedron. For any given face (i.e., a triangle) of the tetrahedron, obviously the three arrows never form an oriented loop.

The second step is called \emph{coloring}, in which one assigns a group element to every edge of the tetrahedron. The coloring must be consistent with certain rules below. Note that an edge already has an arrow, or orientation, associated with it. The assigned group element for a given edge should then be understood in the following way: if we assign the group element $g\in G$ to follow the direction of the arrow, then we automatically assign group element $g^{-1}\in G$ to the direction opposite to the arrow. Let's denote the group element assigned to the bond connecting vertices $j$ and $i$ as $g_{ij}$, following the orientation from $j$ to $i$: $j\rightarrow i$. We then automatically assign $g_{ji}=g_{ij}^{-1}$.

In addition, the three assigned group elements for any given face must satisfy the constraint: $g_{ij}\cdot g_{jk}\cdot g_{ki}=\openone$, where $\openone$ is the identity element in group $G$, and $i,j,k$ are the three vertices of the face. We will call this constraint the ``zero-flux rule'' throughout this paper. With this constraint, it is easy to show that among the six group elements for the 6 edges of the tetrahedron, only three are independent. In particular, let's denote the ordered vertices by $1,2,3,4$; then, $g_{43},g_{32},g_{21}$ completely determine all the other group elements.

Given a 3-cocycle $\omega(x,y,z)\in H^3(G,U(1))$, one assigns the complex number $\omega(g_{43},g_{32},g_{21})^{\epsilon}$ to an ordered and colored tetrahedron. (Sometimes we use the $\omega ^{\epsilon}(g_{43},g_{32},g_{21})$ notation.) Here $\epsilon=\pm1$ depending on the chirality of the ordered vertices. One can determine this chirality by the right-hand rule: imagine looking at the face formed by vertices 2-3-4 from the vertex-1; if the vertices 2-3-4 form a counter-clockwise (clockwise) loop, the chirality of the ordering is positive (negative) and $\epsilon=1$ ($\epsilon=-1$).
\begin{figure}
  \centering
\includegraphics[width=0.49\textwidth]{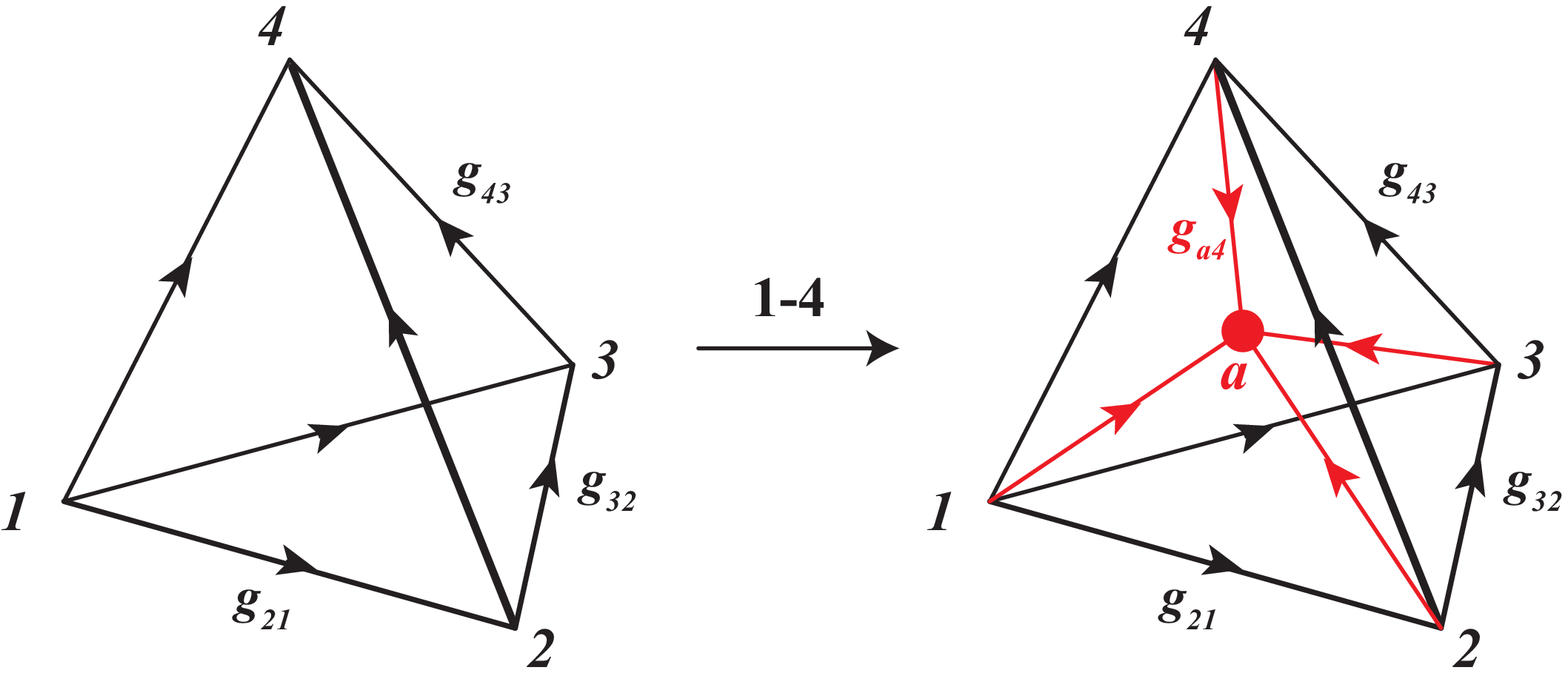}
\caption{The 1-4 move (3 dimensions) changes triangulation but not total product of phases $\prod_I W(\sigma_I)^{\epsilon_I}$ of 3-simplices $\sigma_I$. Left: The initial tetrahedron is assigned the phase $W_0\equiv\ccy(g_{43},g_{32},g_{21}) ^{-1}$ (see Fig.~\ref{fig:order_color}). Right: The vertex $a$ is added, and we choose the ordering such that $1\rightarrow 2\rightarrow 3\rightarrow 4\rightarrow a$ (obvious from chosen orientations of new (red) edges). There are now four smaller tetrahedrons, with phases: 1. Tetrahedron $1\rightarrow 2\rightarrow 4\rightarrow a$: $W_1\equiv\ccy(g_{a4},g_{42},g_{21})$; 2. Tetrahedron $2\rightarrow 3\rightarrow 4\rightarrow a$: $W_2\equiv\ccy(g_{a4},g_{43},g_{32})$; 3. Tetrahedron $1\rightarrow 3\rightarrow 4\rightarrow a$: $W_3\equiv\ccy(g_{a4},g_{43},g_{31})^{-1}$; 4. Tetrahedron $1\rightarrow 2\rightarrow 3\rightarrow a$: $W_4\equiv\ccy(g_{a3},g_{32},g_{21})^{-1}$. The 3-cocycle condition, Eq.~\eqref{eq:3-cocycle}, says the total phase does not change by the move: $W_0=W_1W_2W_3W_4$. Note that only one independent new group element is introduced (we marked the $g_{a4}$), and from zero-flux rule $g_{a3}=g_{a4}\cdot g_{43}$. Changing our choice of ordering for $a$ relative to $1,2,3,4$ would lead to an equivalent 3-cocycle condition.}
\label{fig:1-4_move}
\end{figure}

This assignment of $\omega(g_{43},g_{32},g_{21})^{\epsilon}$ to an ordered and colored tetrahedron allows a simple geometric interpretation of the cocycle condition Eq.(\ref{eq:3-cocycle}), see Fig.\ref{fig:1-4_move}. To see this, consider an ordered and colored tetrahedron and the associated complex number $\omega(g_{43},g_{32},g_{21})^{\epsilon}$. One can now add one more vertex-$a$ inside the tetrahedron. With vertex-$a$, the original tetrahedron can be triangulated into 4 smaller tetrahedra. One can further continue the ordering and coloring procedure for the 4 smaller tetrahedra. Since we already have the ordering and coloring for the large tetrahedron, we only need to assign an order to vertex-$a$, as well as to color the four newly created edges $1a$,$2a$,$3a$ and $4a$. Actually, according to the zero-flux rule, it is easy to show that only one of the four new edges is independent. After we complete ordering and coloring the 4 small tetrahedra, we will have 4 new complex numbers, each of which is associated with a small tetrahedron. It is straightforward to show that, \emph{no matter how one performs the complete ordering and coloring procedure, the cocycle condition Eq.(\ref{eq:3-cocycle}) dictates that the product of the 4 new complex numbers exactly equals the orginal complex number $\omega(g_{43},g_{32},g_{21})^{\epsilon}$.}
\begin{figure}
  \centering
\includegraphics[width=0.49\textwidth]{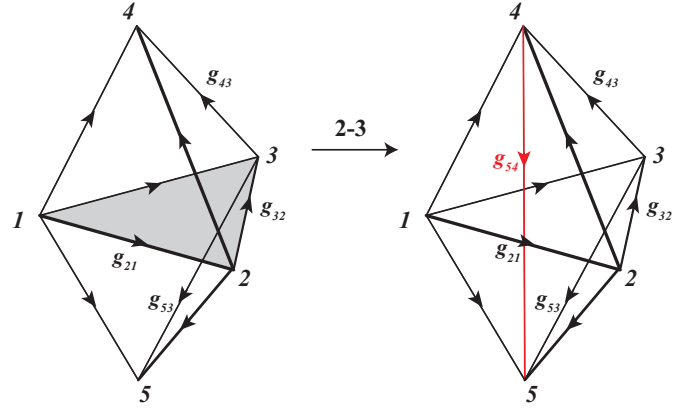}
\caption{The 2-3 move (3 dimensions) changes triangulation but not total product of phases $\prod_I W(\sigma_I)^{\epsilon_I}$. Left: Two initial tetrahedrons, $1234$ and $1235$, are assigned the total phase $W_0\equiv\ccy(g_{43},g_{32},g_{21})\ccy(g_{43},g_{32},g_{21})^{-1}$ (see Fig.~\ref{fig:order_color}). Right: One edge (red) is added, and we choose the ordering $4\rightarrow 5$. The volume is now divided into 3 smaller tetrahedrons, with phases: 1. Tetrahedron $1\rightarrow 2\rightarrow 4\rightarrow 5$: $W_1\equiv\ccy(g_{54},g_{42},g_{21})^{-1}$; 2. Tetrahedron $2\rightarrow 3\rightarrow 4\rightarrow 5$: $W_2\equiv\ccy(g_{54},g_{43},g_{32})^{-1}$; 3. Tetrahedron $1\rightarrow 3\rightarrow 4\rightarrow 5$: $W_3\equiv\ccy(g_{54},g_{43},g_{31})$. The 3-cocycle condition, Eq.~\eqref{eq:3-cocycle}, says the total phase does not change by the move: $W_0=W_1W_2W_3$. Note that the new group element $g_{54}$ is not independent, e.g. $g_{54}=g_{52}\cdot g_{42}^{-1}$.}
\label{fig:2-3_move}
\end{figure}

Such a procedure of completing the triangulation, ordering and coloring of tetrahedra after adding a vertex is called a 1-4 move. A specific example of a 1-4 move is shown in Fig.\ref{fig:1-4_move}.

Similarly, there is a 2-3 move, Fig.~\ref{fig:2-3_move}. Namely, one can consider two face-sharing tetrahedra, both of which have been ordered and colored. There are then two complex numbers, each of which is associated with a tetrahedron. One can now connect the two vertices that are on opposite sides of the shared face, and the volume enclosed by the original two tetrahedra can be triangulated into three tetrahedra. One can continue the ordering and coloring procedure for the three tetrahedra and obtain three new complex numbers. It is also easy to show that, \emph{no matter how one performs the further ordering and coloring procedure, the cocycle condition Eq.(\ref{eq:3-cocycle}) dictates that the product of the three new complex numbers equals the product of the two orginal complex numbers.} A specific example of such a 2-3 move is illustrated in Fig.\ref{fig:2-3_move}.

In this paper we will use ``canonical'' 3-cocycles $\omega$, meaning that $\omega(g_1,g_2,g_3)=1$ if any of $g_1,g_2,g_3$ is equal to $\openone$ (the identity element of group $G$). It is always possible to choose a gauge for $\omega$ such that it becomes canonical~\cite{Chen:2011p6670}. Specifically, the explicit elementary cocycles that we will use in studying examples of our models in Section~\ref{sec:example} are going to be canonical.

\subsubsection{The Dijkgraaf-Witten topological invariants}\label{sec:DW_inv}

The above examples of 1-4 move and 2-3 move suggest that the products of the assigned complex numbers $\omega(g_{lk},g_{kj},g_{ji})^{\epsilon}$ for a given volume may be related to a certain invariant that is independent of the triangulation, ordering and coloring procedure. This is indeed true, as stated by two mathematical theorems presented in the following.

Let us first consider a closed 3-manifold $M$ without a boundary. One can triangulate $M$ by a finite number of 3-simplices (i.e., tetrahedra), and then order the vertices of this triangulation. Next, one can have a coloring $\varphi$ of all the edges in the triangulation obeying the zero-flux rule. Note that under a fixed triangulation and ordering of vertices, there can be many different colorings. Let's denote a 3-simplex of the triangulation, together with the ordering of its vertices, by $\sigma_I$, where $I=1,2,...,S$ labels 3-simplices and $S$ is the total number of 3-simplices. For a given coloring $\varphi$, let's also denote the assigned complex number $\omega(g_{lk},g_{kj},g_{ji})^{\epsilon}$ for the simplex-$\sigma_I$ as $W(\sigma_I,\varphi)^{\epsilon(\sigma_I)}$, and we can further compute the product of all these complex numbers for 3-simplices: $\prod_{I=1}^S W(\sigma_I,\varphi)^{\epsilon(\sigma_I)}$. For each given coloring $\varphi$, we will have one such product.

\emph{Theorem 1: The sum of such products for all possible colorings, with an appropriate normalization factor, is a topological invariant of the closed manifold $M$:\cite{Dijkgraaf:1990p7194}} 
\begin{equation}
 Z_M=\frac{1}{|G|^V}\sum_{\substack{\varphi\in\text{ all}\\\text{possible}\\\text{colorings}}}\prod_{I=1}^S W(\sigma_I,\varphi)^{\epsilon(\sigma_I)}.\label{eq:DW_noboundary}
\end{equation}
Here $|G|$ is the number of elements in group $G$, and $V$ is the number of vertices in the triangulation. Note that without Theorem-1, one would naively expect that $Z_M$ depends on both the triangulation and the ordering of vertices (while different colorings are already summed over). But with Theorem-1, we know that $Z_M$ does not depend on either of them --- it only depends on the topology of the manifold $M$ and the 3-cocyle $\omega\in H^3(G,U(1))$. One can further show that equivalent 3-cocycles (i.e., 3-cocycles differing by a 3-coboundary) give exactly the same topological invariant $Z_M$;\cite{Dijkgraaf:1990p7194} namely, $Z_M$ only depends on inequivalent elements in $H^3(G,U(1))$.

The topological invariant $Z_M$ is exactly the partition function of the Dijkgraaf-Witten (DW) topological quantum field theory (TQFT) for discrete gauge group $G$ in 2+1 dimensions\cite{Dijkgraaf:1990p7194,WAKUI:1992p7120}. In order to have a well-defined TQFT, it turns out that one not only needs to define partition functions for closed space-time manifolds, but one also needs to define quantum transition amplitudes for space-time manifolds with boundaries. This is given by the second theorem.

Consider a 3-manifold $M$ with boundary $\partial M$. $\partial M$ is formed by a collection of closed 2-manifolds. One can triangulate $\partial M$ by a finite number of 2-simplices (i.e., triangles), order the vertices of the 2-simplices, and then color their edges again following the zero-flux rule (i.e., $g_{ij}g_{jk}g_{ki}=\openone$ for all 2-simplices). Let's denote the triangulation, ordering, and coloring of the boundary $\partial M$ by $\tau$.

Next, we fix the coloring $\tau$ and extend it into the bulk of $M$. This means that we consider a triangulation of $M$, an ordering of its vertices, and a coloring $\varphi$ such that they become exactly the same as $\tau$ when limited to the boundary $\partial  M$. In this case, we also say that the bulk triangulation, ordering, and coloring in $M$ are compatible with $\tau$ on $\partial M$. For instance, the triangular faces of a tetrahedron can be viewed as the boundary of a 3-dimensional ball. Then a 1-4 move can be viewed as a specific extension of the boundary $\tau$ into the bulk of the ball.

Now let's fix the bulk triangulation and ordering of vertices in $M$ that is compatible with $\tau$. There are still many possible colorings $\varphi$ in $M$ that are compatible with $\tau$, and they form a set which we denote as $Col(M,\tau)$. As in Theorem-1, with a fixed $\varphi\in Col(M,\tau)$ one can compute the product of complex numbers $\prod_{I=1}^S W(\sigma_I,\varphi)^{\epsilon(\sigma_I)}$ assigned to all the 3-simplices in the bulk of $M$. It turns out the sum of all such products satisfies the following theorem:

\emph{Theorem 2: The complex number $Z_M(\tau)$ does not depend on the triangulation of $M$ nor ordering of its vertices, whenever the topology of $M$ and $\tau$ on $\partial M$ are fixed\cite{Dijkgraaf:1990p7194,WAKUI:1992p7120}:}
\begin{align}
 Z_M(\tau)=\frac{1}{|G|^{V+\frac{V_{\partial M}}{2}}}\sum_{\varphi\in Col(M,\tau)}\prod_{I=1}^S W(\sigma_I,\varphi)^{\epsilon(\sigma_I)}.\label{eq:DW_boundary}
\end{align}
Here $V$ is the total number of vertices inside $M$ (i.e., not including $\partial M$), while $V_{\partial M}$ is the number of vertices in $\partial M$. Obviously $Z_M(\tau)$ becomes $Z_M$ in Eq.(\ref{eq:DW_noboundary}) when $M$ does not have a boundary.
\begin{figure}
  \centering
\includegraphics[width=0.49\textwidth]{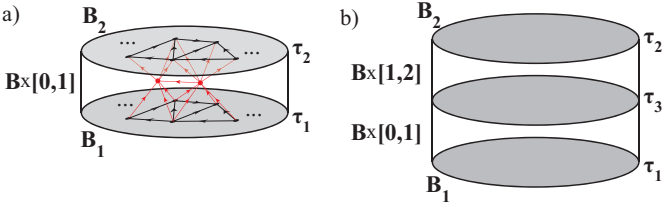}
\caption{The Dijkgraaf-Witten topological invariant $Z_M$ and TQFT for manifold $M$ with boundary. (a) Two copies of 2-manifold $B$: $B_1,B_2$, with colorings $\tau_1,\tau_2$, form the boundary of 3-manifold $M=B\times [0,1]$. The triangulation and ordering in $B_1$ and $B_2$ are chosen identical, as sketched. The triangulation and ordering in the bulk of $M$ (example red vertices and edges) does not influence the value of DW invariant $Z_M(\tau=\tau_1\cup\tau_2)$. A coloring $\tau_i$ of $B$ represents a quantum state $\ket{\tau_i}$, and then $Z_M(\tau)=\bra{\tau_2}P\ket{\tau_1}$ is a quantum amplitude of operator $P$ in the DW TQFT associated with $B$. The image of $P$ is the ground state manifold of the TQFT. (b) The operator $P$ is a projector: The quantum amplitude $\bra{\tau_2}P^2\ket{\tau_1}$ is sketched, representing a product of amplitudes of $P$ from $\tau_1$ to $\tau_3$, and $\tau_3$ to $\tau_2$, with a sum over colorings $\tau_3$. This becomes a sum over internal colorings in space $M=B\times[0,2]$ with boundary $B_1\cup B_2$. The latter is equal to the amplitude $\bra{\tau_2}P\ket{\tau_1}$ from panel (a).}
\label{fig:DW_projector}
\end{figure}

To see the physical meaning of $Z_M(\tau)$, let's consider a special case: $M = \mathbf{B}\times [0,1]$ where $\mathbf{B}$ is a certain closed orientable 2-manifold. $\partial M$ is formed by two disconnected but identical closed 2-manifolds: $\mathbf{B}_1\cong\mathbf{B}$ and $\mathbf{B}_2\cong\mathbf{B}$ corresponding to $0\in [0,1]$ and $1\in[0,1]$ respectively (see Fig.\ref{fig:DW_projector}). We can then triangulate and order the vertices on $\mathbf{B}_1$ and $\mathbf{B}_2$ in the same fashion.

For each edge-$ij$ of the triangulation of $\mathbf{B}_1$, we construct a $|G|$-dimensional local Hilbert space $\mathcal{H}^{DW}_{ij}\equiv \{ g_{ij}\in G \}$; namely each group element labels a quantum state in $\mathcal{H}_{ij}$. Then we consider the tensor product of all such local Hilbert spaces $\mathcal{H}^{DW}\equiv \otimes_{\mbox{all edges}}\mathcal{H}^{DW}_{ij}$. Now we can associate each possible coloring on $\mathbf{B}_1$ with a quantum state in $\mathcal{H}^{DW}$. Because a coloring must satisfy the zero-flux rule, clearly all possible colorings of $\mathbf{B}_1$ span a sub-space $\mathcal{\tilde H}^{DW}\subset \mathcal{H}^{DW}$. All possible colorings of $\mathbf{B}_2$ then also form the exact same Hilbert sub-space $\mathcal{\tilde H}^{DW}$.

Let's choose a coloring $|\tau_1\rangle \in\mathcal{\tilde H}^{DW}$ on $\mathbf{B}_1$, and another coloring $|\tau_2\rangle\in\mathcal{\tilde H}^{DW}$ on $\mathbf{B}_2$. $\tau_1$ and $\tau_2$ completely specify the triangulation, ordering and coloring $\tau$ on $\partial M$. Theorem-2 means that there is a well-defined quantum transition amplitude from the state $|\tau_1\rangle$ to the state $|\tau_2\rangle$:
\begin{equation}
  \label{eq:DWProj}
\langle\tau_2|P|\tau_1\rangle\equiv Z_M(\tau).  
\end{equation}
Because all possible $|\tau_1\rangle$ ($|\tau_2\rangle$) form a basis of $\mathcal{\tilde H}^{DW}$, this equation defines a quantum operator $P$ on $\mathcal{\tilde H}^{DW}$.

Theorem-2 immediately dictates that $P$ is a projector: $P^2=P$. This is because after we insert an identity operator $\sum_{\tau_3}|\tau_3\rangle\langle\tau_3|=\openone$ in $\mathcal{\tilde H}^{DW}$, $\langle\tau_2|P^2|\tau_1\rangle=\sum_{\tau_3}\langle\tau_2|P|\tau_3\rangle\langle\tau_3|P|\tau_1\rangle$ has a simple geometric interpretation (see Fig.\ref{fig:DW_projector}): we can consider two copies of the manifold $M$, $M_1 =\mathbf{B}\times [0,1]$ and $M_2= \mathbf{B}\times [1,2]$, so that $\langle\tau_3|P|\tau_1\rangle$ ($\langle\tau_2|P|\tau_3\rangle$) is the quantum amplitude due to an internal triangulation and ordering of $M_1$ ($M_2$). We can then glue the top boundary of $M_1$ with the bottom boundary of $M_2$. After the gluing, the vertices on the glued boundary $\mathbf{B}\times \{1\}$ become internal vertices. Then $\sum_{\tau_3}\langle\tau_2|P|\tau_3\rangle\langle\tau_3|P|\tau_1\rangle$ can be simply interpreted as the quantum amplitude due to an internal triangulation and ordering of $M_1\cup M_2\cong M$, which must be the same as $\langle\tau_2|P|\tau_1\rangle$ according to Theorem-2.

Because $P$ is a projector, the image of $P$ forms a sub-space in the Hilbert space $Img(P)\subset\mathcal{\tilde H}^{DW}$ in which $P$ acts as identity. $Img(P)$ turns out to be the ground state sector associated with the Dijkgraaf-Witten TQFT for the closed 2-manifold $\mathbf{B}$. One can also prove\cite{Dijkgraaf:1990p7194} that the dimension of $Img(P)$ (i.e. the ground state degeneracy of the TQFT) and the partition function of the closed space-time 3-manifold $\tilde M\equiv \mathbf{B}\times S^1$ are identical: $\mbox{dim}[Img(P)]=Z_{\tilde M= \mathbf{B}\times S^1}$.

At this point, it is useful to introduce an example. Consider the simplest group $G=Z_2$. According to Eqs.~\eqref{eq:Z_U1} and \eqref{eq:HpZnZ}, $H^3(Z_2,U(1))=Z_2$. This means that there are two inequivalent 3-cocycles and let's choose the trivial one: $\omega(x,y,z)=1$, $\forall x,y,z\in Z_2$, which gives rise to a Dijkgraaf-Witten TQFT. This particular TQFT turns out to be a familiar one: the $Z_2$ gauge theory of the toric code model\cite{Kitaev:2003p6185}. We can then use Theorem-1 to compute the partition function $Z_M$ for a closed 3-manifold $M$, and use Theorem-2 to compute the ground state degeneracy via the projector $P$. For instance, for a 3-sphere and a 3-torus, $Z_{S^3}=1/2$ and $Z_{T^3}=4$, respectively. The latter result implies that the ground state degeneracy on a torus is $4$, since $T^3=T^2\times S^1$. More generally, the ground state degeneracy on a closed orientable 2-manifold $\mathbf{B}$ is $\mbox{dim}[Img(P)]=4^g$, where $g$ is the genus of $\mathbf{B}$.

\subsubsection{The generalization to other dimensions}

The above discussion has been limited to 2+1 dimensions. In fact, the geometric interpretation of an n-cocycle can be easily generalized to any $n\geqslant 2$ space-time dimensions. Some aspects of this generalization have been discussed in Ref.\onlinecite{Chen:2011p6670}. Here for the purpose of the current paper, we briefly discuss the $n=2$ and the $n=4$ cases.
\begin{figure}
  \centering
\includegraphics[width=0.49\textwidth]{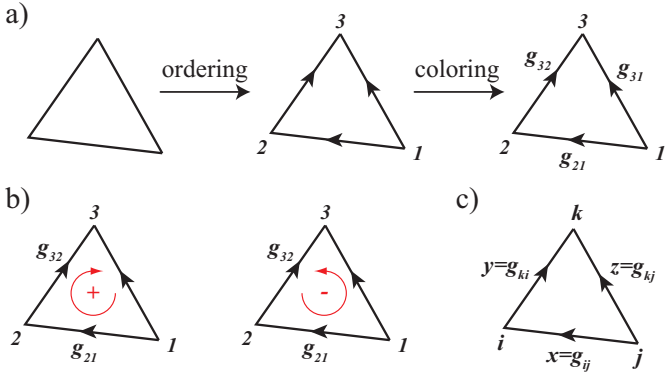}
\caption{The 2-cocycle $\omega(h_1,h_2)$ assigns a $U(1)$ complex number (i.e. a phase) to a 2-simplex (triangle). (a) Left to Center: the ``ordering'' of triangle's three vertices, we choose here $1\rightarrow 2\rightarrow 3$. An edge is oriented from lower to higher vertex, so no triangle forms an oriented loop. (Alternatively, one orients all edges without forming oriented triangle loops, and a consistent underlying vertex ordering is guaranteed.) Center to Right: ``Coloring'' assigns group element $g_{ij}$ to $j\rightarrow i$ edge, with $g_{ji}=g_{ij}^{-1}$. The shown triangle $1\rightarrow 2\rightarrow 3$ is assigned the phase $W=\ccy(g_{32},g_{21})^\epsilon=\ccy(g_{43},g_{32},g_{21})$. The exponent $\epsilon=\pm1$ is determined by: (b) Chirality. For triangle $1\rightarrow 2\rightarrow 3$, (counter-)clockwise loop 123 means $\epsilon=+1$ ($-1$), which is realized in the left (right) triangle. (c) The zero-flux rule applied to a triangle. Generally, $g_{ki}\cdot g_{ij}\cdot g_{jk}=\openone$, the group identity element. Recall that $g_{jk}=g_{kj}^{-1}$. Choosing an ordering and assigning elements to ordered bonds as shown leads to the constraint $z=y\cdot x$.}
\label{fig:color_triangle}
\end{figure}

\emph{Geometric interpretation of a 2-cocycle $\omega(x,y)\in H^2(G,U(1))$.} (See Fig.\ref{fig:color_triangle}.) Let's choose a 2-cocycle $\omega(x,y)\in H^2(G,U(1))$. Consider a 2-simplex (i.e., triangle). Again one needs to perform the ordering and coloring precedure. Let's choose an ordering of the vertices $1\rightarrow 2\rightarrow 3$. We then color the edges by group elements $g_{31},g_{32},g_{21}$ under the zero-flux rule: $g_{31}=g_{32}g_{21}$. Therefore, we can choose $g_{32},g_{21}$ to be the only independent elements. Next, we assign the complex number $\omega(g_{32},g_{21})^{\epsilon}$ to this 2-simplex. Here $\epsilon=\pm1$ depending on the chirality of the ordering of vertices: if $1\rightarrow 2\rightarrow 3$ is clockwise (counter-clockwise), $\epsilon=1$ ($\epsilon=-1$).
\begin{figure}
  \centering
\includegraphics[width=0.4\textwidth]{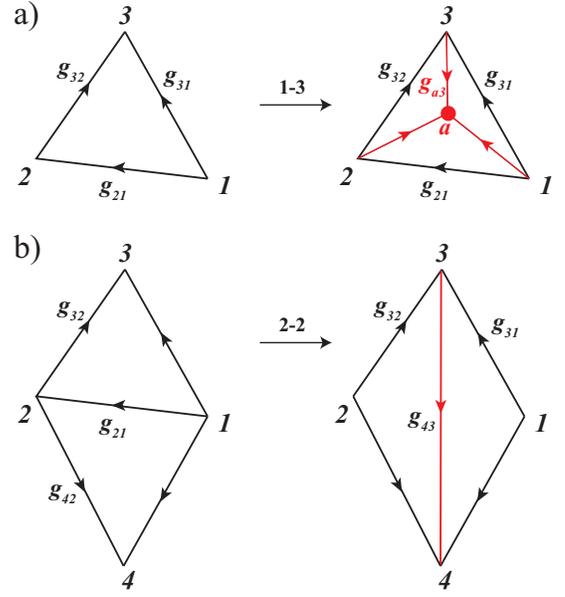}
\caption{The 1-3 and 2-2 moves (in 2 dimensions) change triangulation but not total product of phases $\prod_I W(\sigma_I)^{\epsilon_I}$ of 2-simplices $\sigma_I$. (a) The 1-3 move. Left: The initial triangle is assigned the phase $W_0\equiv\ccy(g_{32},g_{21})$ (see Fig.~\ref{fig:color_triangle}). Right: The vertex $a$ is added, and we choose the ordering such that $1\rightarrow 2\rightarrow 3\rightarrow a$ (obvious from chosen orientations of new (red) edges). There are now three smaller triangles, with phases: 1. Triangle $1\rightarrow 2\rightarrow a$: $W_1\equiv\ccy(g_{a2},g_{21})$; 2. Triangle $2\rightarrow 3\rightarrow a$: $W_2\equiv\ccy(g_{a3},g_{32})$; 3. Triangle $1\rightarrow 3\rightarrow a$: $W_3\equiv\ccy(g_{a3},g_{31})^{-1}$. The 2-cocycle condition, Eq.~\eqref{eq:2-cocycle}, says the total phase does not change by the move: $W_0=W_1W_2W_3$. Note that only one independent new group element is introduced (we marked the $g_{a3}$), and from zero-flux rule $g_{a2}=g_{a3}\cdot g_{32}$. Changing our choice of ordering for $a$ relative to $1,2,3$ would lead to an equivalent 2-cocycle condition. (b) The 2-2 move. Left: Two initial triangles, $123$ and $124$, are assigned the total phase $W_0\equiv\ccy(g_{32},g_{21})\ccy(g_{42},g_{21})^{-1}$. Right: The area is divided into two triangles by the other possible edge (red), and we choose the ordering $3\rightarrow 4$. The phases of new triangles: 1. Triangle $1\rightarrow 3\rightarrow 4$: $W_1\equiv\ccy(g_{43},g_{31})^{-1}$; 2. Triangle $2\rightarrow 3\rightarrow 4$: $W_2\equiv\ccy(g_{43},g_{32})$. The 2-cocycle condition gives $W_0=W_1W_2$. Note that the group element $g_{43}$ is not independent, e.g. $g_{43}=g_{42}\cdot g_{32}^{-1}$.}
\label{fig:1-3_2-4_moves}
\end{figure}

The geometric interpretation of the 2-cocycle condition Eq.(\ref{eq:2-cocycle}) can now be understood as the invariance of the product of these assigned complex numbers in a 1-3 move or a 2-2 move (see Fig.\ref{fig:1-3_2-4_moves}). For instance, in a 1-3 move, we consider an ordered and colored triangle, together with the assigned complex number $\omega(g_{32},g_{21})^{\epsilon}$. Then we add one new vertex inside the triangle. After connecting the new vertex with the original 3 vertices, 3 new edges are created and the original triangle is thus further triangulated into 3 smaller triangles. We now can choose any ordering and coloring of the new vertex and new edges under the zero-flux rule, which then assigns 3 new complex numbers for the 3 smaller triangles. The 2-cocycle condition Eq.(\ref{eq:2-cocycle}) dictates that the product of the 3 new complex numbers equals the original one $\omega(g_{32},g_{21})^{\epsilon}$.

Theorem-1 and Theorem-2 can also be generalized to 2-manifolds and 2-cocycles. For example, let's consider Theorem-2. For a 2-manifold $M$ with boundary $\partial M$, one firstly chooses a triangulation (using 1-simplices, i.e., line segments), an ordering of vertices, and a coloring on $\partial M$, which we denote by $\tau$. Note that now there are no zero-flux rule constraints for $\tau$, because there is no triangle in a 1-simplex. Then one can extend the triangulation, ordering and coloring into the bulk of $M$ (where the zero-flux rule holds). We denote the assigned complex number for a 2-simplex $\sigma_I$ in $M$ as $W(\sigma_I,\varphi)^{\epsilon(\sigma_I)}$ as before, where $\varphi$ denotes the bulk coloring. With a fixed bulk triangulation and ordering, there will be many possible colorings that are compatible with $\tau$. Theorem-2 for 2-manifolds and 2-cocyles states that Eq.(\ref{eq:DW_boundary}) defines a complex number $Z_M(\tau)$ which is independent of the choice of bulk triangulation and ordering of vertices, as long as $\tau$ is fixed.

Following the discussion from the previous section, Theorem-1 and Theorem-2 suggest that a cocyle $\omega(x,y)\in H^2(G,U(1))$ may define a 2D TQFT. This is indeed true and was discussed in a mathematical context\footnote{see, for instance, the webpage by John Baez at:\\ http://math.ucr.edu/home/baez/qg-winter2005/w05week06.pdf}. In 2+1d, we know that different topological orders can be characterized by different TQFTs. One may ask: does this mean that there are non-trivial topological orders in 1+1d? However, we also know from previous research that non-trivial topological orders do not exist in 1+1d\cite{PhysRevLett.94.140601}. It turns out that the 2D TQFTs induced by 2-cocycles, do not give rise to physically non-trivial topological order. This is because the ground state degeneracy is not robust, as one can lift it by a local perturbation.\footnote{For simplicity, let's consider $G=Z_2$ and use the trivial 2-cocycle $\omega(x,y)=1$, $\forall x,y\in Z_2$. We can construct the projector $P$ for a circle following the discussion in Sec.\ref{sec:DW_inv}, after triangulating it by $N$ line segments. Let's use $g_I=\pm1$ as the group element for the colored line segment $I$. The 2-fold ground state sector $Img(P)$ of the induced TQFT can be easily found: $|\psi_1\rangle=1/2^{(N-1)/2}\sum_{\prod_I g_I=+1}|\{g_I\}\rangle$, and $|\psi_2\rangle=1/2^{(N-1)/2}\sum_{\prod_I g_I=-1}|\{g_I\}\rangle$. Naively the second ground state corresponds to a trapped $Z_2$ gauge flux inside the circle. However, the degeneracy between $|\psi_1\rangle$ and $|\psi_2\rangle$ is not protected. One can consider a local perturbation $\delta H=\epsilon |g_I=-1\rangle\langle g_I=+1|+h.c.$ for a certain line segment $I$. Straightforward perturbative calculation shows that this perturbation lifts the ground state degeneracy by a finite energy gap.}

\emph{Geometric interpretation of a 4-cocycle $\omega(x,y,z,u)\in H^4(G,U(1))$.} Similarly to the $n=3$ case, for a given 4-simplex one can choose an ordering of its vertices $1\rightarrow 2\rightarrow 3\rightarrow 4\rightarrow 5$, color the edges following the zero-flux rule, and assign the complex number $\omega(g_{54},g_{43},g_{32},g_{21})^{\epsilon}$ to it. Here again $\epsilon$ is determined by the chirality of the ordering of vertices. The 4-cocycle condition $\delta \omega=1$ in Eq.(\ref{eq:n-cocycle}) when $n=4$ can be understood as the invariance of the product of the assigned complex numbers to 1-5, 2-4, and 3-3 moves. Theorem-1 and Theorem-2 also hold for 4-manifolds. For any fixed 4-cocycle $\omega(x,y,z,u)\in H^4(G,U(1))$, these theorems give rise to a 4D TQFT. Equivalent 4-cocycles induce the same TQFT. These 4D TQFTs characterize different topological orders in 3+1d.

\subsection{Exactly solvable models}
\label{esmodels}

In this section we define our exactly solvable models. Although we discuss the generalization to other dimensions in Sec.~\ref{sec:discussion_general_dim}, from now on we constrain ourselves to the 2+1 dimensional case. It will become clear that our models exhibit both a global symmetry forming a group $\SG$, as well as topological order described by a discrete gauge group $\GG$. We will explain our models' relation to both the SPT models of Ref.\onlinecite{Chen:2011p6670} and the Dijkgraaf-Witten gauge theories of Ref.\onlinecite{Dijkgraaf:1990p7194}. Through these connections it will also become clear that each inequivalent choice of 3-cocycle in our models leads to a model with specific topological and symmetry properties, as labeled by the classification in Sec.~\ref{sec:classification}.

\begin{figure}
  \centering
\includegraphics[width=0.49\textwidth]{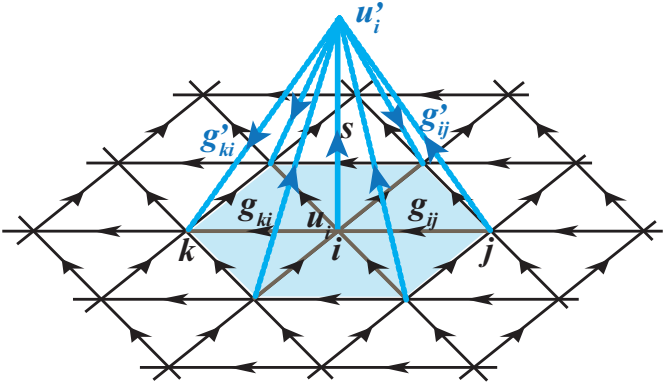}
  \caption{Action of the operator $\hat{B}^s_p$, $\gG=h_\gG\cdot \tilde{\gG}$, with $h_\gG\in\GG$, $\tilde{\gG}\in\SG$, on plaquette $p$ centered on site $i$: For six elements $\hG_{il}\in\GG$ on edges $il$ it resembles a local gauge transformation preserving zero-flux through all triangles, e.g. $\hG_{ij}'=h_s\cdot \hG_{ij}$ and $\hG_{ki}'=\hG_{ki}\cdot h_s^{-1}$, while on-site element transforms as $\gsg_i'=\tilde{\gG}\gsg_i$, leading to e.g. $g'_{ij}=s\cdot g_{ij}$ and $g'_{ki}=g_{ki}\cdot s^{-1}$; additionally, there is an overall phase factor which is the product of complex numbers assigned by the 3-cocycle $\ccy\in Z^3(\SG\times\GG,\ccygroup)$ to the six 3-simplices (tetrahedrons) forming the 'pyramid' (see Fig.~\ref{fig:order_color} and Eq.~\eqref{eq:2}). Note that fixing the initial and final state (i.e. the values $u_i,u_i'$, etc.) leads to a unique value of $s$ for which the operator matrix element does not vanish. In that sense, this picture also describes the action of the plaquette operator $\hat{B}_p$.}
\label{fig:1}
\end{figure}

We consider a triangular (two dimensional) lattice with oriented edges (bonds), where these orientations are compatible with an ordering of lattice sites, i.e. each edge oriented from a lower to higher ordered site and no triangle edges form an oriented loop, just as we discussed in Section~\ref{sec:DW} and Figure~\ref{fig:color_triangle}. For concreteness, in Fig.~\ref{fig:1} we show our choice of edge orientations on the triangular lattice. We next introduce the ``coloring'' $\colo$ by assigning an element $\hG_{ij}\in\GG$ to each oriented edge $j\rightarrow i$, again as discussed in Section~\ref{sec:DW}, \textit{however, we now also assign a group element $\gsg_i\in\SG$ to each lattice site $i$.}

Actually, we further introduce the group element
\begin{equation}
  \label{eq:53}
  \gGG_{ij}\equiv\hG_{ij}\cdot \gsg_{i}\cdot \gsg^{-1}_{j}\in\G
\end{equation}
as our variable on edge $ij$. This definition might seem somewhat redundant, since the $u_i$ elements appear both on sites, and on edges through $\gGG_{ij}$. It however has important meaning. As discussed in Section~\ref{sec:connection}, it is already known that some cohomology based classifications of phases with symmetry and phases with topological order can be explicitly connected by duality. Due to the direct product structure of the group $\G=\SG\times\GG$ we consider here, it is simple to dualize either $\SG$ or $\GG$ (entire groups or their subgroups) without having to change the formalism. We will in fact use dualization of $\SG$ explicitly when considering symmetry protected degeneracy in examples, Section~\ref{sec:multiple-vison-pairs}.

Let us then briefly consider how $\SG$ is dualized. The group variables $\gsg_{i}$ defined on lattice sites $i$ are \textit{replaced} by group variables $\gsg_{ij}$ living on oriented edges $j\rightarrow i$ according to the rule\footnote{As usual with duality, the flux, i.e. $\prod_{ij}\gsg_{ij}$, automatically vanishes through a closed loop $ij$ on the lattice. The dual $\widetilde{\SG}$ therefore has no flux particles. Also, note that the duality transformation is not a 1---to---1 mapping, since configurations $\{u_i\}$ and $\{u_i\cdot \tilde{s}^{-1}\}$ are mapped to the same configuration $u_{ij}$ in $\widetilde{\SG}$, with $\widetilde{\SG}$ understood as the group $\SG$ acting in the dual theory.}
\begin{equation}
  \label{eq:73}
  \gsg_{ij}\equiv\gsg_{i}\cdot \gsg^{-1}_{j}.
\end{equation}

Due to importance of duality, we want to ensure that all gauge degrees of freedom present in a theory are treated equally. This is our motivation for using the edge variables $\gGG_{ij}$ defined in Eq.~(\ref{eq:53}) as degrees of freedom on equal footing with $\hG_{ij}$.

An arbitrary quantum state in the Hilbert space $\mathcal{H}$ of our model is therefore labeled by $\ket{\text{i}}=\ket{\{\gsg_i\},\{\gGG_{ij}\}}$, or by $\ket{\text{i}}=\ket{\{\gsg_i\},\{\hG_{ij}\}}$.

The elementary building block for the theory is the operator $\hat{B}^\gG_p$ labeled by a group element $\gG\in\G\equiv \SG\times\GG$, and a plaquette $p$ containing six triangles sharing a lattice site $i$ at the center. The plaquette operator therefore acts on seven group elements, one at the site $i$ and six on the edges that share this site. To define its action, we need an additional edge oriented vertically up into the third dimension at site $i$, to which we assign the element $\gG\in\G$ which can always be uniquely factored as
\begin{equation}
  \label{eq:54}
\gG=h_\gG\cdot \tilde{\gG},
\end{equation}
with $h_\gG\in\GG$, $\tilde{\gG}\in\SG$ (Fig.~\ref{fig:1}).
The operator $\hat{B}^\gG_p$ transforms the seven values of elements in the plaquette by $\gG$, preserving the orientation of edges, and these new values are represented on edges lifted above the original ones, see Fig.~\ref{fig:1}. With $p$ centered on site $i$, we have:
\begin{align}
  \label{eq:19}
  \gsg_i&\rightarrow \tilde{\gG}\cdot \gsg_i\\\notag
  \hG_{ij}&\rightarrow h_\gG\cdot \hG_{ij}\\\notag
  \hG_{ki}&\rightarrow \hG_{ki}\cdot h_\gG^{-1},
\end{align}
leading to
\begin{align}\label{eq:Bedgeaction}
\gGG_{ij}&\rightarrow \gG\cdot \gGG_{ij}\\\notag
\gGG_{ki}&\rightarrow \gGG_{ki}\cdot \gG^{-1}.
\end{align}

Further, non-zero matrix elements of $\hat{B}^\gG_p$,
\begin{equation}
  \label{eq:6}
  B^\gG_p=\bra{\text{f}(\gG)}\hat{B}^\gG_p\ket{\text{i}},
\end{equation}
are assigned the following quantum amplitude
\begin{equation}
  \label{eq:2}
  B^\gG_p\equiv\prod_{I=1}^6 W(\sigma_I,\colo)^{\ccysgn(\sigma_I)},
\end{equation}
where the six 3-simplices $\sigma_I$ are built using the six triangles of the plaquette $p$ (with the initial group element values assigned), the vertical edge (assigned the group element $\gG$), and the six lifted edges (assigned the final element values according to Eq.~(\ref{eq:Bedgeaction})). This action is shown in Fig.~\ref{fig:1}. The orientation of new (lifted) edges is chosen to match the orientation of original edges upon downward projection.

It is important to note that the zero-flux rule (discussed in Sec.~\ref{sec:DW}) is by construction satisfied on all faces (triangles) of the six tetrahedra, \textit{if it is satisfied} in the 6 triangles of the plaquette $p$. The zero-flux rule must hold on all faces of the tetrahedra for which we are calculating the phase $W$. The operator $B^\gG_p$ is therefore defined only in a Hilbert subspace $\mathcal{K}_p$ which consists of states having the zero-flux rule satisfied in all six triangles of the plaquette $p$. Finally, note that choosing a final state $\ket{\text{f}}$ in a non-zero matrix element fixes a unique value of $\gG$.

We finally define the plaquette operators $\hat{B}_p$ as having matrix elements
\begin{equation}
  \label{eq:3}
  B_p=\frac{1}{|\G|}\sum_{\gG\in\G}B^\gG_p.
\end{equation}

To explicitly illustrate the plaquette operator (as well as the full Hamiltonian defined below) through examples, in Appendix\ref{sec:exampleH_Z1Z2} we will consider the models for two $Z_2$ topologically ordered phases, i.e. the well-known ``toric code'' \cite{Kitaev:2003p6185} and ``double semion'' theory\cite{Levin:2005p3468}.

Returning to the most general case, our plaquette operators $\hat{B}_p$ turn out to be projectors. Namely, using the properties of 3-cocycles, one finds that applying a $\hat{B}^\gG_p$ operator twice at the same plaquette leads to a group multiplication in the amplitude,
\begin{equation}
  \label{eq:7}
\bra{\text{f}}\hat{B}^\gG_p \hat{B}^{\gG'}_p\ket{\text{i}}=B^{\gG\cdot\gG'}_p.
\end{equation}
Using the normalization in Eq.~\eqref{eq:3} it follows that
\begin{equation}
  \label{eq:8}
  \bra{\text{f}}\hat{B}_p \hat{B}_p\ket{\text{i}}=B_p,
\end{equation}
and also that $\hat{B}_p$ is a projector.

Crucially, we will show further below that the plaquette operators commute:
\begin{equation}
  \label{eq:4}
  [B_p,B_{p'}]=0,\quad\forall p,p'.
\end{equation}

Let us next introduce the operator $Q_t$, which projects flux in a triangle $t$ to zero, i.e. it enforces the zero-flux rule discussed in Section~\ref{sec:DW}. In other words, $Q_t$ is non-zero (and equal to 1) only when acting on a triangle $t$ made out of lattice sites $i,j,k$ such that
\begin{align}
  \label{eq:9}
  \hG_{ij}\cdot\hG_{jk}\cdot\hG_{ki}&=\openone_{\GG}\\\notag
    \gGG_{ij}\cdot\gGG_{jk}\cdot\gGG_{ki}&=\openone,
\end{align}
where $\openone$ is the group identity in $\G$, and the second line follows directly from the definition Eq.~(\ref{eq:53}).

We can at last define the Hamiltonian as
\begin{equation}
  \label{eq:5}
  H=-\sum_t Q_t - \sum_p \hat{B}_p\prod_{t\in p}Q_t,
\end{equation}
where the label $t\in p$ enumerates the six triangles making up the plaquette $p$. As mentioned above, the factor $\prod_{t\in p}Q_t$ is actually crucial to ensure that $H$ is well-defined: it ensures that $\hat{B}_p$ acts within the subspace $\mathcal{K}_p$ on which it is defined (see discussion after Eq.~\eqref{eq:2}).

Further, it is easy to see that plaquette operator term $\hat{B}_p\prod_{t\in p}Q_t$ actually commutes with the projectors $Q_{t'}$. Namely, the transformation rule by $\gG$ in operator $\hat{B}^\gG_p$, as introduced above, preserves the product rule Eq.~\eqref{eq:9} on all triangle faces of simplices in Fig.~\ref{fig:1}, if it is satisfied in either the upper or lower triangles, i.e. either in the $\ket{\text{i}}$ or $\ket{\text{f}}$ state. Obviously then the zero-flux rule enforced by action of $Q_{t'}$ commutes with the action of $\hat{B}_p\prod_{t\in p}Q_t$ even when $t'$ belongs to the plaquette $p$.

Our model has the global symmetry group $\SG$, following from the fact that the Hamiltonian commutes with the global symmetry operations
\begin{equation}
  \label{eq:46}
\gsg_i\rightarrow \gsg_i\cdot\tilde{\gG}^{-1},\;\forall i,  
\end{equation}
and $\tilde{\gG}\in\SG$. The symmetry operation obviously does not influence the zero-flux rule in Eq.~(\ref{eq:9}), and therefore commutes with every $Q_t$. Considering a plaquette operator, the symmetry operation leaves the edge elements $g_{ij}$ invariant, and also the final value of site elements $u_i$ is the same no matter the order in which apply $\hat{B}_p$ and $\tilde{\gG}$, due to the group property.

Next, our model $H$ is exactly solvable: All terms in the Hamiltonian $H$ commute with each other (we still have to prove the commutation of $\hat{B}_p$, $\hat{B}_{p'}$), so the model is exactly solvable.
Let us now consider the ground state manifold of our model. Since all the terms in $H$ are also projectors, the ground state manifold is the image of the projector $P=\prod_p\hat{B}_p\prod_{t\in p}Q_t$. Actually, it is also easy to see that $\hat{B}_p^s\hat{B}_p=\hat{B}_p$ due to Eq.~\eqref{eq:7} and the group property. This means that for a ground state it also holds that $\hat{B}_p^s=1, \forall p,s$.

First, let's consider the special cases in which $SG=Z_1$ is trivial. In this case, the projector $P=\prod_p\hat{B}_p\prod_{t\in p}Q_t$ is exactly the projector in the Dijkgraaf-Witten theory, Eq.~(\ref{eq:DW_boundary}). Namely, applying all $\hat{B}_p$ operators in the plane creates a lifted copy of the plane, leaving the volume between them triangulated by tetrahedrons; the transition amplitude for this operation is equal to the product of $\prod_{I=1}^6 W(\sigma_I,\colo)^{\ccysgn(\sigma_I)}$ phases contributed by all the $\prod_p\hat{B}_p$. The initial and final states fix the coloring $\tau$ on the two planes, so the transition amplitude exactly equals the Dijkgraaf-Witten topological invariant $Z_M(\tau)$ (Eq.~(\ref{eq:DW_boundary})) evaluated on the manifold having the two planes as boundaries. (Note that there are no vertices inside the volume, and the number of plaquettes $p$ is equal to $V_{\partial M}/2$ since there are two planes in $\partial M$, leading to correct prefactor from Eq.~(\ref{eq:DW_boundary}).) We can therefore conclude that when $SG=Z_1$, the ground state sector of our model, to which $P$ projects with eigenvalue 1, is also the ground state sector of the Dijkgraaf-Witten TQFT\cite{Dijkgraaf:1990p7194} defined on the triangular plane. We will generally study the topological order of our models in Appendix \ref{app:extended_ribbon}.

On the other hand, we can consider the opposite situation where the gauge group is trivial $\GG=Z_1$, so that $\G=\SG$. In that case, our models become equal to the exactly solvable models for symmetry protected topological phases constructed by Chen \textit{et al.} in Section IIF of Ref.\onlinecite{Chen:2011p6670}. Namely, since the only degrees of freedom on the edges are from $\SG$, see Eq.~(\ref{eq:53}), the zero-flux rule is automatic so $Q_t=1$. The Hamiltonian is just a sum of the $\hat{B}_p$ plaquette operators, and these are obviously identical to the plaquette operators forming the Hamiltonian in Ref.\onlinecite{Chen:2011p6670}. We can conclude, as claimed in the introduction, that our exactly solvable models in the case of trivial gauge group $\GG$ become the models for symmetry protected topological phases classified by $H^3(\SG,U(1))$.

Our exactly solvable model is in-between these two extreme situations (the DW theory and the SPT model), and it can be understood as a partially dualized version of either of them.

In general, the ground states of our models do not break the physical symmetry $SG$ so that they describe symmetric quantum phases. The simplest way to convince oneself of this is by noting that the $\hat B^s_p$ operators in our models (see Eq.(\ref{eq:6})) create/annihilate small domain walls in a quantum state when $s\in SG$. Since in a ground state it holds that $\hat B^s_p=1$, $\forall p$, the ground state is a domain wall condensate --- i.e., the symmetric phase.

\begin{figure*}
  \centering
\includegraphics[width=1\textwidth]{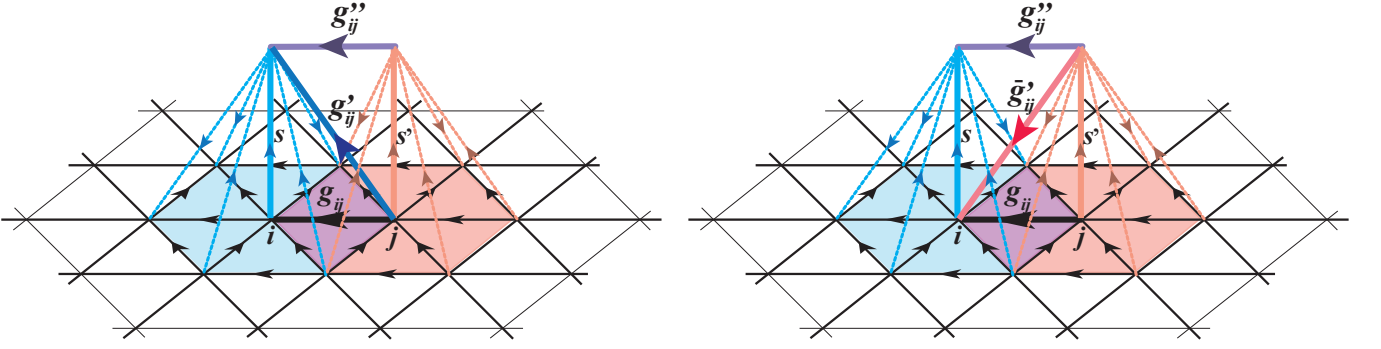}
  \caption{Overlapping plaquete operators commute. As in Fig.~\ref{fig:1}, the action and matrix elements of $BB_1=\bra{\text{f}}\hat{B}_p \hat{B}_{p'}\ket{\text{i}}$ (right) and $BB_2=\bra{\text{f}}\hat{B}_{p'} \hat{B}_p\ket{\text{i}}$ (left) are shown. Note that fixing $\ket{\text{f}}$ and therefore $g_{ij}''$ in both cases is consistent, giving $g_{ij}''=s\cdot g_{ij}\cdot {s'}^{-1}$. Concerning the overall phase factor due to the 3-cocycle factors in Eq.~\eqref{eq:2}, the two images differ only in the choice of the single internal edge (dark blue on left, pink on right) of the triangulation of the tent-shaped object. The summation over elements $s,s'$ inherent in the plaquette operators, Eq.~\eqref{eq:3}, amounts  to the sum over internal colorings, while $\ket{\text{i}},\ket{\text{f}}$ fix the surface coloring of the 'tent'. The phase factor in both quantities $BB_{1,2}$ becomes just the Dijkgraaf-Witten topological invariant (Eq.~\eqref{eq:DW_boundary}) of the 'tent' calculated with different triangulation choices, therefore having the same value.}
\label{fig:2}
\end{figure*}

Let us now prove Eq.~\eqref{eq:4} by using the Dijkgraaf-Witten topological invariant from Eq.~(\ref{eq:DW_boundary}). Consider the picture of action of two overlapping plaquette operators, described by matrix elements
\begin{equation}
  \label{eq:10}
BB_1=\bra{\text{f}}\hat{B}_p \hat{B}_{p'}\ket{\text{i}},
\end{equation}
and
\begin{equation}
  \label{eq:11}
BB_2=\bra{\text{f}}\hat{B}_{p'} \hat{B}_p\ket{\text{i}},  
\end{equation}
where $p$ and $p'$, the plaquettes centered on sites $i,j$ respectively, necessarily share two triangles, while only the $ij$ edge is acted upon by both operators, Fig.~\ref{fig:2}. The operator product $\hat{B}_p \hat{B}_{p'}$ is obviously defined within the Hilbert subspace $\mathcal{K}_{pp'}=\mathcal{K}_p\cap\mathcal{K}_{p'}$.

First note that because the final state is the same in both $BB_{1,2}$ cases (Fig.~\ref{fig:2}a,b), the final values on the $ij$ bond, $g''$ and $\bar{g}''$ respectively, have to be equal (the initial value is $g\equiv\gGG_{ij}$). Applying the rule Eq.~\eqref{eq:9} on triangular faces created by $\gG,\gG',g$ edges shows that indeed $g''=\bar{g}''=s\cdot g\cdot {s'}^{-1}$, with conventions as in Figs.~\ref{fig:1},\ref{fig:2}. Next, note that fixing the initial and final states amounts to choosing a coloring on the surface of the tent-shaped body formed on top of the $p,p'$ plaquettes. The only unconstrained internal edges are $\gG,\gG'$. By construction of the model and the properties of the ground state manifold, the edge orientations and the constraints on elements are consistent with a triangulation and coloring of the tent-shaped manifold as required in the definition Eq.~(\ref{eq:DW_boundary}) of $Z_\manif(\tau)$. The surface coloring $\tau$ is fixed by the choice of initial and final state, while the sum over $\gG,\gG'$ in the expressions for $BB_1,BB_2$ is the sum over internal colorings in the definition of $Z_\manif(\tau)$, Eq.~\eqref{eq:DW_boundary}. The $BB_1$ and $BB_2$ are therefore equal to the DW invariant $Z_\manif(\tau)$ of the tent-like object in Fig.~\ref{fig:2}, and they differ from each other only in the choice of triangulation, i.e. the position of one internal edge (notice that the value of element on this edge is also different in the two cases, but in both consistent with general coloring demands from subsection~\ref{sec:DW}). According to the properties of the DW invariant expressed by Theorem-2 (Eq.~(\ref{eq:DW_boundary})), this difference in triangulation does not change its value, meaning that $BB_1=BB_2$.

\section{Elementary Excitations}
\label{sec:elem-excit}

In this section we introduce the low energy excitations in our models, and study their general properties. We define ribbon operators which describe excitations at the ends of open strings in Sec.~\ref{sec:ribbon-operator}, having first introduced the motivation for the definition in Sec.~\ref{sec:intr-ribb-oper}. In Section~\ref{sec:extend-ribb-algebra} we will use the algebra of ribbon operators (extended by some local operators) to study the general structure of these excitations. The 3-cocycles present in our models introduce a ``twist'' into this extended ribbon algebra and therefore play a key role.

Further, we will explicitly show in Sec.~\ref{sec:example} on examples that excitations in our models can have distinctive symmetry protected properties. Appendix~\ref{app:extended_ribbon} presents in detail the general braiding and fusion of quasiparticles based on the twisted extended algebra.

Up to now, the $SG$ and the $GG$ groups in our models were either Abelian or non-Abelian. From now on, for simplicity \emph{we assume both the $SG$ and the $GG$ to be Abelian.}

\subsection{Towards ribbons: Loop operators}
\label{sec:intr-ribb-oper}

We study closed-string (loop) operators in this subsection, which will motivate the subsequent expression for open string (ribbon) operators. The loop operators we will describe commute with the Hamiltonian in Eq.~\eqref{eq:5}. The open-string operators will inherit this property locally along their string, except at the string ends, where the excitations are located.

\begin{figure*}
  \centering
\includegraphics[width=1\textwidth]{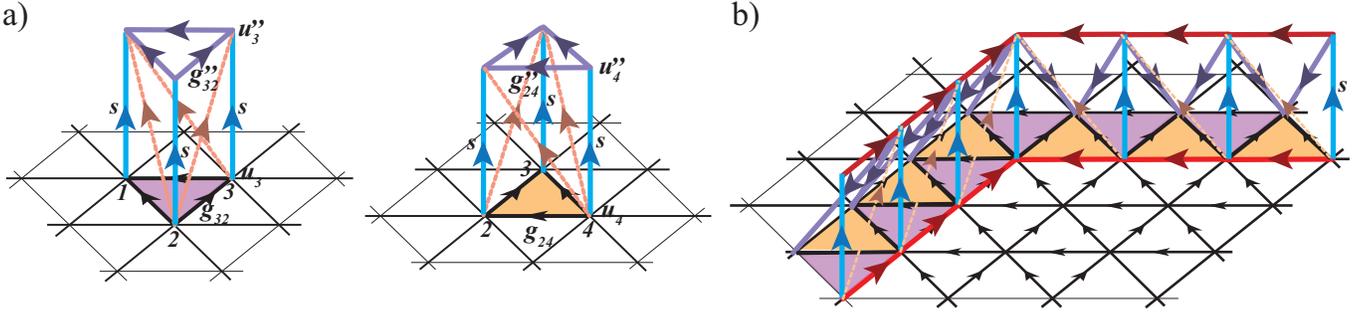}
  \caption{Creating a loop operator $\hat{\bstring}^\gG_{\loopc}=\prod_{p\in\area}\hat{B}^\gG_p$ by multiplying plaquette operators centered on sites within area $\area$ bounded by a lattice loop $\loopc$. This leads to two types of phase factors in the bulk of $\area$, coming from: (a) Left: down triangles; and (a) Right: up triangles. The $\hat{B}_p$'s in the product are chosen to be ordered according to order of vertices $i$ on which they are centered ($p\equiv i$), with highest $p$ being rightmost (i.e. applied first). For an Abelian group and the choice $\gG\equiv h_\gG\cdot\openone_{\SG}$ the elements within $\area$ and $\loopc$ do not change, i.e. $u_4''=u_4$, $g''_{32}=g_{32}$, etc., and phase factors from all tetrahedrons cancel up to a total phase due to a 2-cocycle evaluated on edges along $\loopc$, Eq.~(\ref{eq:23}). (b) The loop operator only acts on triangle edges (thicker black) lying outside $\area$ but sharing one site with $\loopc$ (red edges).}
\label{fig:3}
\end{figure*}

To define a loop operator, let us consider a contiguous area $\area$ of the lattice. This area is bounded by a sequence of connected edges on the triangular lattice forming the lattice loop $\loopc$. Next, if a lattice site $i$ is inside the area $\area$, or is lying on its boundary $\loopc$, we define the plaquette $p$ centered on $i$ to be ``inside $\area$'', i.e., $p\in\area$. Now, the loop operator is just a product of plaquette operators $\hat{B}^\gG_p$ inside the area:
\begin{equation}
  \label{eq:12}
  \hat{\bstring}^\gG_{\loopc}=\prod_{p\in\area}\hat{B}^\gG_p,
\end{equation}
where the ordering of the product is defined below, although it is physically irrelevant since the plaquette operators $\hat{B}^\gG_p$ operators commute for $p\neq p'$. Obviously the Hilbert subspace on which the loop operator is defined has the zero-flux rule obeyed in all triangles belonging to all plaquettes $p\in\area$. This space is given by $\mathcal{K}_{p_1}\cap\ldots \cap\mathcal{K}_{p_a}$, with $p_1,\ldots,p_a$ the plaquettes in $\area$. For the purpose of this subsection, we can for simplicity consider only states which satisfy the zero-flux rule in all triangles of the lattice.

The loop operator will, for $\gG\equiv h_\gG \tilde{\gG}\equiv h_\gG\cdot\openone_{\SG}$, $h_\gG\in\GG$, have an action only on the boundary $\loopc$ of the area, and therefore we label $\hat{\bstring}_\loopc^\gG$ by $\loopc$ only.

To prove this basic property of the loop operator, start by considering the bulk of $\area$, meaning the sites, edges and triangles within $\area$ including its boundary $\loopc$. We now need to fix the choice of ordering the operators $\hat{B}^\gG_p$ in the product Eq.~\eqref{eq:12} according to their plaquettes $p$. A natural choice is according to the order of lattice sites $i\equiv p$ on the lattice, putting highest $p$ rightmost in the product. (As before, $i\equiv p$ means the site $i$ on which the plaquette $p$ is centered.) This choice turns out to be the simplest and most convenient for calculations. The action of $\hat{\bstring}^\gG_{\loopc}$ in the bulk of $\area$ is then given by the expressions presented in Fig.~\ref{fig:3}a. The $\SG$ elements on the sites are unchanged due to $\tilde{\gG}\equiv\openone$, while the $\GG$ elements on the edges get conjugated by $\gG$ (contrast to Fig.~\ref{fig:1}). Since we focus on Abelian groups, the loop operator acts trivially on edges lying inside $\area$, including its boundary $\loopc$.

The non-trivial action of the loop operator is therefore limited to the edges which lie outside $\area$ but still share a site with the loop $\loopc$, and this action is due to the $\hat{B}^\gG_p$ operators having $p$ centered on a site on the loop $\loopc$, see Fig.~\ref{fig:3}b.

Returning to the contribution from the bulk of $\area$, it reduces to the phase factors of 3-simplices (tetrahedrons) lying on top of the bulk of $\area$, according to the action of operators $\hat{B}_p^s$. To evaluate these, we need to consider further the chosen ordering of $p$'s in the product. Consider an oriented edge $j\rightarrow i$ inside $\area$ (including $\loopc$), so that its arrow points towards $i$. By our choice $\hat{B}_{p\equiv i}^s$ is \textit{applied before} $\hat{B}_{p\equiv j}^s$. The action of $\hat{B}_{p\equiv i}^s$ assigns the orientation $j\rightarrow i$ to the lifted edge as in Fig.~\ref{fig:1}, so the lifted edge must have its arrow pointing ``upwards'' (towards the lifted $i$ vertex). This property applies to all edges in the plane, so every lifted edge, connecting a vertex to a lifted vertex, will have its arrow pointing upwards, i.e. towards the lifted plane.

Having determined this fact, it is simple to determine the tetrahedrons formed by action of $\hat{\bstring}^\gG_{\loopc}$ in bulk of $\area$, Fig.~\ref{fig:3}a. It becomes obvious that the bulk of the area is spanned by only two different types of triangles.
\begin{figure*}
  \centering
\includegraphics[width=1\textwidth]{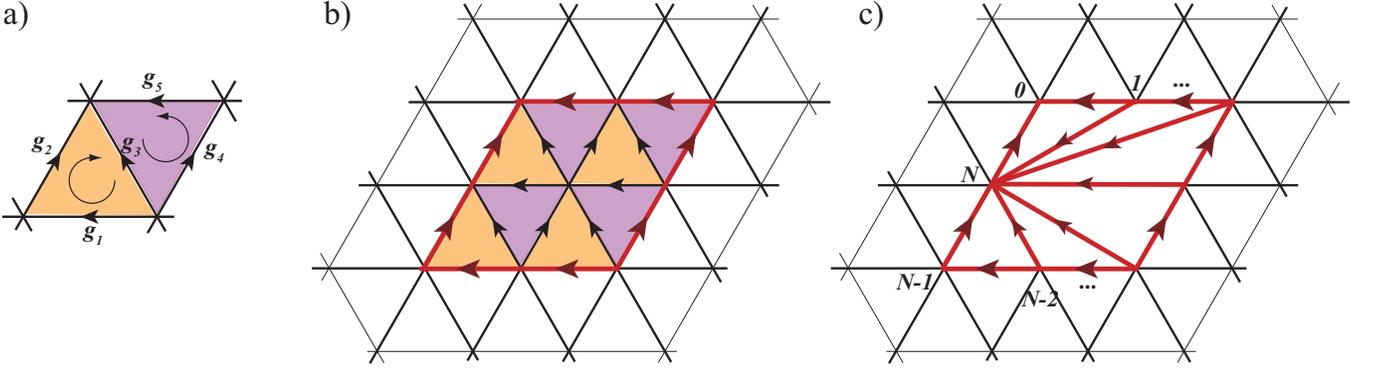}
  \caption{(a) The loop operator $\hat{\bstring}^\gG_{\loopc}$, with $\loopc$ the loop of lattice edges bounding area $\area$, has total contribution from triangles within $\area$ which equals the product of 2-cocycle $\ccyc_\gG$ (Eq.~(\ref{eq:15})) evaluated on the triangles as oriented 2-simplices (compare to Figs.~\ref{fig:color_triangle} and~\ref{fig:3}). (b) An example loop $\loopc$ (red edges). (c) The phase can be calculated by re-triangulating the area $\area$ inside the loop $\loopc$. The resulting expression for the phase $\phasec_\loopc^\gG$, Eq.~(\ref{eq:17}), is given in terms of elements along edges on $\loopc$ with clockwise ordered vertices $i=0,\ldots,N$. The expression is used for the definition of an open string, Section~\ref{sec:ribbon-operator}.}
\label{fig:4}
\end{figure*}
For the up ($\bigtriangleup$) and down ($\bigtriangledown$) triangles on the lattice, as labeled in Fig.~\ref{fig:3}a, the resulting phases are
\begin{align}
  \label{eq:13}
  \phase^\gG_\upt&=\ccyc_\gG(g_{32},g_{24}),\\\notag
    \phase^\gG_\downt&=\ccyc^{-1}_\gG(g_{13},g_{32}),
\end{align}
where using the 3-cocycle $\ccy$ we have introduced the function:
\begin{equation}
  \label{eq:14}
\ccyc_\gG (g_1,g_2)\equiv\frac{\ccy(\gG,g_1,g_2)\,\ccy(g_1,g_2,\gG)}{\ccy(g_1,\gG,g_2)},
\end{equation}
with $g_1,g_2\in\G$. The function $\ccyc_\gG (g_1,g_2)$ is most generally parametrized by an arbitrary group element $\gG\in\G$, even though here $\gG\in\GG$, and it can be directly shown that this function satisfies the 2-cocycle condition introduced in Eq.~\eqref{eq:2-cocycle}.

On the other hand, the 2-cocycle ($\ccyc_\gG$) value appearing in the phases Eq.~\eqref{eq:13} exactly corresponds to the phase assigned to the 2-simplices (i.e. triangles) on our ordered and colored lattice, according to the general considerations from Section~\ref{sec:DW}. More precisely, just as in that section, a 2-simplex $\sigma$ is defined by a triangle with ordered vertices and group elements assigned to its edges. The 2-simplex is assigned the complex number
\begin{equation}
  \label{eq:18}
\phasec^\gG (\sigma,\colo)=W(\sigma,\colo)^{\ccysgn (\sigma)},
\end{equation}
where $W(\sigma,\colo)\equiv\ccyc_\gG (g_{kj},g_{ji})$ for a triangle with ordered vertices $i\rightarrow j\rightarrow k$. The sign $\ccysgn=\pm 1$ is given by the chirality, see Fig.~\ref{fig:color_triangle} and Fig.~\ref{fig:4}.

The bulk of $\area$ is formed by the 2-simplices $\sigma_I$, so the total phase contributed by the bulk of $\area$ is:
\begin{equation}
  \label{eq:20}
  \phase^\gG_\area=\prod_{I\in\area} \phasec^\gG(\sigma_I,\colo),
\end{equation}
which one can calculate by changing the triangulation of this area as in Fig.~\ref{fig:4}. (Recall the this re-triangulation will not change the total phase due to the allowed ``moves'' from Fig.~\ref{fig:1-3_2-4_moves}.) Namely, we make all the triangles share the vertex $i=N\in\loopc$. Labeling the vertices $i$ on the lattice loop in CW order $\loopc=\{i\mid i=0,1,\ldots ,N\}$, see Fig.~\ref{fig:4}c, the phase $\phase^\gG_\area\equiv\phasec_\loopc^\gG$ becomes
\begin{widetext}
\begin{equation}
  \label{eq:17}
  \phasec_\loopc^\gG=\ccyc_\gG(g_{N,N-1},g_{N-1,N-2})\, \ccyc_\gG(g_{N,N-1}\cdot g_{N-1,N-2},g_{N-2,N-3})\cdots\ccyc_\gG(g_{N,N-1}\cdots g_{21},g_{10}),
\end{equation}
\end{widetext}
where we took into account the zero-flux rule as well as $g_{ij}=g^{-1}_{ji}$. Eq.~(\ref{eq:17}) shows explicitly that the contribution of bulk of area $\area$ depends only on the edges along its boundary loop $\loopc$.

This expression will be useful for us later on, but let us now consider in more detail the case when the 2-cocycle $\ccyc_\gG$ is trivial, which means that it can be rewritten in the form
\begin{equation}
  \label{eq:15}
  \ccyc_\gG (g_1,g_2)=\frac{\ccye_\gG(g_1)\,\ccye_\gG(g_2)}{\ccye_\gG(g_1\cdot g_2)}.
\end{equation}
Because in expression Eq.~(\ref{eq:15}) the three elements $g_1,g_2,g_1\cdot g_2$ belong to the three sides of the triangle, it is easy to check that contributions to $\phasec^\gG$ from an edge shared by an up and down triangle cancel. The total phase then becomes a simple expression obtained by going around the loop:
\begin{multline}
  \label{eq:21}
  \phasec_\loopc^\gG=\ccye_\gG(g_{N,N-1})\, \ccye_\gG(g_{N-1,N-2})\cdots\ccye_\gG(g_{20})\,\ccye_\gG(g_{10})=\\
 =\prod_{\substack{i\in\loopc\\ CW}}\ccye_\gG(g_{i+1,i}).
\end{multline}

Let us now return to the general case of arbitrary 2-cocycle $\ccyc_s$. Having dealt with the bulk contribution of $\area$, the action of the loop operator is presented in Fig.~\ref{fig:3}b, where a segment of the loop is marked by red edges, as in Fig.~\ref{fig:4}. Note again that along the loop $i,i'\in\loopc$, elements $u_i,g_{ii'}$ are unchanged between the initial and final states, just as they are not changed inside the loop.

\subsubsection{Summary for loop operators}
\label{sec:summ-loop-oper}

In summary, the loop operator $\hat{\bstring}^\gG_{\loopc}$ only acts on two edges per triangle that has a vertex or edge on the loop $\loopc$. This action follows from the action of $\hat{B}^\gG_p$ (see Eq.~(\ref{eq:19})) with $p$ centered on the loop $\loopc$. (Such edges are marked thick black in Fig.~\ref{fig:3}c.) Additionally, the loop operator has a phase factor contribution from the 3-cocycle of tetrahedrons $\sigma_t$ on top of each such triangle (see shaded triangles in Fig.~\ref{fig:3}c), which we do not explicate as they cannot be simplified further; this phase factor we label succinctly by
\begin{equation}
  \label{eq:22}
  \phasew^\gG_\loopc\equiv\prod_{t\in\loopc}W(\sigma_t)^{\ccysgn (\sigma_t)}.
\end{equation}
Finally, there is the overall phase $\phasec_\loopc^\gG$, giving the total amplitude for the loop operator:
\begin{align}
  \label{eq:23}
  \bstring^\gG_{\loopc}&=\phasec_\loopc^\gG\phasew^\gG_\loopc=\\\notag
&=\prod_{\substack{i=1}}^{N-1}\ccyc_\gG\left(\prod_{\substack{j=i}}^{N-1}g_{j+1,j},g_{i,i-1}\right)\prod_{t\in\loopc}W(\sigma_t)^{\ccysgn (\sigma_t)}=\\\notag
  &=\prod_{\substack{i\in\loopc\\ CW}}\ccye_\gG(g_{i+1,i})\prod_{t\in\loopc}W(\sigma_t)^{\ccysgn (\sigma_t)},
\end{align}
where the last line holds only for trivial 2-cocycles $\ccyc_\gG$ and the CW loop $\loopc$ contains vertices $\loopc=\{i\mid i=0,1,\ldots ,N\}$.

One can show, using a direct method as in Fig.~\ref{fig:2}, that the plaquette operator $\hat{B}_p$ overlapping with the loop $\loopc$, e.g. $p\in\loopc$, commutes with the loop operator.

The final expression Eq.~(\ref{eq:23}) motivates a definition of an open string operator, to which we now turn.

\subsection{Ribbon operator}
\label{sec:ribbon-operator}

In this subsection we introduce the open-string (ribbon) operators, which describe the low energy excitations in our models.

Let us start by defining a geometric object: the open ribbon $\pathg$. The ribbon has two ends, end-A and end-B, which we need to define first. Choosing two neighboring vertices on the triangular lattice, vertex-$i_A$ and vertex-$i'_A$, there is a unique 2-simplex (triangle) formed by vertex-$i_A$ and vertex-$i'_A$ and another vertex that is \emph{not} contained in $\pathg$. We denote this 2-simplex by $t_A$. We then use the label $A=(i_A,t_A)$, the collection of the vertex-$i_A$ and the triangle $t_A$, to formally define the end-A of the ribbon $\pathg$. Similarly, we will use $B=(i_B,t_B)$ to define the end-B of $\pathg$. The 2-simplices $t_A,t_B$ are \emph{not} inside $\pathg$. We also define the edges of ribbon $\pathg$ at the two ends: for end-A, there is a single 1-simplex that is shared between $\pathg$ and $t_A$, which we denote as edge-$A$. Similarly we can define edge-$B$.

Having defined its ends, the open ribbon $\pathg$ is finally specified by its two ``ribbon edges''. Namely, $\pathg$ has an ``inner-edge'', which is a sequence of connected edges on the triangular lattice, going from the vertex-$i_A$ to the vertex-$i_B$.  Further, $\pathg$ also has an ``outer-edge'', running from the vertex-$i'_A$ to the vertex-$i'_B$, which is displaced from the inner-edge of $\pathg$ by one lattice spacing. (See Fig.\ref{fig:FD} for an illustration of these definitions.) As a geometric object, the ribbon $\pathg$ contains all the vertices on the inner and outer edges, together with all 1-simplices and 2-simplices connecting these vertices (for these simplices, we will say that they are \emph{inside} $\pathg$, or we will write $\in\pathg$).

We now proceed to define the operator $\hat F^{(h,g)}(\pathg)$ for a given open ribbon $\pathg$.

Let us define a Hilbert sub-space $\mathcal{K}(\pathg)\subset\mathcal{H}$, formed by those states which satisfy the zero-flux rule everywhere inside $\pathg$. Namely, $\forall \ket{\psi}\in\mathcal{K}(\pathg)$ and $\forall t\in\pathg$, $Q_t\ket{\psi}=\ket{\psi}$. (Note that however, $\ket{\psi}$ may violate the zero-flux rule at $t_A,t_B$, for instance.) \emph{For the purpose of this paper, we only define the operator $\hat F^{(h,g)}(\pathg)$ in the Hilbert sub-space $\mathcal{K}(\pathg)$.}

Based on the understanding of the loop operator from previous subsection, we define the ribbon operator $\hat F^{(h,g)}(\pathg)$ such that it modifies the gauge degrees of freedom living on the lattice edges connecting the inner and outer ribbon edges of $\pathg$, while leaving all degrees of freedom living elsewhere unchanged. In particular, for a lattice edge $ij$ inside $\pathg$, \textit{such that it connects the inner and outer ribbon edges}, the group element $\hG_{ij}$ living on it is changed into $\hG'_{ij}=h\cdot \hG_{ij}$ [$\hG'_{ij}=\hG_{ij}\cdot h^{-1}$] if the lattice edge is oriented to point towards the inner[outer] ribbon edge. The operator $\hat F^{(h,g)}(\pathg)$ therefore has non-zero matrix element only between states $|\{u_i\},\{\hG'_{ij}\}\rangle$ and $|\{u_i\},\{\hG_{ij}\}\rangle$.

\begin{figure*}
  \centering
\includegraphics[width=1\textwidth]{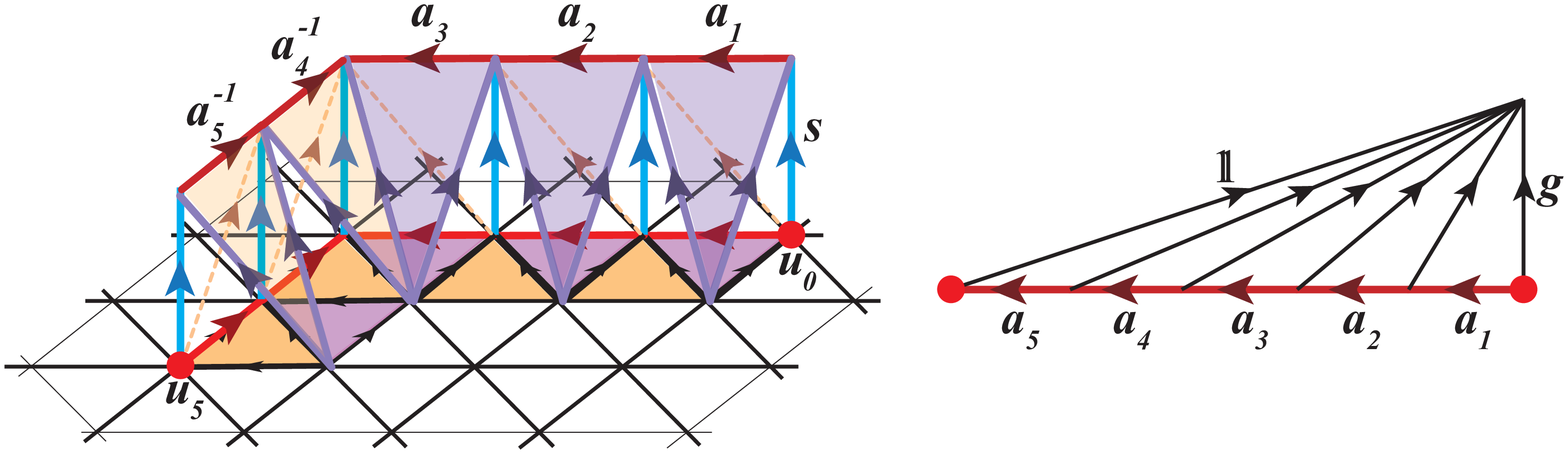}
\caption{The phase contribution $\wstring_\gG^{\pathg}(\gGG)=\phasew^\gG_\pathg\, \phasec_\pathg^\gG\;\delta\left(\prod_{n=1}^N a_n,\gGG\right)$ to matrix element of ribbon operator $F^{(s,g)}(\pathg)$, with $s\in\GG$, on a length $N=5$ example. (Left) The operator acts non-trivially only on elements adjacent to path $\pathg$ (thick black edges)
  Note that elements $a_{i}\equiv g_{i,i-1}\equiv \hG_{i,i-1}u_i u_{i-1}^{-1}$ are defined to be directed along the path, so for $n=4,5$ they are opposite to the standard definition, which is along the edge orientation. Elements at ends $A$, $B$ of the ribbon $\pathg$, i.e. vertex-$i_A$ and vertex-$i_B$, are $u_5\in\SG$ and $u_0\in\SG$, respectively. The phase factor $\phasew^\gG_\pathg$ is the product of 3-simplex phases $W(\sigma_I)^{\ccysgn(\sigma_I)}$ shown on top of shaded triangles. (Right) The phase $\phasec_\pathg^\gG$ is the product of 2-simplex phases $W(\sigma_t)^{\ccysgn(\sigma_t)}$ shown, where $\openone$ is the identity element, and the parameter $g=a_1\cdots a_N\in\G$.}
\label{fig:5}
\end{figure*}

Finally, we need to define the non-zero matrix element of $\hat F^{(h,g)}(\pathg)$. This matrix element has two factors: one chosen in accordance with the closed-loop operator, and the other dependent on the degrees of freedom at the two ends of the open ribbon $\pathg$. We define:
\begin{align}
  \label{eq:Fmatelem}
 \langle \{u_i\},\{\hG'_{ij}\}|F^{(h,g)}(\pathg)|\{u_i\},\{\hG_{ij}\}\rangle= f_A\cdot f_B \cdot f_{AB}\cdot w_h^{\pathg}(g),
\end{align}
where: 1) $w_h^{\pathg}(g)$ is motivated by the closed-loop operator and will be presented shortly; and 2) $f_A,f_B,f_{AB}$ are rather complicated phase factors depending only on the degrees of freedom living on ends of $\pathg$, and we present them in Appendix~\ref{app:extended_ribbon}.
To motivate the expression for $w_h^{\pathg}(g)$, let us start from the expression for loop operator Eq.~(\ref{eq:23}). Let the $\pathg$ ribbon's inner edge go along the sequence of lattice sites $\{i\mid i=0,\ldots,N\}$, where now the sites $i=0$ and $i=N$ are not nearest neighbors, but rather the vertex-$i_B$ and vertex-$i_A$, respectively. For convenience (see Fig.~\ref{fig:5} for the pictorial definition), we define the group elements $a_{n}\in\G$, $n=1,\ldots,N$ by $a_{i}\equiv g_{i,i-1}$. We then define the phase
\begin{widetext}
\begin{align}
  \label{eq:24}
  \wstring_h^{\pathg}(\gGG)&=\phasew^h_\pathg\, \phasec_\pathg^h\;\delta\left(\prod_{i=1}^N g_{i,i-1},\gGG\right)\equiv\\\notag
  &\equiv\left(\prod_{t\in\pathg}W(\sigma_t)^{\ccysgn (\sigma_t)}\right) \ccyc_h(\openone,a_N)\,\ccyc_h(a_N,a_{N-1})\, \ccyc_h(a_N\cdot a_{N-1},a_{N-2})\cdots\ccyc_h(a_N\cdots a_2,a_1)\,\delta(a_1\cdots a_N,\gGG)=\\\notag
    &=\left(\prod_{t\in\pathg}W(\sigma_t)^{\ccysgn (\sigma_t)}\right) \ccyc_h(g\cdot a_1^{-1},a_1)\,\ccyc_h(g\cdot a_1^{-1}\cdot a_2^{-1},a_2)\cdots\ccyc_h(g\cdot a_1^{-1}\cdots a_{N-1}^{-1},a_{N-1})\,\ccyc_h(g\cdot a_1^{-1}\cdots a_N^{-1},a_N)\,\delta(a_1\cdots a_N,\gGG),
\end{align}
\end{widetext}
with $\delta(g,g')$ the Kronecker delta function, which we used in the second line to obtain a simple pictorial definition, Fig.~\ref{fig:5}.
\begin{figure*}
  \centering
\includegraphics[width=1\textwidth]{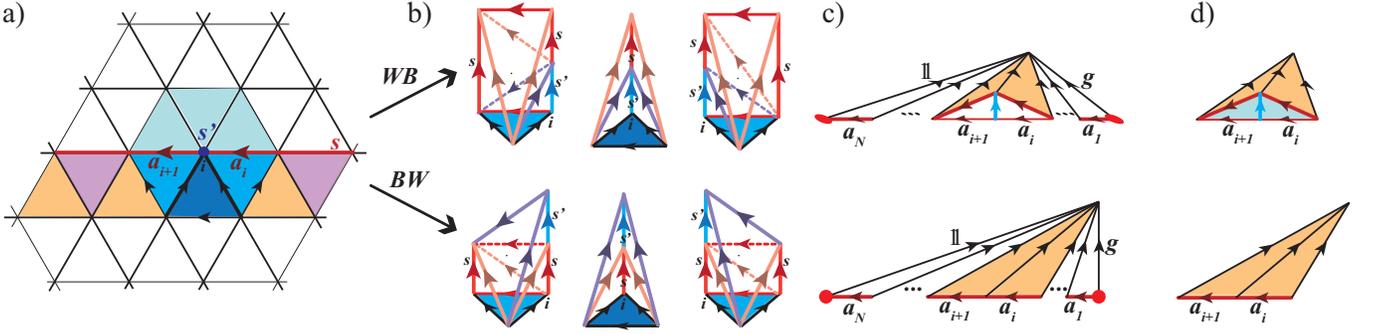}
  \caption{Inner segment of ribbon operator $F^{(s,g)}(\pathg)$, with $s\in\GG$, commutes with plaquette operator $\hat{B}^{\gG'}_p$ having $p$ centered on inner edge of $\pathg$, and therefore also with the Hamiltonian. (a) Only elements on two edges (thick black) are changed, but in the same way for $F\hat{B}$ and $\hat{B}F$ due to Abelian $\G$. (b) Quantum phase for $F\hat{B}$ (top row) and $\hat{B}F$ (bottom row) differ only due to shown 3-simplices which are on top of three blue shaded triangles in (a). (c) The phase contribution due to string phase $\phasec$ is shown. (d) The 3-simplices contribution to phase factor ratio $WB/BW$ (from panel b) equals precisely the phase $W(\sigma_t)^{\ccysgn(\sigma_t)}$ due to two blue 2-simplices (see Eq.~(\ref{eq:27})). The $WB$ (top) and $BW$ (bottom) case are therefore graphically seen to be the same, as interior points of the polygon can be removed in a 1-3 move (Fig.~\ref{fig:1-3_2-4_moves}).}
\label{fig:comm}
\end{figure*}

A way to understand the meaning of phase in Eq.~\eqref{eq:24} is to note that ``cutting open'' the loop to get an open string introduced the parameter $\gGG\in\G$ which is related to the charge carried by the excitations at the string ends. We will consider it in more detail below. (Compared to the loop, there is one extra factor $\ccyc_h(\openone,a_N)$, which is inconsequential for canonical cocycles, see Eq.~(\ref{2cy1}).)

In accordance with the definition of loop operator, the element $h\in\GG\subset\G$, see after Eq.~(\ref{eq:12}). Physically, the loop operator $\hat{\bstring}^h_{\loopc}$ can be seen as a closed, $\loopc$ loop-shaped, domain wall inside which we acted by element $h$ (through action of $\hat{B}_p^h$). For a gauge element $h\in\GG$ there is actually no transformation inside the domain wall. It is further possible to create an 'open domain wall' (i.e. open string $\pathg$) in the gauge transformation $h$, and it defines gauge excitations at the ends of $\pathg$. (On the other hand, one could define $\hat{\bstring}^{\tilde{h}}_{\loopc}$ for a global symmetry operation $\tilde{h}\in\SG$, which would create a closed domain wall inside which the transformation $\tilde{h}\in\SG$ is applied. However, it is physically unsound to try to define an open domain wall of such a transformation $\tilde{h}\in\SG$.)


It is important to realize that the delta function in the ribbon operator, $\delta(a_1\cdots a_N,\gGG)$, with $a_{i}\equiv g_{i,i-1}=\hG_{i,i-1}\cdot u_{i}\cdot u^{-1}_{i-1}$, has different effect on the global symmetry and gauge parts of $\G=\GG\times\SG$. Namely, given the factorization $\gGG=h_\gGG\cdot\tilde{\gGG}$, the delta function separates into
\begin{equation}
  \label{eq:55}
  \delta(a_1\cdots a_N,\gGG)=\delta(\hG_{10}\cdots \hG_{N,N-1},h_\gGG)\,\delta(u_N u_0^{-1},\tilde{\gGG}).
\end{equation}
The delta function therefore constrains the product of gauge degrees of freedom along the inner edge of ribbon to the value $h_\gGG$, while leaving \textit{only one constraint on the undualized elements of $\SG$} at the two ends of the ribbon: $u_N=\tilde{\gGG}\cdot u_0$. (Recall that actually $u_N$ is the element on site vertex-$i_A$, while $u_0$ is on vertex-$i_B$.)

A special case that offers insight occurs when the 2-cocycle $\ccyc_h$ is trivial. In that case, using the property Eq.~(\ref{eq:15}) of trivial cocycles, we get
\begin{equation}
  \label{eq:31}
  \wstring_h^{\pathg}(\gGG)=\left(\prod_{t\in\pathg}W(\sigma_t)^{\ccysgn (\sigma_t)}\right) \delta(a_1\cdots a_N,\gGG)\,\ccye^{-1}_h(\gGG)\prod_{i=1}^N\ccye_h(a_i),
\end{equation}
where $\ccye_h$ is a family of $\ccygroup$ valued functions on $\G$ (i.e. a 1-cochain) parametrized by the element $h$.

We now show that away from its endpoints the ribbon operator $F^{(\gG,\gGG)}(\pathg)$ commutes with $\hat{B}^{\gG'}_p$, and therefore also with the Hamiltonian in Eq.~(\ref{eq:5}). Obviously, the non-trivial situation occurs when the plaquette $p$ is centered on a site $i\in\pathg$ which is positioned on the inner edge of $\pathg$, see Fig.~\ref{fig:comm}a. Note that by definition of the operators, elements $\gG\in\GG$ and $\gG'\in\G$. The product of the two operators is defined in the Hilbert subspace $\mathcal{K}(\pathg)\cap\mathcal{K}_p$. Physically, we do not allow states with a flux-carrying excitation positioned on the ribbon $\pathg$ or inside the plaquette $p$, as we consider this commutator. We also choose $p$ that does not overlap with $t_A$ or $t_B$, i.e. we consider the commutation away from ribbon ends.

It is clear that the final states of the system are the same no matter the order of applying the two operators, since $\G$ is Abelian. The two resulting quantum phases $\bra{\text{f}}F^{(\gG,g)}(\pathg) \hat{B}^{\gG'}_p\ket{\text{i}}$ and $\bra{\text{f}}\hat{B}^{\gG'}_p F^{(\gG,g)}(\pathg) \ket{\text{i}}$ we label by $\wstring B$ and $B\,\wstring$, respectively, and we will show that their ratio is $1$.

The $\wstring B$ and $B\,\wstring$ differ in their phase factor $\phasec_\pathg^\gG$ (which was defined in Fig.~\ref{fig:5}), and these are shown on the right of Fig.~\ref{fig:comm}(b). Recalling that $a_i\equiv \gGG_{i,i-1}=\hG_{i,i-1}\cdot u_{i}\cdot u^{-1}_{i-1}$ and $i=1,\ldots,N$, the difference is due to operator $\hat{B}^{\gG'}_{p}$ sending
\begin{align}
  \label{eq:25}
  a_i&\rightarrow \gG'\cdot a_i\\\notag
  a_{i+1}&\rightarrow a_{i+1}\cdot{\gG}'^{-1},
\end{align}
see the basic definition, Eq.~(\ref{eq:19}), and we remind that $p$ is centered on the site $i$. Due to the product structure in $\phasec_\pathg^\gG$, there are only two factors affected by this change, and these are the shaded 2-simplices in Fig.~\ref{fig:comm}(b)(right). The ratio of quantum amplitudes is
  \begin{align}
    \label{eq:26}
    \frac{\phasec_{\wstring B}}{\phasec_{B\,\wstring}}&=\frac{\ccyc_\gG(A\cdot \gG'^{-1}\cdot a_i^{-1},\gG'\cdot a_{i})\, \ccyc_\gG(A\cdot a_i^{-1}\cdot a_{i+1}^{-1},a_{i+1}\cdot \gG'^{-1})}{\ccyc_\gG(A\cdot a^{-1}_i,a_{i})\, \ccyc_\gG(A\cdot a_i^{-1}\cdot a_{i+1}^{-1},a_{i+1})}=\\\notag
    &=\frac{\ccyc_\gG(\gG'^{-1}\cdot a_{i+1},\gG')}{\ccyc_\gG(,\gG',a_i)},
\end{align}
where $A$ is the product of $a_{i'<i}$ elements, and we used the 2-cocycle condition (see Eq.~(\ref{eq:2-cocycle})).

Further, it is obvious that the phase difference coming from 3-simplices is only due to ones stacked on top of three triangles that are shared by the ribbon and the plaquette, Fig.~\ref{fig:comm}(a), and these simplices are presented in detail in Fig.~\ref{fig:comm}(b). Grouping the phases according to the triangles from left to right, and using the definition of the 2-cocycle Eq.~\eqref{eq:14}, we get
\begin{align}
  \label{eq:16}
      \frac{\phasew^1_{\wstring B}}{\phasew^1_{B\,\wstring}}&=\ccyc_\gG(a_{i+1},h)\, \ccyc^{-1}_\gG(\gG'^{-1}\cdot a_{i+1},\gG'\cdot h)\,\ccy^{-1}(\gG',h,\gG)\\\notag
      \frac{\phasew^2_{\wstring B}}{\phasew^2_{B\,\wstring}}&=\ccy(\gG',\gG,h)\,\ccy^{-1}(\gG,\gG',h)\,\ccy(\gG,\gG',a_i\cdot h) \,\ccy(\gG',\gG,a_i\cdot h)\\\notag
            \frac{\phasew^3_{\wstring B}}{\phasew^3_{B\,\wstring}}&=\ccyc_\gG(\gG', a_{i})\,\ccy(\gG',\gG,a_i\cdot h) \,\ccy^{-1}(\gG,\gG',a_i\cdot h),
\end{align}
where $h$ is the element on the leftmost edge of left triangle in Fig.~\ref{fig:comm}(a). The total phase due to 3-simplices becomes
\begin{equation}
  \label{eq:27}
      \frac{\phasew_{\wstring B}}{\phasew_{B\,\wstring}}=\frac{\ccyc_\gG(\gG',a_i)}{\ccyc_\gG(\gG'^{-1}\cdot a_{i+1},\gG')}.
    \end{equation}
    This phase ratio exactly cancels the contribution from the string of 2-cocycles in Eq.~\eqref{eq:26}, completing the proof that the inner piece of ribbon operator commutes with plaquette operators.

We point out the pictorial interpretation of the result, that the 3-simplex contributions to the phase ratio, Eq.~\eqref{eq:27}, are exactly equal to the phase of the two blue-shaded triangles in Fig.~\ref{fig:comm}(c). The total phase ratio is then equal to the ratio of the upper and lower polygons in Fig.~\ref{fig:comm}(c), which obviously equals $1$ due to the rule that allows removal of internal points in polygons (rule from Fig.~\ref{fig:1-3_2-4_moves}).

\subsection{Local symmetric operators and the twisted extended ribbon algebra}
\label{sec:extend-ribb-algebra}
\begin{figure*}
  \centering
\includegraphics[width=1\textwidth]{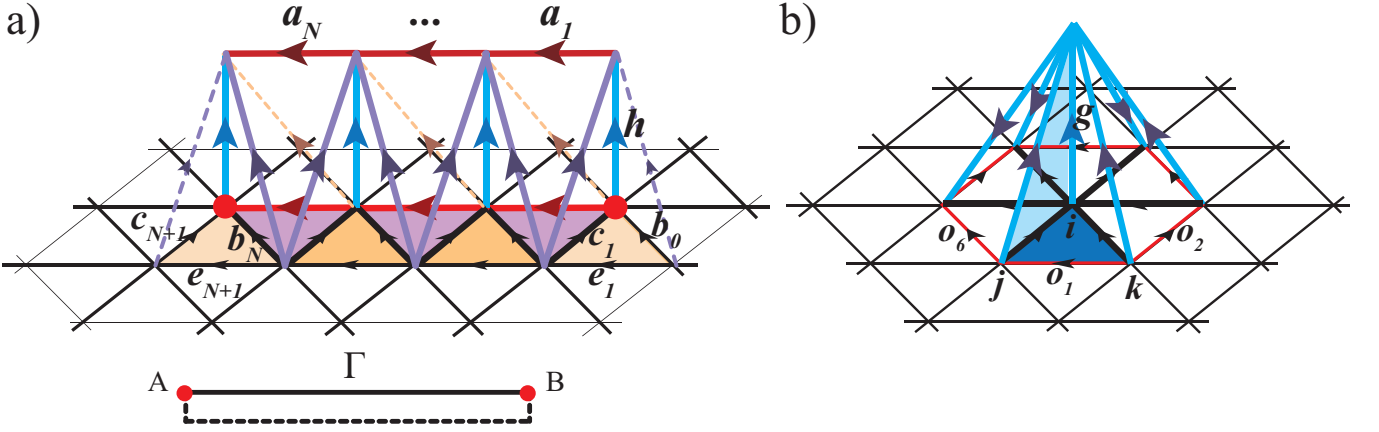}
\caption{(a) Ribbon operator with both ends (red dots), shown schematically in the bottom. It only acts on the thick black edges, consistent with zero-flux rule on all 2-simplices above and within the ribbon. The transition amplitude of the ribbon is given in Fig.~\ref{fig:5}
without the phase factor $f_Af_Bf_{AB}$ local at the string ends; $f$ is involved (Appendix \ref{app:extended_ribbon}), but includes the edges of two triangles $t_A,t_B$ at the ends (shaded light orange). Dashed violet lines indicate edge elements that are not acted on by the operator.
  (b) The local operator at position $B\equiv (i,t)$, with $i$ the lattice site and triangle $t$ (dark blue) having vertices $i,j,k$ (shaded dark blue), has the transition amplitude $\bra{\text{f}}D_{(h,g)}(B)\ket{\text{i}}=\delta(g_{ij}\cdot g_{jk}\cdot g_{ki},h)\,W_h\,W_6$. It acts as the plaquette operator $\hat{B}^g_p$ (Fig.~\ref{fig:1}) with plaquette $p$ centered on $i$ (changing only thick black edge elements), except that it also projects flux in $t$ to value $h$, and also has an additional phase factor $W_h$ which is the phase of the 2-simplex (shaded light blue). (For operator $D(A)$, the triangle $iki'$ is used instead.) The phase factor $W_6$ depends on six drawn 3-simplices (tetrahedrons) having altered elements $o$ (red edges) such that it is well-defined even if the zero-flux rule is violated in the plane (as occurs at ribbon string ends), see after Eq.~(\ref{eq:36}).}
\label{fig:FD}
\end{figure*}

In this subsection we will introduce local symmetric operators which have a non-trivial algebra with the ribbon operators. This will allow us to understand the general structure of excited states. The focus will be on states with a single pair of excitations, or two pairs when discussing braiding. The obtained results will be directly relevant for studying examples in Sec.~\ref{sec:example}. The fully general case of many quasiparticle pairs, including their braiding and fusion properties, will be studied in detail in Appendix~\ref{app:extended_ribbon} by using the extended algebra.

Before giving the formal definitions, let us remark that given the positions of excitations, the ``extended algebra'' contains the set of ribbon operators $F$ (with their strings connecting pairs of excitations), as well as a set of local operators $D$ acting at the positions of excitations. However, the presence of the 3-cocycle $\ccy$ adds a ``twist'' in the algebra and crucially determines the resulting properties of excitations. It has been shown\cite{Propitius:1993CS_ccy} that certain broken gauge theories with Chern-Simons terms lead to discrete (group $\widetilde{H}$) gauge theories having such twisted algebra describing their quasiparticles \cite{Propitius:1993CS_ccy}. In that situation, the cocycle is generated by the Chern-Simons term, and the resulting discrete gauge theory can be classified using $H^3(\widetilde{H},U(1))$,\cite{Propitius:1995p6856} making connection with discrete DW TQFTs. Importantly, a non-trivial cocycle twisting of the algebra can render the resulting theory non-Abelian even though its gauge group $\widetilde{H}$ is Abelian\cite{Propitius:1997ccy_nA}. Our models inherit this interesting property, and we will discuss this briefly concerning properties of multiple pairs of excitations in Section~\ref{sec:multiple-vison-pairs}.
In Appendix~\ref{app:extended_ribbon}  we will also show explicitly how the considered operators of our models form a Hopf algebra (more precisely, a quasi-triangular quasi-Hopf algebra due to the cocycle twist, see Refs.\onlinecite{Propitius:1993CS_ccy,Dijkgraaf:1990p7462}), and how they describe the braiding and fusion of excitations\cite{Propitius:Hopf}.

It is important to emphasize that in the present context the local operators $D$ are also crucial for determining the interplay of symmetry and topological order: After we explicitly construct the $D$ operators, we will show that they are symmetric, i.e. commute with transformations from $\SG$, and further we expect them to span the algebra of all local symmetric operators.\footnote{Although we do not have a general proof, we believe that this is true, which is also confirmed in the examples in Sec.\ref{sec:example} that we solved.}
Thereby the set $D$ will provide us all symmetry allowed local perturbations and therefore the possibility of calculating the symmetry protected degeneracy and other properties of excited states in Section~\ref{sec:example}.


Let us briefly recall some relevant details about the ribbon operator $F^{(h,g)}(\pathg)$ from Sec.~\ref{sec:ribbon-operator}, see Fig.~\ref{fig:FD}. By definition, the element $h\in\GG$ while $g\in\G$. The operator definition demands that the zero-flux rule is satisfied for all triangles in ribbon $\pathg$, i.e. it acts within the Hilbert subspace $\mathcal{K}(\pathg)$. The two ends of ribbon $\pathg$ we label by $A,B$. For this subsection it is important to recall the structure of the ribbon ends (Fig.~\ref{fig:FD}), which are completely determined by a site and a lattice triangle $A=(i_A,t_A)$, $B=(i_B,t_B)$ ($t_A,t_B$ are not considered to be within $\pathg$).

We next define the local operator $D_{(h,g)}(B)$ positioned at $B\equiv (i_B,t_B)$, Fig.~\ref{fig:FD}. (Operator $D(A)$ will be very similar, below.)
Let the triangle $t_B$ have ordered vertices $k\rightarrow j\rightarrow i$ as in Fig.~\ref{fig:FD}. The $D_{(h,g)}(B)$ operator acts on the elements in the plaquette centered on $i_B$ in the same way as the plaquette operator $\hat{B}^g_{p\equiv i_B}$; however, $D_{(h,g)}(B)$ additionally projects the 2-simplex at $t_B$ to having flux $h\in\GG$, and also has an additional phase factor. Actually, in contrast to $\hat{B}^g_{p\equiv i_B}$, we define the operator $D_{(h,g)}(B)$ in the entire Hilbert space $\mathcal{H}$, as will become clear soon. Let us first state the quantum amplitude of the local operator:
\begin{equation}
  \label{eq:36}
\bra{\text{f}}D_{(h,g)}(B)\ket{\text{i}}=\delta(g_{ij}\cdot g_{jk}\cdot g_{ki},h)\,W_h\,W_6(\text{i}),
\end{equation}
where $W_h\equiv W_h(\sigma_{ij},\colo)^{\ccysgn(\sigma_{ij})}$ is the phase of 2-simplex $\sigma_{ij} $ (light blue shaded in Fig.~\ref{fig:FD}) formed by edge $ij$ and vertical edge $g$, so it equals $\ccyc_h(g,g_{ij})$ (see Fig.~\ref{fig:FD}). Of course, by the definition of group elements on edges, Eq.~(\ref{eq:53}), $\delta(g_{ij}\cdot g_{jk}\cdot g_{ki},h)=\delta(h_{ij}\cdot h_{jk}\cdot h_{ki},h)$ depends only on the elements of gauge group $\GG$.

The phase $W_6(\text{i})$ is due to 3-simplices (tetrahedrons) on top of the plaquette; although analogous to the $\hat{B}_p$ operator (Fig.~\ref{fig:1}), in this case operator $D$ has to be well-defined even if the zero-flux rule is violated in the plane, as can occur at the $t_A,t_B$ of the ribbon $\pathg$. On the other hand, the 3-cocycle can assign a phase to a tetrahedron only if the zero-flux rule is satisfied on all its faces. Because of this, \textit{just for the purpose of calculating the phase due to the six tetrahedrons} $\prod_{I=1}^6 W(\sigma_I,\colo')^{\ccysgn(\sigma_I)}$, we redefine the values $o_1,\ldots,o_6$ of elements on six outer edges of the plaquette (red in Fig.~\ref{fig:FD}b) such that the zero-flux rule is satisfied in all six triangles of the plaquette in the plane. The six internal edge elements of the plaquette in the plane are considered unchanged from their value in $\ket{\text{i}}$, and they suffice to fix the redefined values $o'_1,\ldots,o'_6$ on the outer edges of the plaquette according to the zero-flux rule in all six triangles of the plaquette. Formally, this phase contribution is:
\begin{equation}
  \label{eq:76}
W_6(\text{i})=\bra{o'_1,\ldots,o'_6;f_1,\ldots,f_6}\hat{B}_p\ket{o'_1,\ldots,o'_6;i_1,\ldots,i_6},
\end{equation}
with $i_1,\ldots,i_6$ the initial values of elements on six internal edges of plaquette, and $f_1,\ldots,f_6$ their final values. (Note that the redefined values on outer edges $o'_1,\ldots,o'_6$ have to stay the same in the initial and final state.)

It is now clear that operator $D_{(h,g)}(B)$ is well defined in $\mathcal{H}$. Actually, it is also well-defined within the subspace $\mathcal{K}(\pathg)$ because it does not influence the flux in triangles belonging to ribbon $\pathg$ (recall that by definition $t_B$ is not inside $\pathg$).

The definition and properties of the $D(A)$ operator, relevant due to its action at the $A$-end of a string, is almost identical to the $D(B)$ case just described. The only difference is in the phase $W_h$, which in this case is not the phase of the 2-simplex $iji'$ (light blue triangle in Fig.~\ref{fig:FD}b), but the 2-simplex $iki'$ instead.

Obviously, a non-trivial algebra between $D(A),D(B)$ and $F$ is due to their overlap at the triangles $t_A,t_B$. For concreteness, we presented ribbon ends of the form in Fig.~\ref{fig:FD}, i.e. $t_A,t_B$ being bottom up-triangles, omitting versions rotated by multiples of $60^{\circ}$.

The local operators $D_{(h,g)}(C)$, with $C=A,B$, are symmetric. As noted before (see discussion of plaquette operator, Sec.~\ref{esmodels}), a global symmetry transformation $\tilde{s}\in\SG$ does not alter the elements on the edges, leaving the phases in Eq.~\eqref{eq:36} intact. Further, even though the local operator acts on the site $i_C$ by the element $\tilde{g}\in\SG$, where $g=h_g\cdot\tilde{g}$, $h_g\in\GG$, this action automatically commutes with the action of $\tilde{s}$ since we have restricted $\G$ to be Abelian in the present analysis of elementary excitations in our models.

We emphasize again that for local operators $D_{(h,g)}(C)$, with $C=A,B$, by definition $h\in\GG$ while $g\in\G$.

Having the definitions of: ribbon $\pathg$ having ends $C=A,B$; the ribbon operator $F(\pathg)$; and the local operators $D(A)$, $D(B)$ at hand, one can tediously but straightforwardly derive the following algebra:
\begin{widetext}
\begin{subequations}\label{eqFDexpl}
\begin{align}
\label{eqDDD}
D_{(h_2,g_2)}(C)\;   D_{(h_1,g_1)}(C)&=\delta_{h_1,h_2}\,c_{h_1}(g_2,g_1)\,D_{(h_1,g_1g_2)}(C)\\
\label{eqFFF}
F^{(h_2,g_2)}(\pathg)\;   F^{(h_1,g_1)}(\pathg)&=\delta_{g_1,g_2}\,c_{g_1}(h_2,h_1)\,F^{(h_2h_1,g_1)}(\pathg)\\
\label{eqFD} F^{(h_1,g_2g_1)}(\pathg)\;D_{(h_2h_1,g_2)}(A)&=c_{g_2}(h_2,h_1)\,c_{h_1}(g_2,g_1)\,D_{(h_2,g_2)}(A)\;F^{(h_1,g_1)}(\pathg)\\
\label{eqDF}
  D_{(h_2h_1,g_2)}(B) \; F^{(h_1,g_2g_1)}(\pathg)&=c_{g_2}(h_1,h_2)\,c_{h_1}(g_1,g_2)\,F^{(h_1,g_1)}(\pathg)\;D_{(h_2,g_2)}(B),
\end{align}
  \end{subequations}
\end{widetext}
where Eq.~\eqref{eqDDD} holds in the entire Hilbert space $\mathcal{H}$ (and also within $\mathcal{K}(\pathg)$), while the three other relations hold within the subspace $\mathcal{K}(\pathg)$, where the ribbon operators are well-defined.

To understand the implications of the operator algebra in Eq.~\eqref{eqFDexpl}, we need to consider the precise form of the excited state. Starting from a two-particle excited state, having excitations at the two ends $A,B$ of the string $\pathg$, one would expect it to be given by the simple action of the ribbon operator on the ground state:
\begin{equation}
  \label{eq:37}
  \ket{\state^{(h,g)}}=F^{(h,g)}(\pathg)\ket{\gs}.
\end{equation}
These states can be shown to be orthogonal.
However, the space $\mathcal{L}(A,B)$ spanned by these states needs to be specified further. Namely, as Eq.~\eqref{eq:55} shows, the ribbon operator puts only one constraint on the values of two elements at the lattice sites $i_A,i_B$, i.e. $u_{i_A}\cdot u_{i_B}^{-1}=\tilde{g}\in\SG$, with the factorization $g=h_g\cdot\tilde{g}$. We already know that the local operators transform these elements, e.g. under action of $D_{(h,\gd_1)}(A)$, the element $u_{i_A}\rightarrow\gdg_1\cdot u_{i_A}$, where $\gd_1=\gdh_1\cdot\gdg_1$, $\gdh_1\in\GG$, $\gdg_1\in\SG$. We therefore need to specify the value of one element (either of $u_{i_A}$ or $u_{i_B}$) in the excited state. By using the projectors
\begin{equation}
  \label{eq:82}
  \hat{P}_u(A)\ket{\{\gsg_i\},\{\gGG_{ij}\}}=\delta_{u_{i_A},u}\ket{\{\gsg_i\},\{\gGG_{ij}\}},
\end{equation}
we can consider the subspace $\mathcal{L}_{u_A}(A,B)$ of the Hilbert space $\mathcal{H}$ spanned by projected states:
\begin{equation}
  \label{eq:58}
  \ket{\state^{(h,g)}_{u_A}}\equiv\hat{P}_{u_A}(A) \ket{\state^{(h,g)}},
\end{equation}
with a fixed element $u_A\in\SG$. The value of $u_{i_B}$ is then automatically fixed by the action of ribbon operator $u_{i_B}=\gvisg^{-1}\cdot u_A$.

Completely analogously we define the subspace $\mathcal{L}_{u_B}(A,B)$ spanned by $\ket{\state^{(h,g)}_{u_B}}\equiv\hat{P}_{u_B}(B) \ket{\state^{(h,g)}}$, by using the projector $\hat{P}_{u_B}(B)$ at the vertex-$i_B$. Note that the projectors $\hat{P}(A),\hat{P}(B)$ commute with the ribbon operator $F(\pathg)$.
It is also easy to check that $D_{(h,\gd)}(C)\hat{P}_{u_C}(C)=\hat{P}_{\gdg\cdot u_C}(C) D_{(h,\gd)}(C)$, where $C$ is either $A$ or $B$, and the usual group element factorization is $\gd=\gdh_g\cdot\gdg$, $\gdh_g\in\GG$, $\gdg\in\SG$.

The subspaces $\mathcal{L}_{u_C}(A,B)$ were introduced using the action of ribbon operators on the ground state. The end-$A$ and end-$B$ of ribbon $\pathg$ in principle carry excitations, and $\mathcal{L}_{u_C}(A,B)$ does not depend on the particular shape of $\pathg$ (as long as $\pathg$ does not change topological class). It is however not trivial to prove that the subspace $\mathcal{L}(A,B)=\oplus_{u_C\in\SG}\mathcal{L}_{u_C}(A,B)$, with $C$ either $A$ or $B$, actually exhausts all possible excited states with two quasiparticles positioned at $A$ and $B$. In Appendix~\ref{sec:all-two-particle} we prove that the space $\mathcal{L}(A,B)$ indeed contains all such excited states. (We do not however have a proof that all multi-particle states having more than two particles, are also given by the action of appropriate ribbon operators on the ground state.)

Now that we established the appropriate Hilbert space, it is easy to show that the local operators form a unitary projective representation of the group $\G$ within this two-particle Hilbert space. We give the precise definition and explicit proof of this fact in Appendix~\ref{sec:all-two-particle}.

Let us next show that the local operators, as well as the ribbon operators, form Hopf algebras. (The succinct notation we introduce here will be useful
in Appendix~\ref{sec:all-two-particle}.) More precisely, let us denote the algebra formed by local operators $D_{(h,g)}(A)$, with $h\in\GG$, $g\in\G$, by the symbol $\mathbb{D}(A)$. Analogously, the operators $D_{(h,g)}(B)$ form the algebra $\mathbb{D}(B)$. Since $\mathbb{D}(A)$ and $\mathbb{D}(B)$ are formally algebraically the same, we use $\mathbb{D}$ to denote this abstract algebra, keeping in mind that $\mathbb{D}(A)$ and $\mathbb{D}(B)$ act at physically different positions. The algebra of ribbon operators $F^{(h,g)}$, $h\in\GG$, $g\in\G$, having ribbon $\pathg$, will be denoted by $\mathbb{F}$. We now consider the subspace $\mathcal{K}(\pathg)$ of Hilbert space, so that all these operators are simultaneously well-defined, and Eq.~\eqref{eqFDexpl} holds.

To start with, the operator relations in Eqs.~\eqref{eqFDexpl} can be rewritten succinctly using double index notation: the Latin indices $i,j,k,\ldots,r$ are shorthand for group element pairs $i\equiv (h_i,g_i)$, $j\equiv (h_j,g_j)$, etc., with $h_i,h_j,\ldots\in\GG$ and $g_i,g_j,\ldots\in\G$.
We can then write:
\begin{subequations}\label{eqFDten}
\begin{align}
  \label{eq:DD}
  D_m(C)\;D_n(C)&=\Omega^k_{mn}\,D_k(C)\\
    \label{eq:FF}
    F^m(\pathg)\;F^n(\pathg)&=\Lambda_k^{mn}\,F^k(\pathg)\\
          \label{eq:FD}
          F^m(\pathg)\;D_i(A)&=\Lambda_i^{jk}\,\Omega^m_{kl}\,D_j(A)\;F^l(\pathg)\\
                \label{eq:DF}
                D_i(B)\;F^m(\pathg)&=\Lambda_i^{kj}\,\Omega^m_{lk}\,F^l(\pathg) \;D_j(B)\\
  \end{align}
\end{subequations}
where we defined the tensors
\begin{subequations}  \label{eq:38}
\begin{align}
  \Omega^k_{ij}&\equiv\delta_{h_i,h_j}\,\delta_{h_k,h_i}\,\delta_{g_k,g_ig_j}\,c_{h_k}(g_i,g_j)\\
  \Lambda_k^{ij}&\equiv\delta_{g_i,g_j}\,\delta_{g_k,g_i}\,\delta_{h_k,h_ih_j}\,c_{g_k}(h_i,h_j),
\end{align}
\end{subequations}
and summation over repeated double indices is understood as, e.g., $\sum_i\equiv\sum_{h_i\in\GG,g_i\in\G}$.

The tensors in Eq.~(\ref{eq:38})contain the twist due to the cocycle $\ccy$ present in our models. Similar realizations of twisted algebra, but describing broken gauge theories, are analyzed in e.g. Ref.\onlinecite{Propitius:Hopf}. We also note that only when the cocycle is trivial, $\ccy(g_1,g_2,g_3)\equiv 1\Rightarrow\ccyc_g(g',g'')=1$, these tensors reduce to the form occurring in the generalization of toric code to arbitrary finite groups, as given in Ref.~\onlinecite{Kitaev:2003p6185}.

To derive the properties of this algebra, we will have to use several properties of a 2-cocycle $\ccyc$ derived from an arbitrary 3-cocycle $\ccy$ (Eq.~(\ref{eq:14})), listed here for convenience:
\begin{subequations}
\begin{gather}
  \label{2ccy}
  \ccyc_a(g,h)\,\ccyc_a(gh,s)=\ccyc_a(h,s)\,\ccyc_a(g,hs)\\
\label{2cy6}
\ccyc_a(g,h)\,\ccyc_b(g,h)\,\ccyc_g(a,b)\,\ccyc_h(a,b)=\ccyc_{gh}(a,b)\,\ccyc_{ab}(g,h)\\
\label{2cy1}
\ccyc_a(g,h)=1,\text{ if any of }a,g,h\text{ equals }\openone\\
\label{2cyinv}
    \ccyc_a(g,g^{-1})=  \ccyc_a(g^{-1},g).
  \end{gather}
\end{subequations}
(We note that all these properties actually hold generally, i.e. for arbitrary elements $a,b,g,h,s\in\G$.)
The identity (\ref{2ccy}) is the general condition for a 2-cocycle, (\ref{2cy6}) can be derived directly using the definition Eq.~(\ref{eq:14}), identity (\ref{2cy1}) follows from our choice to use only canonical cocycles $\ccy$ ($\openone$ is the identity group element), and finally the useful relation (\ref{2cyinv}) follows from (\ref{2ccy}) and (\ref{2cy1}).

Using these identities we can prove that the ribbon and local operators form Hopf algebras.
  A basic axiom of Hopf algebra is the associativity of multiplication with identity, which holds for $F$ and $D$ algebras, as shown by the following two relations, respectively:
\begin{subequations}
\begin{align}
  \label{LLLL}
  \Lambda_i^{lm}\Lambda_k^{in}&=\Lambda_k^{lj}\Lambda_j^{mn}, \quad \epsilon_i\Lambda_k^{im}=\epsilon_j\Lambda_k^{mj}=\delta^m_k\\
  \label{OOOO}
  \Omega^i_{lm}\Omega^k_{in}&=\Omega^k_{lj}\Omega^j_{mn}, \quad e^i\Omega^k_{im}=e^j\Omega^k_{mj}=\delta^k_m,
\end{align}
\end{subequations}
where for double indices the Kronecker delta function $\delta^i_j\equiv\delta_{h_i,h_j}\delta_{g_i,g_j}$, and the functions $\epsilon_i\equiv\delta_{h_i,\openone}$, $e^i\equiv\delta_{g_i,\openone}$ define the unit and counit of the algebras, $\openone_\mathbb{F}=\epsilon_kF^k$, $\openone_\mathbb{D}=e^kD_k$, $\hat{e}(F^k)=e^k$ and $\hat{\epsilon}(D_k)=\epsilon_k$. Both Eq.~(\ref{LLLL}) and Eq.~(\ref{OOOO}) hold due to Eq.~(\ref{2ccy}) and Eq.~(\ref{2cy1}).

The comultiplication in the Hopf algebra is physically related to fusion, and instead of presenting a formal definition here, we show in Section~\ref{sec:braiding-matrix} and Appendix~\ref{app:extended_ribbon} that the braiding and fusion properties of excitations are consistent, and contain the twist by the 3-cocycle characteristic of a quasi-Hopf algebra first introduced in Ref.\onlinecite{Dijkgraaf:1990p7462}.

Having established the Hopf algebra relations in Eqs.~\eqref{LLLL},~\eqref{OOOO}, we are in a position to prove that all two-particle excited states in our models are indeed within the Hilbert subspace $\mathcal{L}(A,B)$ we defined in this subsection. The proof is presented in Appendix~\ref{sec:all-two-particle}, and is based on a derivation given in Ref.~\onlinecite{Kitaev:2003p6185}. This Hilbert space for an excitation pair is especially important here since it will be studied explicitly for several example groups in Section~\ref{sec:example}.

In summary, the local operators are the set of non-trivial operators acting on excitations, and they are also symmetric and form a projective representation of the group. We will use these properties to study the symmetry protected properties of our models in Section~\ref{sec:example}.

We close this subsection by briefly introducing the Hilbert space of many-particle excitations. (It is studied in more detail in Appendix~\ref{app:extended_ribbon}.) Let us consider a system having $n$ quasiparticles at positions $A_1,\ldots,A_n$, and one quasiparticle at position $B$, and no other excitations. Such a system is described by the Hilbert space $\mathcal{\tilde L}(A_1,\ldots,A_n,B)$. To study these states, we consider a space $\mathcal{L}(A_1,\ldots,A_n,B)$ spanned by the action of ribbon operators, as described in the following. Let us connect each position $A_i$ through a ribbon $\pathg_i$ having end-$A_i$ to the common isolated position $B$. Therefore all ribbons' end-$B_i$ coincide, and all are equal to $B$. (For further discussion and lattice realization, see Appendix~\ref{app:extended_ribbon}.) Focusing on the subspace $\mathcal{K}(\pathg_1,\ldots,\pathg_n)$ in which all ribbon operators having ribbons $\pathg_i$ are simultaneously well-defined (i.e. zero-flux rule obeyed inside all ribbons), we can now define its subspaces $\mathcal{L}_{u_B}(A_1,\ldots,A_n,B)$ spanned by states of the form:
\begin{equation}
  \label{eq:48}
\ket{\state^{k_1,\ldots,k_n}_{u_B}}=\hat{P}_{u_B}(B)F^{k_1}(\pathg_1)\cdots F^{k_n}(\pathg_n)\ket{\gs}.  
\end{equation}
The states in Eq.~\eqref{eq:48} form the subspaces with $u_B$ fixed. We actually expect that the space $\mathcal{\tilde L}(A_1,\ldots,A_n,B)$ coincides with the space $\mathcal{L}(A_1,\ldots,A_n,B)=\oplus_{u_B\in\SG}\mathcal{L}_{u_B}(A_1,\ldots,A_n,B)$. (We have a proof of this only in the two-particle case, see Appendix~\ref{sec:all-two-particle}, but as mentioned this is hard to prove in general.)

At the same time, the extended algebra of each given ribbon $F(\pathg_i)$ contains the local operators $D^{(i)}_k\equiv D_k(B)$ which we define to affect only that (i.e., the $i$-th) ribbon. These operators, being local at $B$, commute with all local operators $D_m(A_j)$, $j=1\ldots n$ at the excitations $A_j$. The $D^{(i)}_k$ operators therefore act, in view of Eq.~\eqref{eq:DF}, as
\begin{equation}
  \label{eq:45}
  D^{(1)}_{j_1}\otimes\cdots\otimes D^{(n)}_{j_n}\ket{\state^{k_1,\ldots,k_n}_{u_B}}=\Omega^{k_1}_{m_1j_1}\cdots \Omega^{k_n}_{m_nj_n}\ket{\state^{m_1,\ldots,m_n}_{u_B}}.
\end{equation}
In this definition we constrain the elements in $D(B)$ strictly to $\GG$, e.g. the double index $j=(h_{j},h_{g_j}\cdot\openone_{\SG})$ with $h_{j},h_{g_j}\in\GG$. This constraint to $\GG$ elements is exactly encountered when describing braiding in the next subsection.

Because the $D^{(i)}_k$ operators are non-local with respect to the excitations at $A_i$ while commuting with $\mathbb{D}(A_i)$ they are called ``topological operators''\cite{Kitaev:2003p6185}.

Topological operators and braiding in many-particle states are analyzed in detail in Appendix~\ref{app:extended_ribbon}.

\subsection{Braiding matrix}
\label{sec:braiding-matrix}
\begin{figure}
  \centering
\includegraphics[width=0.49\textwidth]{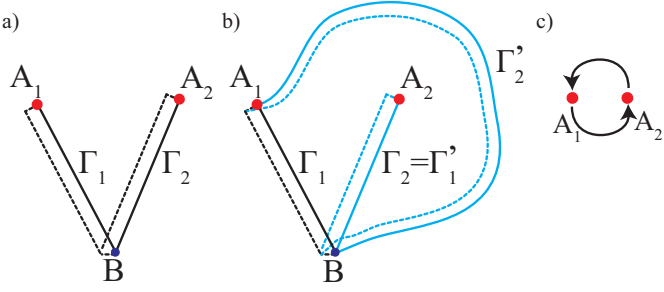}
  \caption{a) Ribbon operators for two-particle excited state, and braiding two excitations at $A_1,A_2$ (see also Fig.~\ref{fig:FD}). The common endpoint $B$ contains no excitation, but ``topological') operators of $i$-th ribbon $D^{(i)}(B)$, $i=1,2$, local at $B$, can be used to represent braiding. b) Applying different ribbons (blue) than in original state shown in (a) leads to a braided state (see Section~\ref{sec:braiding-matrix}). c) The resulting counter-clockwise braid of particles 1 and 2.}
\label{fig:ribbons}
\end{figure}

In this subsection we calculate the braiding matrix of two quasiparticles, restricting ourselves to a system that has quasiparticles at most at three positions, $A_1,A_2,B$ (see previous subsection). An alternative and more explicit way to obtain the braiding properties of quasiparticles is by considering ribbons of strings that cross, but this approach is applicable to Abelian quasiparticles only; we present it in Appendix~\ref{sec:part-stat-triv}, and show there that the results of the two approaches agree for Abelian quasiparticles. The braiding as well as fusion in states with many quasiparticles introduce additional subtleties, and this situation is presented in detail in Appendix~\ref{app:extended_ribbon}. Here we will focus only on braiding of two particles, and show that the braiding properties are entirely determined by the topological order, i.e. the gauge group $\GG$. The effects of interplay of topological order and symmetry are revealed in the concrete examples in Section~\ref{sec:example}.

Recall that we define the ribbon operator matrix element (see Eq.~\eqref{eq:Fmatelem}) such that the operator algebra in Section~\ref{sec:extend-ribb-algebra} is obtained. (The matrix element is fully presented in Appendix~\ref{app:extended_ribbon}.) That definition also leads to the following algebra for two ribbon operators having strings $\pathg_1,\pathg_2$ which share their $B$-end:
\begin{widetext}
\begin{equation}
  \label{eqFFFF}
  F^{(h_2,g_2)}(\pathg_2)\;   F^{(h_1,g_1)}(\pathg_1)=c_{h_1}(g_1h^{-1}_2,h_2)\,F^{(h_1,g_1h_2^{-1})}(\pathg_1)\;F^{(h_2,g_2)}(\pathg_2)\frac{\ccy(h_2,h_1,h_B)}{\ccy(h_1,h_2,h_B)},
\end{equation}
\end{widetext}
where $h_B$ is the value of flux in triangle $t_B$. The Eq.~\eqref{eqFFFF} is well-defined only in the subspace $\mathcal{K}(\pathg_1,\pathg_2)$ of the Hilbert space, i.e. when zero-flux is obeyed in every triangle inside $\pathg_1$ or $\pathg_2$ (note that $t_B$ is outside both ribbons). Fig.~\ref{fig:ribbons}a sketches this situation, which will enable us to determine the braiding of the two quasiparticles at the end-$A_{1}$ and end-$A_2$. (Appendix~\ref{app:extended_ribbon} considers in detail the case when the end-$A$ is shared by the ribbons.)

The 3-cocycle factor on the right hand side of Eq.~(\ref{eqFFFF}) has to be considered with care in a many-particle situation, as is done in Appendix~\ref{app:extended_ribbon}; in this subsection we however focus on a system with only the two ribbons $\pathg_1,\pathg_2$. Because of this, a product of two ribbon operators, such as appearing on both sides of Eq.~(\ref{eqFFFF}), can only act on the ground state, i.e. there are no other excitations in the system. Because of this, the value of flux in triangle $t_B$ on the right-hand side of Eq.~\eqref{eqFFFF} is necessarily zero, which means that the element $h_B=\openone$ (see also Appendix~\ref{app:extended_ribbon}). Due to our choice of canonical cocycles, the two $\ccy$ factors on the right of Eq.~(\ref{eqFFFF}) therefore disappear.

Let us now formally introduce the braiding matrix relevant for ribbons sharing their end-$B$ by rewriting Eq.~(\ref{eqFFFF}) in a compact form:
\begin{equation}
  \label{eq:FFR}
F^j(\pathg_2)\;F^l(\pathg_1)=R^{qr}\Omega^l_{mr}\Omega^{j}_{nq} F^m(\pathg_1)\;F^n(\pathg_2),
\end{equation}
where according to Eq.~(\ref{eqFFFF}), the $R$-matrix of our model equals $R^{ik}=\delta_{h_i,g_k}\delta_{g_i,\openone}$.

We now calculate the operator $\mathcal{R}_{CC}$ describing the counter-clockwise braiding operation by $180^{\circ}$ of the two excitations positioned at $A_1$ and $A_2$. This operator is defined by setting $\ket{\state^{ab}}_{braid}=\mathcal{R}_{CC}\ket{\state^{ab}}$.
We consider the original state, Fig.~\ref{fig:ribbons}a, and the one where excitations are exchanged (Fig.~\ref{fig:ribbons}b) by braiding particle 2 counter-clockwise around particle 1:
\begin{align}
  \label{eq:47}
  &\ket{\state^{ab}_{u_B}}=\hat{P}_{u_B}(B)F^a(\pathg_1)F^b(\pathg_2)\ket{\gs}\\  \label{eq:47b} &\ket{\state^{ab}_{u_B}}_{braid}=\hat{P}_{u_B}(B)\mathcal{R}_{CC}\ket{\state^{ab}}\\
  &=\hat{P}_{u_B}(B)F^a(\pathg'_1)F^b(\pathg'_2)\ket{\gs},
\end{align}
where we used the double index notation, i.e. $a\equiv(h_a,g_a),b\equiv(h_b,g_b)$. Note that we choose to project the states by fixing the value of $u_{i_B}$ equal to $u_B$ --- the ribbon operators then determine the values of $u_{i_{A_1}},u_{i_{A_2}}$ (see the definition in Eq.~\eqref{eq:58}).

According to the definition of states in Eq.~(\ref{eq:48}), the order of applying ribbon operators reflects the order of particles, while the new strings $\pathg_1',\pathg_2'$ compared to original ones $\pathg_1,\pathg_2$ represent the braiding movement. Figs.~\ref{fig:ribbons}a,b illustrate this. We see that the new string of particle 1 coincides with the old string of particle 2, $\pathg'_1=\pathg_2$, while the new string $\pathg'_2$ is topologically equivalent (and the operator therefore the same) to $\pathg_1$ only \textit{if there is no particle at original position of 2 at the time these two strings are compared}, see Fig.~\ref{fig:ribbons}b. Therefore the new ribbon operator on $\pathg'_2$ is the same as a ribbon on $\pathg_1$ \textit{if it is applied before} the new $\pathg'_1$. This is actually true in Eq.~(\ref{eq:47b}), and we therefore have
\begin{equation}
  \label{eq:49}
  F^a(\pathg'_1)F^b(\pathg'_2)=  F^a(\pathg_2)F^b(\pathg_1).
\end{equation}
We now only need to commute the ribbon operators to make a comparison to Eq.~(\ref{eq:47}). The commutation relation in Eq.~(\ref{eq:FFR}) directly gives
\begin{align}
  \label{eq:50}
  \hat{P}_{u_B}(B)\mathcal{R}_{CC}\ket{\state^{ab}}&=  R^{qr}\Omega^b_{mr}\Omega^a_{nq}\ket{\state^{mn}_{u_B}}=\\
  &=R^{qr}D^{(1)}_r\otimes D^{(2)}_q\ket{\state^{ba}_{u_B}}.
\end{align}
We note that the $R^{qr}$ matrix in this expression constrains the group elements in $D^{(1,2)}$ to be strictly in $\GG$, e.g. $r\equiv(h_r,h'_r\cdot\tilde{g}_r)$, $h_r,h'_r\in\GG$, is constrained to $\tilde{g}_r=\openone_{\SG}$; this implies that the site element $u_{i_B}$ is not changed from its value $u_B$ by the action of $D^{(1)},D^{(2)}$.
The explicit form of the braiding operation is therefore
\begin{equation}
  \label{eq:51}
  \mathcal{R}_{CC}\ket{\state^{ab}_{u_B}}=\ccyc_{h_b}(g_bh_a^{-1},h_a) \ket{\state^{(h_b,g_bh_a^{-1})(h_a,g_a)}_{u_B}},
\end{equation}
which one can of course obtain directly using the commutation in Eq.~(\ref{eqFFFF}), without first defining the $R^{ik}$ matrix. The result non-trivially involves the cocycle $\ccy$ of our model.

Applying $\mathcal{R}_{CC}$ twice, we obtain for the $360^{\circ}$ braiding
\begin{align}\notag 
  &\mathcal{R}^2_{CC}\ket{\state^{ab}_{u_B}}=\\  \label{eq:R2pi}
  &=\ccyc_{h_a}(g_ah_b^{-1},h_b)\,\ccyc_{h_b}(g_bh_a^{-1},h_a) \ket{\state^{(h_a,g_a h_b^{-1})(h_b,g_b h_a^{-1})}_{u_B}}=\\\notag
  &=D^{(1)}_{(h_a,h_b)}\otimes D^{(2)}_{(h_b,h_a)}\ket{\state^{ab}_{u_B}},
\end{align}
where we remind that $D^{(i)}_g\equiv D_g(B)$ is acting on the $i$-th ribbon operator in the product state Eq.~(\ref{eq:48}).

The case when the braided quasiparticles are positioned at $B$-ends of their ribbons which share the end-$A$, is similar and discussed in detail in Appendix~\ref{app:extended_ribbon}. Here we just quote the result for the $2\pi$ braiding of particles at end-$B_1$ and end-$B_2$:
\begin{equation}
  \label{eq:R2piB}
\bar{\mathcal{R}}^2_{CC}\ket{\bar{\state}^{ab}_{u_A}}=D^{(1)}_{(h^{-1}_a,h^{-1}_b)}\otimes D^{(2)}_{(h^{-1}_b,h^{-1}_a)}\ket{\bar{\state}^{ab}_{u_A}},
\end{equation}
where we use the bar over symbols to signify that the state and braiding concern particles at end-$B$'s of strings.

Eqs.~(\ref{eq:R2pi}) and (\ref{eq:R2piB}) explicitly show that the braiding of quasiparticles is described by the action of ``topological operators'' (see Eq.~(\ref{eq:45})), which act only on the gauge degrees of freedom, obviously since they are labeled only by elements $h_a,h_b\in\GG$. This fact means that the braiding properties follow directly from the topological order in our models, which is described by the gauge group $\GG$. In Appendix~\ref{app:extended_ribbon} we show explicitly that the topological operators form an algebra called ``quasi-quantum double''\cite{Dijkgraaf:1990p7462,Propitius:Hopf}, which is mathematically a realization of a quasi-Hopf algebra. The prefix ``quasi'' denotes the presence of the twist by cocycle $\ccy$, which is here restricted to elements of $\GG$. Appendix~\ref{app:extended_ribbon} also clarifies how representations of the quasi-quantum double label the quasiparticle species, while the multiplication and comultiplication operations in the algebra directly determine the braiding and fusion of quasiparticles.

The quasi-quantum double construction, i.e. the presence of a cocycle ``twist'' in the algebra of braiding and fusion operators, appeared in the description of excitations in gauge theories broken to a discrete subgroup\cite{Propitius:1993CS_ccy}. This is to be expected since both those theories and our models describe topological order classified by DW TQFTs\cite{Dijkgraaf:1990p7194,Propitius:1993CS_ccy}.

In Appendix~\ref{app:extended_ribbon} we also discuss in detail the situation with an arbitrary number of quasiparticles (either end-$A$ or end-$B$ ones), describing their braiding and fusion as well as the quasi-quantum double mathematical structure.

\section{Examples}\label{sec:example}

As will become clear through this section, the gauge charge and gauge flux quasiparticles in our theories do not behave in the same way regarding the phenomena of symmetry fractionalization. We will therefore focus on the gauge fluxes, which have non-trivial properties, while leaving the case of gauge charges, which behave trivially, to subsection~\ref{sec:gauge-charges}.

At the end of this section, we will also discuss the degeneracy of states with multiple quasiparticle pairs by considering the dualization of the global symmetry $SG$.

\subsection{Hilbert space for a pair of gauge fluxes}
\label{sec:gener-struct-excit}

We will now analyze the properties of excitations in our exactly solvable models, as announced in Sec.~\ref{sec:classification}. We will later study in detail examples where the groups $\SG$ and $\GG$ are products of $Z_2$. In general, we study a pair of gauge fluxes positioned at the ends $A$ and $B$ (which will also label the lattice sites there) of a string $\pathg$, created by the action of a ribbon operator
\begin{equation}
  \label{eq:57}
\ket{\vis,\gvis;u_A}\equiv\ket{\state^{(\vis,\gvis)}_{u_A}}=\hat{P}_{u_A}(A) F^{(\vis,\gvis)}(\pathg)\ket{\gs},
\end{equation}
where we introduced more succinct notation. These projected states span $\mathcal{L}(A,B,\vis)$. According to the general definition of ribbon operators, $\vis\in\GG$ and $\gvis\in\G=\SG\times\GG$, and we label the unique factors of the latter element as $\gvis=\gvish\cdot\gvisg$, with $\gvish\in\GG$, $\gvisg\in\SG$. The value of $u_B$ is automatically fixed by the action of ribbon operator $u_B=\gvisg^{-1}\cdot u_A$, with $u_A,u_B\in\SG$. We will focus on a pair of gauge fluxes, so that $\vis$ is chosen as a fixed non-trivial element of $\GG$. (In $Z_2$ gauge theory, such a flux quasiparticle is called ``vison''.)

Recalling the discussion in Section~\ref{sec:extend-ribb-algebra}, $D_{(\vis,\gd)}(A)\hat{P}_{u_A}(A)=\hat{P}_{\gdg\cdot u_A}(A) D_{(\vis,\gd)}(A)$ because the local operators transform these elements, e.g. $u_A\rightarrow\gdg\cdot u_A$ with $\gd=\gdh\cdot\gdg$, $\gdh\in\GG$, $\gdg\in\SG$. The action of local operators on the Hilbert space of the gauge flux pair having flux $\vis$ is therefore given by:
\begin{align}\notag
    D_{[\gd_1]}(A)\ket{\vis,\gvis;u_A}&=\ccyc^{-1}_{\gd_1}(\vis,\vis^{-1})\ccyc_{\vis}(\gd_1,\gvis) \ket{\vis,\gd_1\cdot\gvis;\gdg_1\cdot u_A}\\\label{eq:DABloc}
  D_{[\gd_1]}(B)\ket{\vis,\gvis;u_A}&=\ccyc_{\vis}(\gd^{-1}_1\cdot\gvis,\gd_1) \ket{\vis,\gd^{-1}_1\cdot\gvis;u_A},
\end{align}
where we remind that $\gd_1=\gdh_1\cdot\gdg_1$, $\gdh_1\in\GG$, $\gdg_1\in\SG$, and since the flux $\vis$ is fixed we introduced the shorthand notation $D_{[\gd]}(A)\equiv D_{(\vis^{-1},\gd)}(A)$ and $D_{[\gd]}(B)\equiv D_{(\vis,\gd)}(B)$. Our analysis of the examples will use Eqs.~\eqref{eq:DABloc} to explicitly construct the matrices of local operators in this basis.


\subsection{Construction of local symmetry operators in symmetry-fractionalized models}
\label{sec:constr-local-symm}

Let us now, from a general viewpoint, consider the possibility of symmetry fractionalization in our models. According to Eq.~\eqref{eq:s_f_assumption}, we seek to factorize the global symmetry transformation $U(\symg\in\SG)$, when acting in the flux-quasiparticle Hilbert space $\mathcal{L}(A,B,\vis)$, into two local factors:
\begin{equation}
  \label{eq:59}
U(\symg)=U_\symg(A)\cdot U_\symg(B),\quad \symg\in\SG.
\end{equation}
First of all, we observe that the global symmetry transformation $\symg$ acts by transforming elements on each lattice site $i$ by sending $u_i\rightarrow u_i\cdot\symg ^{-1}$, while the operators $D$ perform a similar operation on a single site, i.e. \textit{locally}, for example $D_{(h,\gd)}(A): u_{iA}\rightarrow \gdg\cdot u_{iA}$, where $\gd=\gdh_g\cdot\gdg$, $\gdh_g\in\GG$, $\gdg\in\SG$. (Recall that we deal with Abelian groups $\G$ here.) This is reasonable, since for any local operator in $\mathcal{L}(A,B,\vis)$, such as the tentative $U_\symg(C)$, we expect that it is representable in terms of the local operators $D(C)$.

We therefore see that whenever symmetry fractionalization occurs, it should be possible to find a local phase $\varphi$ such that $U_\symg(C)=e^{i\varphi(\vis,\symg,C)}D_{(h,\symg^{-1})}(C)$, where $\symg\in\SG$ and $C=A,B$. (Note that although the phase $\varphi(C)$ can depend on local variables at $C$, e.g. on $u_C$, the functional form cannot depend on the position $C$, since the local symmetry operation cannot depend on the spatial position at which it is applied.) We will now show under which conditions (i.e. for what kind of 3-cocycle $\ccy$) it is possible to find such a phase $\varphi$.

We can effectively use the demand on the global symmetry transformation $U(\sigma)$ which says that $U(\sigma)$ commutes with the ribbon operator creating the quasiparticles; more precisely, only $u_A,u_B$ of the projected basis vectors (Eq.~\eqref{eq:57}) in $\mathcal{L}(A,B,\vis)$ are transformed by $u_C\rightarrow \sigma^{-1}\cdot u_C$. Tentatively writing $U_\sigma(A)=D_{(\vis^{-1},\sigma^{-1})}(A)$, $U_\sigma(B)=D_{(\vis,\sigma^{-1})}(B)$ (note that the gauge fluxes at end-A and end-B are $\vis^{-1}$ and $\vis$ respectively), with $\sigma\in\SG$, we get
\begin{multline}
  \label{eq:60}
  U_\sigma(A)   U_\sigma(B) \ket{\vis,\gvis;u_A,u_B}=\\\epsilon_{\vis,\sigma^{-1},\gvis}\,\ccyc^{-1}_{\sigma^{-1}}(\vis,\vis^{-1}) \ket{\vis,\gvis;\sigma^{-1}\cdot u_A,\sigma^{-1}\cdot u_B},
\end{multline}
where we have introduced the 1-cocycle
\begin{equation}
  \label{eq:61}
\epsilon_{x,y,z}\equiv\frac{\ccyc_x(z,y)}{\ccyc_x(y,z)},\quad x,y,z\in\G.
\end{equation}
Obviously, with this choice of $U(C)$, the quasiparticle state is properly transformed only up to a phase, but we can proceed to absorb the resulting phase by a non-trivial $\varphi(C)$. We will need some useful properties of the introduced 1-cocycle:
\begin{align}
  \label{eq:antisymmepsilon}
&\epsilon_{x,y,z\cdot w}=\epsilon_{x,y,z}\,\epsilon_{x,y,w},\\\notag
&\epsilon_{x,y,z}=\epsilon_{y,z,x}=\epsilon_{z,x,y}=\epsilon^{-1}_{y,x,z}=\epsilon^{-1}_{x,z,y}=\epsilon^{-1}_{z,y,x},\\\notag
&\epsilon_{x,y,z}=\epsilon^{-1}_{x,y,z^{-1}}.
\end{align} 
Using the fact that $\gvis=\gvish\cdot u_Au_B^{-1}$ and the listed properties of $\epsilon$, we can define a valid phase $\varphi$ by:
\begin{align}
  \label{eq:62}
  U_\sigma(A)&\equiv\sqrt{\ccyc_{\sigma^{-1}}(h,h^{-1})}\,\epsilon_{h,\sigma^{-1},u_A}D_{(h,\sigma^{-1})}(A)\\\notag
  U_\sigma(B)&\equiv\sqrt{\ccyc_{\sigma^{-1}}(h,h^{-1})}\,\epsilon_{h,\sigma^{-1},u_B}D_{(h,\sigma^{-1})}(B).
\end{align}
Note that the $h$ in $U_\sigma(A)$($U_\sigma(B)$) is defined to be the gauge flux at the end-$A$(end-$B$), which can be measured locally.
This leads to
\begin{align}
  \label{eq:63}
 & U_\sigma(A)   U_\sigma(B)\ket{\vis,\gvis;u_A,u_B}\notag\\
=&\epsilon_{\vis,\sigma^{-1},\gvish}\ket{\vis,\gvis;\sigma^{-1}\cdot u_A,\sigma^{-1}\cdot u_B}\notag\\
=&\epsilon_{\vis,\sigma^{-1},\gvish}U(\sigma)\ket{\vis,\gvis;u_A,u_B}.
\end{align}
Symmetry fractionalization $U(\sigma)=U_\sigma(A)U_\sigma(B)$ can therefore occur if
\begin{equation}
  \label{eq:64}
\epsilon_{\vis,\sigma,\gvish}=1,\;\forall\,\vis,\gvish\in\GG, \sigma\in\SG.
\end{equation}
It is not physical to absorb this phase factor into the local operators $U_\sigma(C)$. The reason is that the phase depends on the element value $\gvish$ carried by the quasiparticles (recall the definition of ribbon operators Eq.~\eqref{eq:55}: $\gvish=\hG_{1,0}\cdots \hG_{N,N-1}\in\GG$ for a ribbon on lattice sites $0,\ldots,N$), while on the other hand the symmetry transformation should not depend on such specific properties of the quasiparticles. The presented argument at least gives an indication when symmetry fractionalization should be impossible. We will see that for $\GG=Z_2\times Z_2$ and $\SG=Z_2$ there is one 3-cocycle, $\ccy^{(123)}$ (see Table~\ref{tab:2}), for which Eq.~\eqref{eq:64} is violated, in accordance with our claim about a single $Z_2$ index beyond SFC (see example ``3'' at the end of section~\ref{sec:connection}). For that example, we will additionally show that global symmetry \textit{exchanges} the two species of excitations, thus giving a strong physical argument against the possibility of representing symmetry by local operators.

In Appendix~\ref{sec:symmfracmulti} we show that this construction of fractionalized symmetry operators is generally correct for multiple quasiparticles in the examples below.

Returning to the present single ribbon case, let us for a moment assume that symmetry fractionalization is explicitly possible in a given model, i.e. Eq.~\eqref{eq:64} is satisfied. We can then give general formulas for the transformation of a single quasiparticle at position $C$ under fractionalized symmetry transformations $U_\sigma(C)$, where again $C=A,B$ is either end of ribbon. The fractionalized symmetry transformations physically only need to be projective, as discussed in section~\ref{sec:psg}. In the examples of the following subsections, these general expressions will be used to check whether the symmetry group generators obey the SFC relations shown at the end of section~\ref{sec:psg}. (The results will be corroborated by explicit constructions of $U_\sigma(C)$). Because our examples will be based on the group $Z_2$ and on direct products of the group $Z_2$, we can make some simplifications. Namely, we can set $\vis^{-1}=\vis$. Further, considering arbitrary symmetry operations $\symg_0,\symg_1,\symg_2\in\SG$, we have that $\symg_0^2=\symg_1^2=\symg_2^2=\openone$, with $\openone$ the group identity element. Finally, for such groups the inequivalent cocycles $\ccy$ can be chosen to only take values $\pm1$, see Tables~\ref{tab:1},\ref{tab:2}. Now, using Eq.~\eqref{eq:DABloc} and~\eqref{eq:62}, as well as properties of the 2-cocycle (see Eqs.~\eqref{2ccy}-\eqref{2cyinv}), it is easy to show that in the Hilbert space of a $\vis$ flux pair:
\begin{align}
  \label{eq:65a}
  U_{\symg_2}(C)   U_{\symg_1}(C)&= \epsilon_{\vis,\symg_2,\symg_1}U_{\symg_1}(C)   U_{\symg_2}(C)\\  \label{eq:65b} U_{\symg_0}(A)^2&=\ccyc_{\vis}(\symg_0,\symg_0)\,\ccyc_{\symg_0}(\vis,\vis) \\  \label{eq:65c}
 D_{[\symg_0]}(A)^2&=\ccyc_{\vis}(\symg_0,\symg_0),
\end{align}
where the fluxes are positioned at ribbon ends $C=A,B$. In Eq.~\eqref{eq:65c} we show the result for the standard local operator $D$, to contrast it with the fractionalized symmetry operator in Eq.~\eqref{eq:65b} (in both cases the operator on the right-hand side is just the identity operator).

Equipped with this formalism, we can proceed to solve the instructive examples introduced in sections~\ref{sec:psg},~\ref{sec:connection}.

\subsection{The simplest example with symmetry fractionalization (PSG): Vison pair in $\GG=Z_2$ and $\SG=Z_2$}
\label{sec:exampl-gg=z_2-sg=z_2}
\begin{table}[t]
  \centering
  \begin{tabular}[c]{l|c}
    $\ccy(x,y,z)=-1$ & $x,y,z$\\
    \hline
    $\ccy^{(1)}$ & $(1,*),(1,*),(1,*)$\\
    $\ccy^{(2)}$ & $(*,1),(*,1),(*,1)$\\
    \hline
    $\ccy^{(12)}$& $(1,*),(*,1),(*,1)$\\
  \end{tabular}
  \caption{Inequivalent basic 3-cocycles $\ccy^{(I)}$ for $\G\equiv\GG\times\SG=Z_2\times Z_2$ which label classes in $H^3(\G,U(1))=Z_2^3$. A cocycle $\ccy(x,y,z)$ has value $1$, \textit{except} when the group elements $x,y,z\in\G$ take the special values shown in the table, in which case $\ccy(x,y,z)$ is equal to $-1$. The element notation $x=(g_1,g_2)$ signifies the direct product factorization, i.e. $g_1\in\GG$ and $g_2\in\SG$, and the elements of $Z_2$ are $0,1$ (additive group action). The star symbol $*$ stands for ``any value of element''.}
  \label{tab:1}
\end{table}

\begin{table*}[t]
  \centering
  \begin{tabular}[c]{c||c|c|c|c||c|c|c|c||c|c|c}
    & $D_{[(0,0)]}$ & $D_{[(0,1)]}$ & $D_{[(1,0)]}$ & $D_{[(1,1)]}$ & $D'_{[(0,0)]}$ & $D'_{[(0,1)]}$ & $D'_{[(1,0)]}$ & $D'_{[(1,1)]}$ & $U_\sigma(A)$ & $U_\sigma(B)$ & $U(\sigma)$\\
        \hline\hline
    $\ccy^{(1)}$ & $\openone$ & $\mu_x\tau_x$ & $i\rho_y$ & $\mu_x i\rho_y\tau_x$ & $\openone$ & $\mu_x$ & $i\rho_y$ & $\mu_xi\rho_y$ & $\mu_x\tau_x$ & $\mu_x$ & $\tau_x$\\
  \hline
    $\ccy^{(2)}$ & $\openone$ & $\mu_x\tau_x$ & $\rho_x$ & $\mu_x \rho_x\tau_x$ & $\openone$ & $\mu_x$ & $\rho_x$ & $\mu_x\rho_x$ & $\mu_x\tau_x$ & $\mu_x$ & $\tau_x$\\
  \hline
    $\ccy^{(12)}$ & $\openone$ & $-i\mu_y\tau_x$ & $\rho_x$ & $-i\mu_y \rho_x\tau_x$ & $\openone$ & $i\mu_y$ & $\rho_x$ & $i\mu_y\rho_x$ & $-i\mu_y\tau_x$ & $i\mu_y$ & $\tau_x$\\
  \end{tabular}
  \caption{The values of local operators and fractionalized symmetries for $\GG=Z_2$, $\SG=Z_2$ for (the only possible) vison pair $\vis=(1,0)=1\in\GG$. $D$ and $D'$ are shorthand for $D(A)$ and $D(B)$, respectively. The cocycle $\ccy^{(12)}$ is the only one leading to non-trivial projective realization of local symmetry, $U_\sigma(C)^2=-\openone$, and is therefore identified as the cocycle in $H^3(\G,U(1))$ which generates the non-trivial symmetry fractionalization classes, i.e. the index $p_{12}$ labels the $SFC(Z_2\times Z_2,U(1))=Z_2$.
    (The action of the matrices in the vison pair Hilbert space is given in the text.)}
  \label{tab:g2s2}
\end{table*}

Let us from now on use the additive notation for our groups, i.e. any element $g\in Z_2$ takes values $g=0,1$, with $0$ the identity element, and the group product becomes addition $g_1\cdot g_2\equiv g_1+g_2\,\text{(mod 2)}$.

The different classes of phases described by our models are classified by the inequivalent 3-cocycles $\ccy$. An explicit expression for all representatives of inequivalent 3-cocycles in case where the group is a product of $Z_n$ factors is given in Ref.~\onlinecite{Propitius:1995p6856}, and the result for the case $\G=Z_2\times Z_2$ is shown in Table~\ref{tab:1}. This means that there are three indices $p_1,p_2,p_3\in Z_2$, labeling the elements of $H^3(\SG\times\GG,U(1))=Z_2^3$. The trivial phase has $p_1=p_2=p_3=0$, and can be described by the trivial cocycle $\ccy(x,y,z)=1,\forall x,y,z\in\G$. When an index $p_I$ is $1$, the representative 3-cocycle $\ccy\in H^3(\SG\times\GG,U(1))$ of that phase is chosen to satisfy the property described in the definition of $\ccy^{(I)}$; when the index $p_I$ is $0$, this index does not affect the value of cocycle. In other words, the elementary cocycle properties $\ccy^{(I)}$ in Table~\ref{tab:1} generate all representative cocycles that are elements of $H^3$: The three basic 3-cocycles $\ccy^{(I)}$ generate a total of $2^3$ 3-cocycles in $H^3(\SG\times\GG,U(1))=Z_2^3$, each of them leading to a physically different model, as stated at the end of section~\ref{sec:connection}.

Let us now consider a pair of visons $\vis=1\in\GG\equiv (1,0)\in\G$, which are the only flux quasiparticles for this group $\G$.

Since the value of $\vis$ is fixed, we will omit it from the label of local operators, i.e. we will use the labels $D_{[g\equiv(h,\symg)]}$ with $h\in\GG,\symg\in\SG$ (this notation was introduced after Eq.~\eqref{eq:DABloc}). Let us introduce Pauli matrices $\mu_i,\rho_i,\tau_i$ acting in the Hilbert space of the visons on the basis $\ket{\vis=1,\gvis;u_A}\equiv \ket{(\gvish,\gvisg);u_A}$ from Eq.~\eqref{eq:57}, with the usual $\gvis\in\G\equiv(\gvish\in\GG,\gvisg\in\SG)$. The action of the matrices is naturally defined through setting $\mu_z$ as $(-1)^{\gvisg}$, $\rho_z$ as $(-1)^{\gvish}$, and finally $\tau_z$ as $(-1)^{u_A}$. The local operators, Eq.~\eqref{eq:DABloc}, together with the global and fractionalized symmetry operators, Eq.~\eqref{eq:62}, are presented in Table~\ref{tab:g2s2} for all basic 3-cocycles $\ccy^{(I)}$.

We can see that the global symmetry $U(\sigma\equiv 1\in\SG)$, which sends $u_C\rightarrow u_C+1$ is in this basis equal to $\tau_x$, and as expected commutes with all local operators, for any choice of cocycle.

The phase in the definition of fractionalized symmetry, Eq.~\eqref{eq:62}, is for all cocycles trivial (note that the cocycle $\epsilon_{x,y,z}$ contains one factor of $\ccy$ for each permutation of $x,y,z$). The fractionalized symmetry is then just equal to the local operator $U_\sigma(C)=D_{[(0,1)]}(C)$, and clearly for each cocycle it is true that $U(\sigma)=U_\sigma(A) U_\sigma(B)=D_{[(0,1)]}(A) D_{[(0,1)]}(B)=\tau_x$.

There is a non-trivial symmetry-fractionalized phase in case of $\ccy^{(12)}$, since $U_\sigma(C)^2=-\openone$. (One can also confirm this from the general expression Eq.~\eqref{eq:65b}.) That means precisely that the index $p_{12}$ labels a non-trivial realization of symmetry fractionalization, where the local symmetry operation on a single vison at position $C$ is projectively realized $U_\sigma(C)^2=(-1)^{p_{12}}\openone$, exactly as claimed in Eq.~\eqref{eq:sgZ2_ggZ2}). The index $p_{12}$ is obviously the only index classifying the $SFC$, $SFC(Z_2\times Z_2,U(1))=Z_2$, as presented in Eq.~\eqref{eq:H3Z2Z2}.

We can now consider the influence of interplay between global symmetry and topological order on the physical system with excitations. Let us assume that the excitations are a vison pair at $A,B$. The topological nature of the state must be robust against arbitrary \textit{local} perturbations, i.e. we have to consider adding arbitrary local perturbation terms to the Hamiltonian (Eq.~\eqref{eq:5}). The local perturbations however also have to be \textit{symmetric}, i.e. commute with the global symmetries. The set of all local perturbation terms is formed by the local operators, i.e. $\tilde{D}=\{D(A),D(B)\}$, so that
\begin{equation}
  \label{eq:52}
  H_{perturb}=H+\sum_{\alpha}a_\alpha\tilde{D}^\alpha,
\end{equation}
where the index $\alpha$ labels all operators in the set $\tilde{D}$, and $a_\alpha$ are arbitrary coefficients. The operators $\tilde{D}$ are by construction symmetric, which is easily explicitly checked by showing that their matrices commute with the global symmetry operation $\tau_x$. For any fixed elementary cocycle in Table~\ref{tab:g2s2}, one can see that there are no other matrices that commute with the perturbations (except the global symmetries of course). Further, all the local operators, for any fixed cocycle, commute with each other. Therefore, the algebra of conserved observables is trivial, and there are no degeneracies protected by symmetry. (This will change in the next examples.)

Physically it is important to note that the eigenvalues of the local operator $D_{[(1,0)\in\GG]}$ are actually the values of gauge charge of the vison excitations (note that they act on the $\gvish$ element in the ribbon operators, where from Eq.~\eqref{eq:55}: $\gvish=\hG_{1,0}\cdots \hG_{N,N-1}\in\GG$ for a ribbon with inner edge on lattice sites $0,\ldots,N$). These operators are indeed the same on both visons, $D_{[(1,0)\in\GG]}(A)=D_{[(1,0)\in\GG]}(B)$.
The gauge charge value is an important quantity that will be non-trivial when the $\GG$ is enlarged in our third example.


\subsection{Example with symmetry protected degeneracy: Vison pair in $\GG=Z_2$ and $\SG=Z_2\times Z_2$}
\label{sec:exampl-gg=z_2-sg=z_2-1}

In this example, we again have only one type of vison, $\vis=1\in\GG$, but two $Z_2$ global symmetry generators, $\sigma=(1,0)\in\SG\equiv (0,1,0)\in\G$ and $\tau=(0,1)\in\SG\equiv (0,0,1)\in\G$.

We use the elementary 3-cocycles for the group $\G=Z_2\times Z_2\times Z_2$ according to the results of Ref.~\onlinecite{Propitius:1995p6856}, and present them in Table~\ref{tab:2}. This means that there are seven indices $p_I=0,1$, labeling the elements of $H^3(\G,U(1))=Z_2^7$.
The seven elementary 3-cocycles $\ccy^{(I)}$ generate a total of $2^7$ cocycles in $H^3(\G,U(1))=Z_2^7$, each of them leading to a physically different model, as stated at the end of section~\ref{sec:connection}. However, the physical properties of the models will be very different depending on the particular choice of $\SG$ and $\GG$, even though $\G$ is the same in this and the next example.
\begin{table}[t]
  \centering
  \begin{tabular}[c]{l|c}
    $\ccy(x,y,z)=-1$ & $x,y,z$\\
    \hline

    $\ccy^{(1)}$ & $(1,*,*),(1,*,*),(1,*,*)$\\
    $\ccy^{(2)}$ & $(*,1,*),(*,1,*),(*,1,*)$\\
    $\ccy^{(3)}$ & $(*,*,1),(*,*,1),(*,*,1)$\\
    \hline
    $\ccy^{(12)}$& $(1,*,*),(*,1,*),(*,1,*)$\\
    $\ccy^{(23)}$& $(*,1,*),(*,*,1),(*,*,1)$\\
    $\ccy^{(13)}$& $(1,*,*),(*,*,1),(*,*,1)$\\
    \hline
    $\ccy^{(123)}$& $(1,*,*),(*,1,*),(*,*,1)$\\
  \end{tabular}
  \caption{Inequivalent elementary 3-cocycles $\ccy^{(I)}$ for $\G\equiv\GG\times\SG=Z_2\times Z_2\times Z_2$ which label classes in $H^3(\G,U(1))=Z_2^7$. A cocycle $\ccy(x,y,z)$ has value $1$, \textit{except} when the group elements $x,y,z\in\G$ take the special values shown in the table, in which case $\ccy(x,y,z)$ is equal to $-1$. The element notation $x=(g_1,g_2,g_3)$ signifies the direct product factorization, and so in the example $\GG=Z_2$, $\SG=Z_2\times Z_2$, $g_1\in\GG$, while in the example $\GG=Z_2\times Z_2$, $\SG=Z_2$, $g_3\in\SG$. The elements of $Z_2$ are $0,1$ (additive group action). The star symbol $*$ stands for ``any value of element''.}
  \label{tab:2}
\end{table}

After fixing $\vis$, it is straightforward to construct the matrices for all local operators by using Eqs.~\eqref{eq:DABloc}, as well as to construct the fractionalized symmetry operators using their definition, Eq.~\eqref{eq:62}, for each cocycle in Table~\ref{tab:2}. We use the tensor product of Pauli matrices which are defined through setting $\tau_z$ as $(-1)^{\gvish}$, $\mu_z$ as $(-1)^{\gvisg_1}$, $\sigma_z$ as $(-1)^{\gvisg_2}$, $\nu_z$ as $(-1)^{u_{A1}}$, and $\rho_z$ as $(-1)^{u_{A2}}$ in our standard basis $\ket{\gvish,\gvisg;u_A}\equiv \ket{\gvish\times\gvisg_1\times\gvisg_2;u_{A1}\times u_{A2}}$, with $\gvish\in\GG$, $(\gvisg_1,\gvisg_2)\in\SG$, $(u_{A1},u_{A2})\in\SG$.
\begin{table}[t]
  \centering
  \begin{tabular}[c]{r||c|c|c||c|c|c||c}
    \begin{tabular}[c]{c} $\sigma=(0,1,0)$\\$\tau=(0,0,1)$\end{tabular} & $\ccy^{(1)}$ & $\ccy^{(2)}$ & $\ccy^{(3)}$ & $\ccy^{(12)}$ & $\ccy^{(23)}$ & $\ccy^{(13)}$ & $\ccy^{(123)}$\\
            \hline\hline
$D_{[\tau]}(A)$ & $\sigma_x\rho_x$ & $\sigma_x\rho_x$ & $\sigma_x\rho_x$ & $\sigma_x\rho_x$ & $\sigma_x\rho_x$ & $-i\sigma_y\rho_x$ & $\sigma_x\rho_x$\\
$U_\tau(A)$ & $\sigma_x\rho_x$ & $\sigma_x\rho_x$ & $\sigma_x\rho_x$ & $\sigma_x\rho_x$ & $\sigma_x\rho_x$ & $-i\sigma_y\rho_x$ & $\sigma_x\nu_z\rho_x$\\
\hline
$D_{[\tau]}(B)$ & $\sigma_x$ & $\sigma_x$ & $\sigma_x$ & $\sigma_x$ & $\sigma_x$ & $i\sigma_y$ & $\mu_z\sigma_x$\\
$U_\tau(B)$ & $\sigma_x$ & $\sigma_x$ & $\sigma_x$ & $\sigma_x$ & $\sigma_x$ & $i\sigma_y$ & $\sigma_x\nu_z$\\
\hline
$D_{[\sigma]}(A)$ & $\mu_x\nu_x$ & $\mu_x\nu_x$ & $\mu_x\nu_x$ & $i\mu_y\nu_x$ & $\mu_x\nu_x$ & $\mu_x\nu_x$ & $\mu_x\sigma_z\nu_x$\\
$U_\sigma(A)$ & $\mu_x\nu_x$ & $\mu_x\nu_x$ & $\mu_x\nu_x$ & $i\mu_y\nu_x$ & $\mu_x\nu_x$ & $\mu_x\nu_x$ & $\mu_x \sigma_z\nu_x\rho_z $\\
\hline
$D_{[\sigma]}(B)$ & $\mu_x$ & $\mu_x$ & $\mu_x$ & $i\mu_y$ & $\mu_x$ & $\mu_x$ & $\mu_x$\\
$U_\sigma(B)$ & $\mu_x$ & $\mu_x$ & $\mu_x$ & $i\mu_y$ & $\mu_x$ & $\mu_x$ & $\mu_x\sigma_z\rho_z$\\
\hline
$D_{[(1,0,0)]}(A)$ & $i\tau_y$ & $\tau_x$ & $\tau_x$ & $\tau_x$ & $\tau_x$ & $\tau_x$ & $\tau_x$\\
$D_{[(1,0,0)]}(B)$ & $i\tau_y$ & $\tau_x$ & $\tau_x$ & $\tau_x$ & $\tau_x$ & $\tau_x$ & $\tau_x$\\
  \end{tabular}
  \caption{Relevant local operators and fractionalized symmetries for $\GG=Z_2$, $\SG=Z_2\times Z_2$, for (the only possible) vison pair $\vis=(1,0,0)\equiv 1\in\GG$ located at $A,B$.
    The symmetry generators are $\sigma=(1,0)\in\SG$, $\tau=(0,1)\in\SG$, and they act as $U(\sigma)=\nu_x$ and $U(\tau)=\rho_x$. There is symmetry fractionalization $U(s)=U_s(A)U_s(B)$, non-trivial (projective) for cocycles $\ccy^{(12)}$, $\ccy^{(13)}$, and $\ccy^{(123)}$, which ``mix'' $\GG$ with $\SG$ (Table~\ref{tab:2}).
    The fractionalized symmetries differ (by a local operation) from local operators $D_{[\sigma]}$, $D_{[\tau]}$ only for the $\ccy^{(123)}$ cocycle (see Eq.~\eqref{eq:62}). The definitions of all matrices in the vison pair Hilbert space are given in the text.}
  \label{tab:g2s22}
\end{table}

Direct inspection of the obtained matrices, which are presented in Table~\ref{tab:g2s22}, reveals the results quoted in sections~\ref{sec:psg},~\ref{sec:connection}. Namely, for all cocycles one indeed finds that both global symmetries, $U(\sigma)=\nu_x$ and $U(\tau)=\rho_x$, are \textit{fractionalized}: $U(s)=U_s(A)U_s(B)$. Although $U(\sigma)^2=U(\tau)^2=\openone$ and $U(\sigma)U(\tau)=U(\tau) U(\sigma)$, the fractionalized symmetry can be realized projectively:
\begin{itemize}
\item For cocycle $\ccy^{(12)}$ (which ``mixes'' the $\GG$ and the first $Z_2$ of $\SG$), one finds the only non-trivial case $U_\sigma(A)^2=U_\sigma(B)^2=-\openone$.
\item For cocycle $\ccy^{(13)}$ (which ``mixes'' the $\GG$ and the second $Z_2$ of $\SG$), one finds the only non-trivial case $U_\tau(A)^2=U_\tau(B)^2=-\openone$.
  \item For cocycle $\ccy^{(123)}$ (which ``mixes'' all three subgroups), one finds the only non-trivial case $U_\sigma(C) U_\tau(C)=-U_\tau(C) U_\sigma(C)$, for both $C=A,B$.
\end{itemize}

One should note that only in the case $\ccy^{(123)}$ the local operators had to be modified to obtain the fractionalized symmetry operators, as clearly seen in the last column of Table~\ref{tab:g2s22}. We can directly confirm that these are \textit{local} redefinitions by noticing that the operators $(-1)^{u_{A1}}=\nu_z$, $(-1)^{u_{A2}}=\rho_z$ are local at $A$, while operators $(-1)^{u_{B2}}=\sigma_z\rho_z$, $(-1)^{u_{B1}}=\mu_z\nu_z$ are local at $B$.

We can now consider the influence of interplay between global symmetry and topological order on the physical system with excitations by repeating the analysis of local symmetric perturbations as in the previous example. (In this example it is also easy to explicitly check that the local, symmetric perturbation terms $\tilde{D}=\{D(A),D(B)\}$ indeed commute with the global symmetry operations $\nu_x$, $\rho_x$.)

We can now ask: What operators can be used to label this state? Let us for concreteness consider the state having the cocycle $\ccy^{(123)}$. The matrices that commute with the entire $\tilde{D}$ algebra (which includes $D_{[g\in\G]}$ operators not shown explicitly in Table~\ref{tab:g2s22}), and therefore with the arbitrarily locally perturbed Hamiltonian (Eq.~(\ref{eq:52}), are:
\begin{itemize}
\item $\rho_x$, $\nu_x$, which are just the global symmetries.
\item $\tau_x$, which measures the gauge charge, as will be explained further below.
  \item $\sigma_x\nu_z$, $\mu_x\sigma_z\rho_z$, $\sigma_x\rho_x\nu_z$, $\mu_x\sigma_z\nu_x\rho_z$, which are just the fractionalized symmetries $U_\tau(B)$, $U_\sigma(B)$, $U_\tau(A)$, and $U_\sigma(A)$, respectively. (Note that these operators are not symmetric themselves, as a consequence of $U_\sigma(C) U_\tau(C)=-U_\tau(C) U_\sigma(C)$.)
\end{itemize}
In the algebra formed from these conserved operators, there are exactly two disjunct pairs, $U_\tau(B),U_\sigma(B)$ and $U_\tau(A),U_\sigma(A)$, respectively, which anticommute, as we already learned through the symmetry fractionalization for $\ccy^{(123)}$. Each anticommuting pair, acting on one of the visons, forces a two-fold degeneracy on the state. (We checked this degeneracy numerically by considering an arbitrary perturbation term from Eq.~(\ref{eq:52}).) We arrive at the physical signature of this phase:

\textbf{The symmetry protects a 2-fold degeneracy \textit{per vison} of the pair in the $p_{123}=1$ phase of $\GG=Z_2$, $SG=Z_2\times Z_2$ model.} The degeneracy due to a vison at $C$ is labeled by the fractionalized symmetry operators $U_\sigma(C),U_\tau(C)$.

Let us now discuss the gauge charge observables $\tau_x$. These are just the local operators $D_{[1\in\GG]}$, listed in Table~\ref{tab:g2s22}. Physically, the operators $D_{[1\in\GG]}(B)$ appear in the braiding matrix $\mathcal{R}^2_{CC}$, Eq.~\eqref{eq:R2pi}, describing the braiding of two visons at $A$-ends of two ribbons which share a $B$-end, see Fig.~\ref{fig:ribbons}, Section~\ref{sec:braiding-matrix} and Appendix~\ref{app:extended_ribbon}.

Finally, let us emphasize that, at least in the case when $\G$ is a product of simple $Z_2$ factors, there is no need to explicitly construct all the matrices as was done here. One can directly resort to Eqs.~\eqref{eq:63}, \eqref{eq:65a}-\eqref{eq:65c} to test whether a cocycle allows SF, and then whether this SF is non-trivial. For this purpose, we note that the cocycle $\epsilon$ is easy to use since it contains 6 factors of $\ccy(x,y,z)$, such that each permutation of $x,y,z$ appears exactly once; importantly, when $\G$ has only $Z_2$ factors, the generating 3-cocycles in $H^3(\G,U(1))$ (Ref.\onlinecite{Propitius:1995p6856}) can only take values $\pm 1$, as is the case in Tables~\ref{tab:1},~\ref{tab:2}.

\subsection{The simplest example beyond symmetry fractionalization: Vison pairs in $\GG=Z_2\times Z_2$ and $\SG=Z_2$}
\label{sec:exampl-visons-gg=z_2}

In this example there are two fundamental vison types. We will consider in parallel a pair of $\vis=(1,0)\in\GG\equiv (1,0,0)\in\G$ visons and a pair of $\vis=(0,1)\in\GG\equiv (0,1,0)\in\G$ visons.

We again construct the matrices of all local operators by using Eqs.~\eqref{eq:DABloc}. Before proceeding, one should notice that the condition stated in Eq.~\eqref{eq:64} is violated for the cocycle $\ccy^{(123)}$: only in that case does $\epsilon_{\vis,\sigma,\gvish}=-1$ become possible since $\vis$ and $\gvish$ can differ while both being non-trivial elements of $\GG$. According to Eq.~\eqref{eq:63}, we then expect that in that case the symmetry fractionalization will fail.

Let us nevertheless continue with the explicit construction of fractionalized symmetry operators. Using their definition, Eq.~\eqref{eq:62}, for each cocycle in Table~\ref{tab:2} including $\ccy^{(123)}$, each fractionalized symmetry operator is just equal to its corresponding local operator $D$. We therefore avoid repetition in Table~\ref{tab:g22s2}, and one should keep in mind that $U_\sigma(C)\equiv D_{[\sigma]}(C)$, where the only global symmetry generator is $\sigma=1\in\SG$.
\begin{table}[t]
  \centering
  \begin{tabular}[c]{r||c|c|c||c|c|c||c}
$\sigma=(0,0,1)$& $\ccy^{(1)}$ & $\ccy^{(2)}$ & $\ccy^{(3)}$ & $\ccy^{(12)}$ & $\ccy^{(23)}$ & $\ccy^{(13)}$ & $\ccy^{(123)}$\\
\hline\hline
\multicolumn{7}{c}{Vison pair: $\vis=(1,0,0)$}\\
\hline
$D_{[\sigma]}(A)$ & $\sigma_x\rho_x$ & $\sigma_x\rho_x$ & $\sigma_x\rho_x$ & $\sigma_x\rho_x$ & $\sigma_x\rho_x$ & $-i\sigma_y\rho_x$ & $\sigma_x\rho_x$\\
$D_{[\sigma]}(B)$ & $\sigma_x$ & $\sigma_x$ & $\sigma_x$ & $\sigma_x$ & $\sigma_x$ & $i\sigma_y$ & $\mu_z\sigma_x$\\
\hline
$D_{[(1,0,0)]}(A)$ & $i\tau_y$ & $\tau_x$ & $\tau_x$ & $\tau_x$ & $\tau_x$ & $\tau_x$ & $\tau_x$\\
$D_{[(1,0,0)]}(B)$ & $i\tau_y$ & $\tau_x$ & $\tau_x$ & $\tau_x$ & $\tau_x$ & $\tau_x$ & $\tau_x$\\
\hline
$D_{[(0,1,0)]}(A)$ & $\mu_x$ & $\mu_x$ & $\mu_x$ & $-i\mu_y$ & $\mu_x$ & $\mu_x$ & $\mu_x\sigma_z$\\
$D_{[(0,1,0)]}(B)$ & $\mu_x$ & $\mu_x$ & $\mu_x$ & $i\mu_y$ & $\mu_x$ & $\mu_x$ & $\mu_x$\\
\hline
\multicolumn{7}{c}{Vison pair: $\vis=(0,1,0)$}\\
\hline
$D_{[\sigma]}(A)$ & $\sigma_x\rho_x$ & $\sigma_x\rho_x$ & $\sigma_x\rho_x$ & $\sigma_x\rho_x$ & $-i\sigma_y\rho_x$ & $\sigma_x\rho_x$ & $\sigma_x\rho_x$\\
$D_{[\sigma]}(B)$ & $\sigma_x$ & $\sigma_x$ & $\sigma_x$ & $\sigma_x$ & $i\sigma_y$ & $\sigma_x$ & $\tau_z\sigma_x$\\
\hline
$D_{[(0,1,0)]}(A)$ & $\mu_x$ & $i\mu_y$ & $\mu_x$ & $\mu_x$ & $\mu_x$ & $\mu_x$ & $\mu_x$\\
$D_{[(0,1,0)]}(B)$ & $\mu_x$ & $i\mu_y$ & $\mu_x$ & $\mu_x$ & $\mu_x$ & $\mu_x$ & $\mu_x$\\
\hline
$D_{[(1,0,0)]}(A)$ & $\tau_x$ & $\tau_x$ & $\tau_x$ & $-\tau_x$ & $\tau_x$ & $\tau_x$ & $\tau_x\sigma_z$\\
$D_{[(1,0,0)]}(B)$ & $\tau_x$ & $\tau_x$ & $\tau_x$ & $\tau_x$ & $\tau_x$ & $\tau_x$ & $\tau_x$\\
  \end{tabular}
  \caption{Relevant local operators for $\GG=Z_2\times Z_2$, $\SG=Z_2$: (top half) for a vison pair $\vis=(1,0,0)\equiv (1,0)\in\GG$ located at $A,B$; (bottom half) for a vison pair $\vis=(0,1,0)\equiv (0,1)\in\GG$ located at $A,B$.
    The only global symmetry generator is $\sigma=(0,0,1)\equiv 1\in\SG$, acting as $U(\sigma)=\rho_x$. Except for $\ccy^{(123)}$, there are fractionalized symmetry operators $U_\sigma(C)\equiv D_{[\sigma]}(C)$, with $C=A,B$, (see Eq.~\eqref{eq:62}), such that $U(\sigma)=U_\sigma(A)U_\sigma(B)$. The fractionalized symmetries are non-trivial (projective) for cocycles $\ccy^{(13)}$ and $\ccy^{(23)}$, which ``mix'' $\GG$ with $\SG$ (Table~\ref{tab:2}).
    Phase having $\ccy^{(123)}$ is beyond SFC, since for it $D^{\vis=(010)}_{[\sigma]}(A) D^{\vis=(010)}_{[\sigma]}(B)=\tau_z\rho_x$, $D^{\vis=(100)}_{[\sigma]}(A) D^{\vis=(100)}_{[\sigma]}(B)=\mu_z\rho_x$, and no local redefinition of $D_{[\sigma]}$ can remove the $\tau_z,\mu_z$ factors. The definitions of all matrices are given in the text.}
  \label{tab:g22s2}
\end{table}

The Pauli matrices are here defined through setting $\tau_z$ as $(-1)^{\gvishone}$, $\mu_z$ as $(-1)^{\gvishtwo}$, $\sigma_z$ as $(-1)^{\gvisg}$, and $\rho_z$ as $(-1)^{u_A}$ in our standard basis $\ket{\gvish,\gvisg;u_A}\equiv \ket{\gvishone\times\gvishtwo\times\gvisg;u_A}$, with $(\gvishone,\gvishtwo)\in\GG$, $\gvisg\in\SG$, $u_A\in\SG$.

Direct inspection reveals that the global symmetry, represented by $U(\sigma)=\rho_x$, is fractionalized for both visons, i.e. $U(\sigma)=U^{\vis}_\sigma(A)U^{\vis}_\sigma(B)=\rho_x$, \textit{except for the cocycle $\ccy^{(123)}$}. In the case of this cocycle, we get
\begin{equation}
  \label{eq:65}
  \begin{aligned}
  U^{\vis=(1,0)}_\sigma(A)U^{\vis=(1,0)}_\sigma(B)&=\mu_z\rho_x\\ U^{\vis=(0,1)}_\sigma(A)U^{\vis=(0,1)}_\sigma(B)&=\tau_z\rho_x    
\end{aligned}
\qquad\text{(cocycle }\ccy^{(123)}\text{)},
\end{equation}
for the vison pairs $\vis=(1,0)$ and $\vis=(0,1)$, respectively. We will further discuss this below.

Considering all the cocycles (\textit{except} $\ccy^{(123)}$), we find phases with non-trivial SF, i.e. where the fractionalized symmetry is realized projectively:
\begin{itemize}
\item For cocycle $\ccy^{(13)}$ (which ``mixes'' the first $Z_2$ in $\GG$ with $\SG$), one finds the only non-trivial case $U^{\vis=(1,0)}_\sigma(A)^2=U^{\vis=(1,0)}_\sigma(B)^2=-\openone$.
\item For cocycle $\ccy^{(23)}$ (which ``mixes'' the second $Z_2$ in $\GG$ with $\SG$), one finds the only non-trivial case $U^{\vis=(0,1)}_\sigma(A)^2=U^{\vis=(0,1)}_\sigma(B)^2=-\openone$.
  \end{itemize}
  These results match the claims in section~\ref{sec:psg}. There are obviously two symmetry-fractionalization $Z_2$ indices, $p_{13}$ and $p_{23}$, in accordance with the general results: $H^2(SG,GG)=Z_2^2$ and, for this group, $SFC(SG,GG)=H^2(SG,GG)$ (see section~\ref{sec:connection}).
  
We can try to use our explicit matrix expressions to ``force'' symmetry fractionalization in the case of cocycle $\ccy^{(123)}$. This attempt will fail, as expected from general arguments above, and therefore this state is beyond SFC, with the index $p_{123}$ labeling the $\text{EXTRA}(SG,GG)$ class. What we would need to achieve is the removal of $\mu_z$ and $\tau_z$ factors in Eq.~\eqref{eq:65}. Inspecting the operators
  \begin{align}
    \label{eq:66}\notag
    (-1)^{u_A}&=\rho_z\\
    (-1)^{u_B}&=\sigma_z\rho_z,
  \end{align}
  the first being local at $A$, and the second local at $B$, we see that there is no hope in manipulating the $\mu$ and $\tau$ subspaces. Physically, $\mu_z$ and $\tau_z$ measure the $\GG$ degrees of freedom, and these are connected to the vison gauge charge which is non-local.
The local operators $D_{[\vis]}$ (which we will use to measure the gauge charge shortly) by definition change this group element (i.e., $\tau_x$) and cannot provide us with $\tau_z,\mu_z$.

We next ask: which operators can be used to physically label this state? Let us consider the especially interesting cocycle $\ccy^{(123)}$. As explained in the previous example, we need to find matrices that commute with the entire $\tilde{D}$ algebra (which includes all $D_{[g\in\G]}$ operators). These are:
\begin{itemize}
\item $\rho_x$, which is just the global symmetry.
\item $\gcop_1$, $\gcop_2$, which measure the two $Z_2$ gauge charges in $\GG$, as we explain further below. For the $\vis=(1,0)$ vison pair, $\gcop_1\equiv\tau_x$, $\gcop_2\equiv\mu_x\sigma_z\rho_z$, while for $\vis=(0,1)$ vison pair, $\gcop_1\equiv\tau_x\sigma_z\rho_z$, $\gcop_2\equiv\mu_x$.
\end{itemize}
In the algebra of these conserved operators, exactly one pair anticommutes: the global symmetry anticommutes with the gauge charge of $Z_2$ complementary to the visons' flux $\vis$. For example, considering a vison pair with flux set by $\vis=(1,0)\in\GG$, the \textit{second} $Z_2$ gauge charge of $\GG$ ($\gcop_2$) anticommutes with the global symmetry. A signature of this phase is then:

\textbf{The symmetry protects a 2-fold degeneracy of a \textit{single vison pair} in the $p_{123}=1$ phase of $\GG=Z_2\times Z_2$, $SG=Z_2$ model.}

This result for the cocycle $\ccy^{(123)}$ reveals its physical interpretation. To understand it better, we need to explicitly consider the consequences of anticommutation between symmetry and gauge charge. Let us proceed with that goal, and leave for later the detailed explanation of why $\gcop_1$, $\gcop_2$ can indeed be understood as gauge charge operators.

The quasiparticles of this theory are labeled by the values of flux and charge, and we consider the elementary ones: 1) Flux particle $m_1=1$[$m_2=1$] has flux in the first[second] $Z_2$ of $\GG$, which is set by the values of $\vis=(*,0)$$[\vis=(0,*)]$, and no gauge charge; 2) Charge particle $e_1=1$[$e_2=1$] has charge in the first[second] $Z_2$ of $\GG$, which is set by the values of $\gvish=(*,0)$$[\gvish=(0,*)]$, and has no flux. A general quasiparticle will have arbitrary values (0 or 1) for these four numbers, so we label it $(e_1,e_2,m_1,m_2)$. For instance, $(1,1,0,0)$ is particle bound state of charge and flux in the first $Z_2$ of $\GG$. When symmetry anticommutes with $\gcop$, that means it switches the two eigenvalues of $\gcop$, and therefore switches the particles' $\gcop$ number between trivial (0) and non-trivial (1). As we established, symmetry anticommutes with $\gcop_1$ when the particle pair in question is $m_2$, while it anticommutes with $\gcop_2$ when the particle pair is $m_1$. This leads to the following transformation of elementary quasiparticles under the action of symmetry:
\begin{equation}
  \label{eq:74}
  (e_1,m_1,e_2,m_2) \overset{\sigma}\rightarrow (e_1+m_2,m_1,e_2+m_1,m_2),
\end{equation}
where (mod 2) algebra is understood. As expected, the symmetry changes the charges of particles, but we can further assign a deeper meaning to this transformation. Namely, in $Z_2$ theories the charge and flux are physically equivalent: they are dual to each other, and we can rename them at will without changing the physics (e.g. the braiding properties) of particles. We therefore assign new fluxes $\tilde{m}$ and charges $\tilde{e}$ according to the rules: $(\tilde{e}_1,\tilde{m}_1,\tilde{e}_2,\tilde{m}_2)\equiv(e_1+m_2,m_1,m_2,e_2+m_1)$. Note that the quasiparticle statistics is indeed conserved, i.e. bosonic except between $e_i$ and $m_i$ when it is fermionic. This leads to the elegant transformation rule:
\begin{equation}
  \label{eq:75}
(\tilde{e}_1,\tilde{m}_1,\tilde{e}_2,\tilde{m}_2) \overset{\sigma}\rightarrow (\tilde{e}_2,\tilde{m}_2,\tilde{e}_1,\tilde{m}_1),
\end{equation}
so that the global symmetry transformation only exchanges the particle types 1 and 2! (We note that this physical interpretation might carry on to higher dimensions. For instance in 3d, charge excitations are point-like while fluxes are string-like, so they cannot be interchanged among each other; however it is still well defined to change their type, e.g. $1\leftrightarrow 2$.)

We have therefore found the physical nature of this state:
\textbf{The global symmetry operation exchanges the two quasiparticle types in $p_{123}=1$ phase of $\GG=Z_2\times Z_2$, $SG=Z_2$ model. This shows that symmetry performs a non-local transformation on the excitation pair, and the state is beyond symmetry fractionalization classification.}

Let us now discuss in detail the gauge charge observables $\gcop_1$, $\gcop_2$, as promised above. We focus on a particular vison pair $\vis$. As in the previous example, one naively expects the gauge charge observables to just equal the appropriate local operators: $D_{[(1,0)\in\GG\equiv (1,0,0)]}$ and $D_{[(0,1)\in\GG\equiv (0,1,0)]}$, which are listed in Table~\ref{tab:g22s2}. Focusing on the cocycle $\ccy^{(123)}$, we however see that these local operators differ at the two quasiparticles, e.g. $D_{[(1,0,0)]}(A)\neq D_{[(1,0,0)]}(B)$, which would mean that measuring the gauge charge of a vison and anti-vison in the pair would give differing answers. This is physically wrong, but it is easy to resolve the problem. Namely, the physical operators $\gcop_1,\gcop_2$ differ from the corresponding local operators $D$ by a simple local operation. Using the local transformations in Eq.~\eqref{eq:66}, one immediately obtains that
\begin{itemize}
\item For vison pair $\vis=(1,0)$:
  \begin{align}\notag
  \gcop_1&=D_{[(1,0,0)]}(A)=D_{[(1,0,0)]}(B)=\tau_x\\
  \gcop_2&=(-1)^{u_A}D_{[(0,1,0)]}(A)=(-1)^{u_B}D_{[(0,1,0)]}(B)=\mu_x\sigma_z \rho_z.     \label{eq:68}
\end{align}
\item For vison pair $\vis=(0,1)$:
  \begin{align}    \label{eq:69}
    \gcop_1&=(-1)^{u_A}D_{[(1,0,0)]}(A)=(-1)^{u_B}D_{[(1,0,0)]}(B)=\tau_x\sigma_z \rho_z\\\notag
    \gcop_2&= D_{[(0,1,0)]}(A)=D_{[(0,1,0)]}(B)=\mu_x.
\end{align}
\end{itemize}
These $\gcop_1,\gcop_2$ operators exactly match the ones we found to commute with the entire local $\tilde{D}$ algebra.

We already know that the local operators $D_{[\gvish]}(B)$ appear in the braiding matrix $\mathcal{R}^2_{CC}$, Eq.~\eqref{eq:R2pi}, describing the braiding of two visons at $A$-ends of two ribbons which share a $B$-end, see Fig.~\ref{fig:ribbons}. Physically, we expect these local operators to equal the gauge charge operators and thereby transparently provide the braiding rules of anyons in $Z_2$ topologically ordered theory. A question that arises then is: If we \textit{redefine} the local operators so that they become exactly equal to the gauge charge operators (following Eqs.~\eqref{eq:68},~\eqref{eq:69}), do the braiding properties change? The answer is negative. Namely, if we choose to braid to visons of same type $\vis^1=\vis^2\equiv\vis$, the braiding operator (Eq.~\eqref{eq:R2pi}) will apply the same $D_{(\vis,\vis)}(B)$ on both visons, and these local operators are not even redefined. When we braid two differing vison types, e.g., $\vis^1=(1,0)$, $\vis^2=(0,1)$, the braiding will apply $D_{(\vis^1,\vis^2)}$ and $D_{(\vis^2,\vis^1)}$, \textit{both of which are redefined by $(-1)^{u_B}$} according to Eqs.~\eqref{eq:68},~\eqref{eq:69}. As pointed out earlier, our analysis of vison Hilbert space revealed that such braiding operations should be considered after choosing, and keeping fixed, some value of $u_B$. Since the two ribbons share the end at $B$, the $(-1)^{u_B}$ operation applied on both braided quasiparticles cancels. Therefore, we can freely redefine the considered local operators to be equal to the physical gauge charge operators $\gcop_1,\gcop_2$.

\subsection{Gauge charges}
\label{sec:gauge-charges}

In the previous subsections we considered in detail pairs of gauge fluxes, i.e. visons. The other fundamental type of excitations is a pair of gauge charge excitations, which is created by the ribbon operator:
\begin{equation}
  \label{eq:gaugecharge}
\ket{\gvis}\equiv F^{(h=\openone,\gvis)}(\pathg)\ket{\gs},
\end{equation}
i.e., the special case of the flux $h\in\GG$ being trivial. As always, $\gvis\in\G=\SG\times\GG$ uniquely factorizes as $\gvis=\gvish\cdot\gvisg$, with $\gvish\in\GG$ being related to the gauge charge of the excitations, and $\gvisg\in\SG$. (Recall again the definition of ribbon operators Eq.~\eqref{eq:55}: $\gvish=\hG_{1,0}\cdots \hG_{N,N-1}\in\GG$ for a ribbon on lattice sites $0,\ldots,N$.)

Recall that in this paper we use ``canonical'' 3-cocycles $\omega$, meaning that $\omega(g_1,g_2,g_3)=1$ if any of $g_1,g_2,g_3$ is equal to $\openone$. Specifically, the elementary cocycles for a group $\G$ which is a product of $Z_n$ factors are written in Ref.\onlinecite{Propitius:1995p6856}, and are all canonical.

Since $\vis=\openone$ for gauge charges, it then follows immediately from Eqs.~(\ref{eq:65a}),~(\ref{eq:65b}),~(\ref{eq:62}) that for these excitations the symmetry fractionalization is always trivial. Further, Eq.~(\ref{eq:51}) shows that the braiding matrix between various gauge charges is also trivial in our models.

As a particular demonstration, let us consider the examples from the previous subsections. Eq.~(\ref{eq:DABloc}) reveals that the action of local operators $D$ in our basis of the excitation pair Hilbert space $\mathcal{L}(A,B)$ does not involve any phase factors. It is as if the chosen cocycle is always the trivial one, $\omega=1$. To avoid repetition in writing explicitly the local operators in the present case of a gauge charge pair, we refer the reader to just look at the case of: $\omega^{(2)}$ for the $\GG=Z_2$, $\SG=Z_2$ model; $\omega^{(2)}$ for the $\GG=Z_2$, $\SG=Z_2\times Z_2$ model; and finally $\omega^{(3)}$ for the $\GG=Z_2\times Z_2$, $\SG=Z_2$ model. The lack of non-trivial phase factors in the action of local operators leads to a trivial algebra: Repeating the analysis from above, one finds that in all three examples all the $D$ operators commute, and the only matrices commuting with them are the global symmetry operations, which of course commute amongst each other in these direct product groups. Therefore, there are no symmetry protected degeneracies for the gauge charges in our present examples, in sharp contrast to the case of fluxes.

\subsection{Multiple vison pairs and dualization}
\label{sec:multiple-vison-pairs}

Let us briefly consider a system with multiple vison pairs. For simplicity, we focus on the case where all these quasiparticles are of the same type. Since the calculation even for this case becomes very complicated, we will try to give general arguments and speculate about the symmetry protected degeneracy of a general state with $N$ pairs of visons.

One way to analyze this situation is to consider the dualization of $\SG$. By this we mean the standard replacement of lattice site degrees of freedom $u_i$ by edge degrees of freedom $u_{ij}\equiv u_i\cdot u_j^{-1}$, as we also mentioned in section~\ref{esmodels}. The total group becomes a pure gauge group, $\G=\GG\times\widetilde{\SG}\equiv\widetilde{\G}$, where $\widetilde{\SG}$ denotes the \textit{gauge group} $\SG$. The visons in the $\G$ model are mapped to gauge flux particles in the $\widetilde{\G}$ model. Let us denote the protected degeneracy of $N$---particle-pair state by VDEG$_{\G}(N)$ and VDEG$_{\widetilde{\G}}(N)$ for the models with $\SG$ and the dualized $\widetilde{\SG}$, respectively.

One should note that this is a many-to-one mapping, in the sense that $|\SG|$ states obtained by multiplying all $u_i$ by an arbitrary element $s\in\SG$ (in the $\G$ theory) get mapped to a single state (in the dualized $\widetilde{\G}$ theory). In general therefore, we expect that the state with visons can only have the same or smaller degeneracy upon dualization to $\widetilde{\G}$:
\begin{equation}
  \label{eq:67}
  \text{VDEG}_{\G}(N) \geq \text{VDEG}_{\widetilde{\G}}(N).
\end{equation}
Particularly, the mentioned many-to-one nature of the mapping indicates that the change in degeneracy (from $\G$ to $\widetilde{\G}$) might involve the factor $|\SG|$.

In fact, we can recall that the Hilbert space $\mathcal{L}(A,B)$ of an excitation pair in the $\SG$ description (section~\ref{sec:gener-struct-excit}) had to be specified by keeping track of value of the element $u_A$, or $u_B$, which belong to $\SG$. In the most general description of the $2N$ visons, one connects all their ribbons to a common point on the lattice $x_0$, which contains no excitation (see discussion in section~\ref{sec:braiding-matrix}). We then have to keep track of $u_{x_0}$, which takes $|\SG|$ different values. Since this degree of freedom becomes obsolete upon dualization to $\widetilde{\G}$, one is again lead to the assumption that the degeneracy of the $2N$-vison state in the dualized theory ($\widetilde{\G}$) is reduced by a factor of $|\SG|$.

Let us test these assumptions on the results of Section~\ref{sec:example}. In our first example $\SG=Z_2$, $\GG=Z_2$, a single vison pair had no symmetry protected degeneracy. Upon dualization, this model becomes the gauge theory $\widetilde{\G}=Z_2\times Z_2$. This is an Abelian theory, which implies that its excitations are anyons with quantum dimension equal to 1. (The quantum dimension is defined as the degeneracy per excitation in the limit of infinite number of excitations.) In this case, there is no degeneracy. Therefore, the inequality~(\ref{eq:67}) is saturated, and there is no factor $|\SG|=2$.

The Abelian case, having non-degenerate anyon states, might be a too special case for our general considerations. Let us instead focus on the examples with $\G=Z_2^3$, and the especially interesting cocycle $\omega^{(123)}$. In those cases we found that a single vison pair has degeneracy 4 and 2, in the $\SG=Z_2\times Z_2$ and $\SG=Z_2$ examples, respectively. The topological order described upon dualization of either of these example groups becomes very interesting due to the presence of the cocycle $\ccy^{(123)}$. Namely, the topological order realized in that way can be described by the \textit{non-Abelian} theory having the gauge group $\widetilde{\G}_{nA}=D_4$ and no twist by a cocycle\cite{Propitius:1997ccy_nA}. It is known that the $D_4$ theory has no degeneracy for a single excitation pair, i.e. VDEG$_{D_4}(1)$=1.

For a single excitation pair and the case where the topological order in the dual theory is described by $\widetilde{\G}_{nA}=D_4$, we therefore established the anticipated
\begin{equation}
  \label{eq:70}
  \text{VDEG}_{\G}(N) =|\SG|\cdot\text{VDEG}_{\widetilde{\G}_{nA}}(N),\quad(\text{for $\widetilde{\G}_{nA}=D_4,N=1$}).
\end{equation}

Let us proceed to the case of multiple pairs of particles, still all being of the same type. The excitations of the $D_4$ gauge theory have quantum dimension $2$, therefore for large $N$ one has the scaling $\text{VDEG}_{D_4}(N\gg 1)\sim 4^{N}$. Since for $N=1$ we already discussed that $\text{VDEG}_{D_4}(1)\sim 4^{0}$, we expect that an interpolation formula $\text{VDEG}_{D_4}(N)\sim 4^{(N-1)}$ should hold for all $N$. Combining this expectation with the result in Eq.~(\ref{eq:70}) leads us to conjecture:
\begin{equation}
  \label{eq:71}
\text{VDEG}_{\G}(N) =|\SG|\cdot\text{VDEG}_{\widetilde{\G}_{nA}}(N)\quad(\text{non-Abelian $\widetilde{\G}_{nA},N\geq 1$)}.
\end{equation}
In this conjecture we also anticipate that the key property which ensured Eq.~(\ref{eq:70}) is that $D_4$ is non-Abelian, since in the previous paragraphs we established that Eq.~(\ref{eq:70}) does not hold (even though $N=1$) when $\widetilde{\G}$ described an Abelian topological order (i.e. $Z_2\times Z_2$).

We leave further study of these questions for future research.

\section{Discussion and Conclusions}\label{sec:conclusion}



In this paper, we present a classification of topologically ordered phases in the presence of an on-site global symmetry, equipped with local bosonic exactly solvable models. The discussion in Sec.\ref{sec:connection} and the solutions of our models in some examples allow us to reveal the physical meaning of this classification, which is given in Eq.(\ref{eq:full_exp}) and Eq.(\ref{eq:SET_exp}). Basically, the classification $H^3(SG\times GG,U(1))$ is a direct product of many finite abelian groups, and each group has a clear physical meaning. Among them $H^3(SG,U(1))$ is the index for SPT phases, $H^3(GG,U(1))$ is the index for Dijkgraaf-Witten topological orders, $SET(SG,GG)$ in Eq.(\ref{eq:full_exp}) is the index labeling the possible interplays between the global symmetry and the topological order (symmetry enriched topological phases).

$SET(SG,GG)$ can be further understood. It is a direct product of two abelian groups, $SFC(SG,GG)$ and $EXTRA(SG,GG)$. $SFC(SG,GG)$ is labeling different possible ways to fractionalize the global symmetry by the topological order, while $EXTRA(SG,GG)$ is labeling the phenomena that the global symmetry transformation could interchange quasiparticle species, which is beyond the symmetry fractionalization scheme. 

Some measurable consequences of the symmetry enriched topological phases are discussed. In particular, we show that under certain conditions, symmetry enriched topological phases have non-trivial symmetry protected degeneracy for excited states. For instance, in one striking example beyond the symmetry fractionalization scheme, we show that symmetry protects a 2-fold degeneracy for a single pair of gauge flux quasiparticles, which cannot be locally associated with either quasiparticle.
This degeneracy of excited states can be used to detect the symmetry enriched topological phases in numerics/experiments. However, we leave the most general diagnosis of symmetry enriched topological phases for future investigation. In any case, the exactly solvable models constructed in the current work can be very useful tools for this purpose.

The SPT phases are known to host gapless edge states, topologically protected by the global symmetry. We have not studied the possible gapless edge states due to SET topological indices in this paper. We also leave this issue as a subject of future investigations.

Below we discuss the generalization and limitations of our classification.

\subsection{Generalization to higher dimensions}
\label{sec:discussion_general_dim}

Although we have been focusing on 2+1 dimensions, where the classification is given by $H^3(SG\times GG,U(1))$, this can be easily generalized to $H^{d+1}(SG\times GG,U(1))$ in general d+1 dimensions ($d\geq 2$). One way to understand this generalization is to dualize $GG$ to be part of global symmetry, after which the system has an on-site global symmetry $\widetilde\SG=\SG\times\GG$. According to Ref.\onlinecite{Chen:2011p6670}, the SPT phases in d+1 dimensions are classified by $H^{d+1}(\tilde SG,U(1))$, and it is natural to expect that these phases may still different before the duality. (In fact, this point of view is not completely correct, which we will discuss shortly in Sec.\ref{sec:completeness}. )

The higher dimension classification is also equipped with local bosonic exactly solvable models. For instance, in 3+1d, one can also construct the exactly solvable models on a 3-dimensional lattice which triangulates the 3-dimensional space. The $\GG$ degrees of freedom live on the edges and the $\SG$ degrees of freedom live on the vertices. Similar to the 2+1d case, the Hamiltonian of the model is a sum of local projectors. Theorem-1 and Theorem-2 (see Sec.\ref{sec:DW_inv}) in 3+1d dictate that these local projectors mutually commute and the model is exactly solvable. We leave the complete solution of such 3+1d models (i.e. the excited states) as a subject of future investigations.

However, at this moment it is still possible to explore the physical meaning of the 3+1d classification using $H^4(SG\times GG,U(1))$. A K\"{u}nneth expansion immediately gives:
\begin{align}
 H^4(SG\times GG,U(1))=&H^4(SG,U(1))\times H^4(GG,U(1))\notag\\
&\times SET^4(SG,GG)
\end{align}
Here clearly, the index $H^4(SG,U(1))$ is labeling the SPT phases, while $H^4(GG,U(1))$ is labeling the different 3+1d topological orders described by the gauge group $GG$. Note that $H^4(GG,U(1))$ is labeling the direct generalization of the Dijkgraaf-Witten discrete gauge theories in 3+1 dimension. Further, $SET^4(SG,GG)$, as explicated below, is labeling the different possible interplays between the global symmetry and the topological order in 3+1d. 

Let us use $SET^3(SG,GG)$ to denote the symmetry enriched indices $SET(SG,GG)$ in Eq.(\ref{eq:SET_exp}), emphasizing that it is for 2+1 dimensions. $SET^4(SG,GG)$ contains a rather different mathematical structure than $SET^3(SG,GG)$:
\begin{align}
 &SET^4(SG,GG)=[H^3(SG,Z)\otimes H^2(GG,Z)]\times \notag\\
&[H^2(SG,Z)\otimes H^3(GG,Z)]\times \mbox{Tor}[H^4(SG,Z),H^2(GG,Z)]\times\notag\\
&\mbox{Tor}[H^2(SG,Z),H^4(GG,Z)]\times \mbox{Tor}[H^3(SG,Z),H^3(GG,Z)].\label{eq:SET4_exp}
\end{align}

One can anticipate the possible non-trivial interplay between the global symmetry and the topological order in 3+1d. Firstly, let's consider the generalization to 3+1d of $EXTRA(SG,GG)$, a part of $SET^3(SG,GG)$. In 2+1d, $EXTRA(SG,GG)$ is describing the interchange of quasiparticle species by $SG$ action. We expect that in 3+1d, part of $SET^4(SG,GG)$ also describes the interchange of excitation species. Note that in 3+1d, the topological excitations can be either point-like gauge charges or loop-like gauge fluxes. More precisely, this part of $SET^4(SG,GG)$ should describe the species-interchange of gauge fluxes and of gauge charges by the global symmetry (but not interchange of gauge flux and gauge charge).

Next, we consider the generalization to 3+1d of the other part of $SET^3(SG,GG)$: $SFC(SG,GG)$. One may wonder, does part of $SET^4(SG,GG)$ label the symmetry fractionalization phenomena in 3+1d? Naively, if we consider the symmetry fractionalization classes of point-like gauge charges, we should have a similar mathematical structure as in 2+1d (see Eq.(\ref{eq:PSG1})). In particular, when $\GG$ is abelian, we should have $H^2(SG,GG)=[H^2(SG,Z)\otimes H^2(GG,Z)]
\times \mbox{Tor}(H^3(SG,Z),H^2(GG,Z))$ indices labeling the different projective representations of the symmetry group. However, this mathematical structure is missing in $SET^4(SG,GG)$ above. 

In fact, we know that symmetry fractionalization classes of gauge charges are actually missing from $SET^3(SG,GG)$ in 2+1d, as discussed in Sec.~\ref{sec:gauge-charges} (see also \ref{sec:connection}). It is not surprising that $SET^4(SG,GG)$ is also missing those indices. However, we know that the symmetry fractionalization classes of gauge fluxes are completely contained in $SET^3(SG,GG)$. We expect that $SET^4(SG,GG)$ also contains the generalized ``symmetry fractionalization classes'' of gauge fluxes --- the topological loop excitations in 3+1d. Therefore, a part of $SET^4(SG,GG)$ should describe the non-trivial action of global symmetry on gauge flux loops, \emph{without changing their species}. But because gauge flux loops are extended objects, the action of global symmetry on them can no longer be implemented by local operators. So, to be precise, we should not call this phenomenon symmetry fractionalization, as it is discussed in Sec.\ref{sec:psg}. We will call it the \emph{extended symmetry fractionalization}.

Although the full understanding of the extended symmetry fractionalization is beyond the scope of this paper, we can intuitively guess the underlying mathematical structure. When $GG$ is abelian, a direct generalization of $H^2(SG,GG)$ to one higher dimension is $H^3(SG,GG)$. By the universal coefficients theorem, we have:
\begin{align}
 H^3(SG,GG)&=[H^3(SG,Z)\otimes H^2(GG,Z)]\notag\\
&\times\mbox{Tor}[H^4(SG,Z),H^2(GG,Z)],
\end{align}
where we used $H^2(GG,Z)=GG$ for finite abelian group $GG$. These two terms indeed appear in Eq.(\ref{eq:SET4_exp}) (the 1st and the 3rd term). We propose that $H^3(SG,GG)$ is at least part of the mathematical structure describing the extended symmetry fractionalization classes when $GG$ is abelian.

\subsection{Generalization to continuous groups, and/or anti-unitary symmetry groups}

Our discussion has been limited to the case in which both $SG$ and $GG$ are finite groups, and $SG$ is assumed to be unitary (i.e., does not contain time-reversal). These constraints are introduced here for simplicity rather than due to difficulty of principle.

First, it is quite straightforward to consider an on-site symmetry group $SG$ containing the anti-unitary time-reversal transformation $\mathcal{T}$. In the work by Xie \textit{et al.} on SPT phases\cite{Chen:2011p6670}, when $SG$ contains $\mathcal{T}$, the classification is given by $H^{d+1}(SG,U_T(1))$. Here $U_T(1)$ means that $\mathcal{T}$ acts non-trivially on the $U(1)$ group; in particular, $\mathcal{T}$ sends the phase $e^{i\theta}\in U(1)$ to its complex conjugate $e^{-i\theta}$. (For the detailed definition and discussion of $H^{d+1}(SG,U_T(1))$, see Ref.\onlinecite{Chen:2011p6670}). We expect that the classification of gapped bosonic quantum phases with topological order described by a $\GG$ gauge group, and in the presence of an on-site global symmetry group $SG$ containing $\mathcal{T}$, is given by $H^{d+1}(SG\times GG,U_T(1))$, in which only the $SG$ part of the cross product acts nontrivially on $U_T(1)$.

Further, Ref.\onlinecite{Chen:2011p6670} considered the classification of SPT phases with a continuous on-site symmetry group $SG$, in which case the Borel group cohomology was used.
It appears to us that the inclusion of a continuous $\SG$ results in an appropriate (but may be subtle) mathematical generalization of $H^{d+1}(\G,U(1))$ for a finite group $G$. We in principle do not expect that this generalization is hindered by difficulties.

\subsection{About the completeness of the classification}\label{sec:completeness}

Finally we comment on the issue of completeness of the classification by $H^{d+1}(SG\times GG,U(1))$. \emph{Is the classification complete, incomplete, or overcomplete?}

First, we want to comment on the following question: Do distinct elements in $H^{d+1}(SG\times GG,U(1))$ necessarily correspond to distinct quantum phases? We believe that the answer is negative, and the classification is generally overcomplete in this sense. One can understand this claim by considering a simple example $SG=Z_1$ and $GG=Z_2\times Z_2$ in 2+1 dimensions. According to $H^3(GG,U(1))=Z_2^3$, it appears that there are 8 different topological orders. But this is overcomplete. Among these three $Z_2$ indices, the first(second) $Z_2$ is labeling the toric-code/double-semion topological order in the first(second) $Z_2$ gauge group, and the third $Z_2$ is labeling a certain extra twist of the topological order involving both $Z_2$ gauge groups. Let's consider two phases labeled by $(1,0,0)$ and $(0,1,0)$, where we group the three $Z_2$ indices into a vector. Clearly $(1,0,0)$ and $(0,1,0)$ physically correspond to the same phase --- they just differ by an ordering of the gauge group. This is analogous to the K-matrix classification of abelian quantum Hall states. For instance $K=\big(\begin{smallmatrix}1&0\\0&2\end{smallmatrix}\big)$ and $K=\big(\begin{smallmatrix}2&0\\0&1\end{smallmatrix}\big)$ are labeling the same physical phase.

The $SET(SG,GG)$ classification of SET phases is also overcomplete in this sense. For instance, the symmetry fractionalization classes when $SG=Z_2$ and $GG=Z_2\times Z_2$ are given by $SFC(SG,GG)=H^2(SG,GG)=H^2(SG,Z_2)\times H^2(SG,Z_2)=Z_2\times Z_2$, using the universal coefficients theorem. If we use $(a,b)$, with $a,b=0,1\in Z_2$, to represent this index, then $(1,0)$ [$(0,1)$] simply means that there is non-trivial symmetry fractionalization in the first[second] $Z_2$ gauge sector, and these are physically the same situation.

Second, our classification is certainly \emph{not} a full classification of all possible gapped bosonic quantum phases with both global symmetry and topological order. There are certainly topological orders that cannot be described by discrete gauge theories, for instance, the chiral fractional quantum Hall states. Even for non-chiral topological orders, there are phases realized by the string-net models\cite{Levin:2005p3468} in which quasiparticle quantum dimensions are not integers, which again cannot be described by discrete gauge theories, where quasiparticle quantum dimensions must be integers.

So let us ask, under the constraint that the topological order is indeed described by a discrete gauge theory, is the classification complete? However, we must firstly specify the condition: what exactly do we mean by ``topological order described by a discrete gauge theory''? We actually mean nothing but phases characterized by $H^{d+1}(GG,U(1))$. Unfortunately this sounds like a circular argument, but we do not know how to translate it into a more physical statement.  For example, $H^4(Z_2,U(1))=Z_1$, meaning that there is only one $Z_2$ gauge theory in 3+1d in our context. However the string-net models in 3+1d allow one to construct two topological phases, both of which look like $Z_2$ gauge theories. The difference is that in one phase the $Z_2$ gauge charge is a boson, while in the other phase it is a fermion. In our classification, the second phase is not included.

Even under this condition, namely the topological order given by $H^{d+1}(GG,U(1))$, it seems that the classification may still be incomplete. For instance, we already mentioned that the gauge charges always have trivial symmetry fractionalization in our classification. Maybe it is better to ask: Is the classification complete under additional physical conditions? And if yes, what are these additional physical conditions? These are difficult questions and we currently do not know the answers. Nevertheless, we can make some comments. Below we address two aspects of these issues.

\emph{Is the classification of symmetry fractionalization classes complete?}
We want to further comment on the \emph{missing} symmetry fractionalization classes for gauge charges. Let us focus on 2+1d, where the gauge charges and gauge fluxes are dual to each other, at least for abelian $\GG$. In this case, we can re-interpret the symmetry fractionalization classes for gauge fluxes as those for gauge charges, after performing the duality. So the real missing part should be those phases with non-trivial symmetry fractionalization for both gauge charges and gauge fluxes.

Naively one may think that it is possible to construct such a phase by coupling two phases together. For instance, consider $\GG=Z_2$. Our models can be used to construct two phases: phase-a (phase-b) in which gauge fluxes $m_a$ (charges $e_b$) have non-trivial symmetry fractionalization (i.e. transform under $\SG$ as a non-trivial projective representation), while $e_a$ ($m_b$) transform trivially under $\SG$. (Phase-b can be constructed by performing $e\rightarrow m$ duality from phase-a.) We can couple these two phases together. When the coupling is weak the topological order is $Z_2\times Z_2$. After a phase transition of condensing $m_a m_b$ bound states (or the $e_a e_b$), the topological order will reduce from a $Z_2\times Z_2$ gauge theory to a $Z_2$ gauge theory. But the condensed $m_am_b$ (or $e_a e_b$) quasiparticles actually transform non-trivially under $SG$! And consequently the new $Z_2$ topologically ordered phase breaks $\SG$.

However, it seems possible to construct an effective lattice gauge theory for the phases with non-trivial symmetry fractionalization (for on-site $\SG$) for both gauge charges and gauge fluxes. So it is feasible to expect that these phases do exist. But the above discussion signals that these phases may not be adjacent to the phases classified by $H^{d+1}(SG\times GG,U(1))$ in some sense. The reason that we miss those phases in $H^{d+1}(SG\times GG,U(1))$ may be because they cannot be described by exactly solvable models.

\emph{Is the classification of the interplay between $\SG$ and $\GG$ beyond symmetry fractionalization complete?}
Let us consider the classes indexed by non-trivial elements in $EXTRA(SG,GG)$ in 2+1d. We show that in the example of $GG=Z_2\times Z_2$ and $SG=Z_2$, the minimal model for a non-trivial $EXTRA(SG,GG)$, such a phase means that $\SG$ can interchange the species of quasiparticles. However, we know that in 2+1d, even for the usual toric code topological order with $\GG=Z_2$, it is fine to imagine that $SG=Z_2$ could interchange $e$ and $m$ quasiparticles, leaving the fusion and braiding algebra invariant. In fact, the translational symmetry (\emph{not} an on-site symmetry) along the 45 degree axis of a square-lattice toric code model indeed interchanges $e$ and $m$. Such an $e$ and $m$ interchange induced by $\SG$ is also missing from $H^{3}(GG,U(1))$. It is however actually reasonable that such phenomena are missing in our classification. That is because our classification can be generalized to arbitrary higher dimensions, while the $e$-$m$ interchange can only occur in 2+1d. For instance, in 3+1d $e$ is point-like and $m$ is loop-like, so they can never be interchanged.

As we prepared this manuscript, we noticed the recent work by Ling-Yan Hung and Xiao-Gang Wen\cite{duality_wen}, which discusses the general duality between SPT phases and the Dijkgraaf-Witten TQFTs. Another recent work, by Andrew M. Essin and Michael Hermele\cite{Hermele_2012}, classifies the general symmetry fractionalization of gapped $Z_2$ quantum spin liquids. Also, Yuting Hu, Yidun Wan, and Yong-Shi Wu\cite{YongShi2012} have very recently analyzed in detail the topological phases described by models similar to ours in absence of global symmetries.

YR thanks helpful discussions with Fa Wang and especially Michael Hermele. This work is supported by the Alfred P. Sloan foundation and National Science Foundation under Grant No. DMR-1151440.

\appendix
\section{Projective symmetry group in parton construction}\label{app:PSG}

Parton construction is a convenient way to obtain quantum states with topological order. The basic idea is to write down a topologically ordered state directly using the anyonic quasiparticle degrees of freedom. For example, in the Schwinger-fermion representation of quantum spin liquids with $Z_2$ topological order, fermionic quasiparticles are used to represent a physical spin-1/2: $\vec S_i=1/2 f_{i\alpha}^{\dagger}\vec \sigma_{\alpha\beta}f_{i\beta}$, where $i$ labels sites and $\alpha,\beta$ label spins. Note that this construction enlarges the Hilbert space from 2 to 4 per site, and one eventually needs to remove the unphysical states (empty and doubly-occupied sites) to obtain a physical spin-1/2 wavefunction. This removal of unphysical states can be accomplished by the so-called Gutzwiller projection: $P_{G}\equiv \prod_i n_i(2-n_i)$, where $n_i=f_{i\alpha}^{\dagger}f_{i\alpha}$ is the fermion number on site-$i$.

In this approach, on the mean-field level, a $Z_2$ quantum spin liquid can be represented as a free fermion state $|\psi_{MF}\rangle$ of $f_{i\alpha}$ fermions, which is the ground state of a spin-singlet mean-field Hamiltonian:
\begin{align}
H_{MF}=\sum_{ij}\chi_{ij}f_{i\alpha}^{\dagger}f_{j\alpha}+\Delta_{ij}\epsilon_{\alpha\beta}f_{i\alpha}^{\dagger}f_{j\beta}^{\dagger}+h.c.\label{eq:z2mf}
\end{align} 
$H_{MF}$ has both hopping and pairing terms on a lattice ($\epsilon_{\alpha\beta}$ term is the spin singlet pairing). The physical spin-1/2 wavefunction of the $Z_2$ QSL can be obtained by Gutzwiller projection: $|\psi_{QSL}\rangle=P_{G}|\psi_{MF}\rangle$. Here the $Z_2$ gauge fluctuations emerge exactly because of this projection: two mean-field states differing by a gauge transformation $\chi_{ij}\rightarrow \epsilon_i \chi_{ij}\epsilon_j, \Delta_{ij}\rightarrow \epsilon_i\Delta_{ij}\epsilon_j$ ($\epsilon_i=\pm 1$) give exactly the same spin wavefunction. Therefore such local $Z_2$ fluctuations correspond to redundancies in the formulation and are gauge fluctuations. The $f_{i\alpha}$ fermions are the quasiparticles carrying the $Z_2$ gauge charge. The low energy effective theory of the state $|\psi_{QSL}\rangle$ is described by $Z_2$ gauge charges $f_{i\alpha}$ coupled with a dynamical $Z_2$ gauge field.

How can we make sure that the QSL wavefunction $|\psi_{QSL}\rangle$ is symmetric under a symmetry group SG, such as lattice translations? Naively one would require the mean-field Hamiltonian $H_{MF}$ to be invariant under SG transformations. In fact, this is not required. Because two mean-field states differing by a $Z_2$ gauge transformation label exactly the same physical state, one only requires $H_{MF}$ to be ``projectively symmetric''. Namely, $H_{MF}$ before and after an SG transformation can differ by a $Z_2$ gauge transformation. This is the key observation underlying the PSG.

For any element $g$ in SG, there will be a certain $Z_2$ gauge transformation $G_{g}$ associated with the $g$ such that the combination $G_g \cdot g$ leaves the mean-field Hamiltonian $H_{MF}$ or the state $|\psi_{MF}\rangle$ invariant. The collection of all such combinations form a group, which is defined to be PSG: $PSG\equiv \{G_g \cdot g: G_g \cdot g \mbox{ leaves $H_{MF}$ invariant}, \forall g\in SG\}$. 

Let's look at the mean-field Hamiltonian Eq.(\ref{eq:z2mf}) again. Clearly, one can do a global gauge transformation: $\epsilon_i=-1,\forall i$ and $H_{MF}$ is invariant, which is also the only non-trivial gauge transformation which leaves $H_{MF}$ invariant. This means that in PSG, there will be two elements corresponding to the identity element in SG: either $\epsilon_i=1,\forall i$, or $\epsilon_i=-1,\forall i$. In general, for any element $g\in SG$ there will be two elements, $G_g\cdot g$ and $\tilde G_g\cdot g$ in PSG corresponding to it. And the gauge transformations $G_g$ and $\tilde G_{g}$ differ by the global $Z_2$ gauge transformation. Mathematically, the algebraic relation between PSG and SG is given by\cite{Wen:2002p162}:
\begin{align}
 PSG/IGG=SG.\label{eq:PSG2}
\end{align}
Here IGG=$Z_2$, the group of global gauge transformations.

Eq.(\ref{eq:PSG2}) is the key mathematical structure underlying PSG. It indicates that PSG is a group extension of the group SG by IGG. When IGG is abelian, which is true in the $Z_2$ case, we know that IGG is in the center of PSG, because global gauge transformations obviously commute with any PSG element. In this case, a PSG is a central extension of SG by IGG. Further, the classification of all different PSGs becomes the classification of all possible central extensions. There is a nice mathematical theorem on central extensions of groups stating that all such central extensions are classified by $H^2(SG,IGG)$(see, for example, Ref\cite{robinson1996coursetheorygroups}).

At this moment, it appears that PSG is a feature of the parton mean-field states only. Whether PSG is physical or not beyond the mean-field formulation is not completely clear. To see the physical meaning of PSG and $H^2(SG,IGG)$ beyond the mean-field formulation, we need to consider the low energy effective theory, which is discussed in the main text.


\section{The operator realization of twisted extended ribbon algebra and the quasi-quantum double}\label{app:extended_ribbon}
In this section we write down the explicit forms of the ribbon operators $F^{(h,g)}(\Gamma)$ and the operators acting on ends of ribbons $D_{(h,g)}(A),D_{(h,g)}(B)$, and demonstrate the algebra satisfied by these operators. As in the main text, we also only consider an Abelian group $SG\times GG$. These operators are defined for $h\in  GG$, $g\in SG\times GG$. In addition, we study the braiding and fusion properties of the quasiparticles created by these ribbon operators, and show that they are mathematically described by the quasi-quantum double.\cite{Dijkgraaf:1990p7462,Propitius:1993CS_ccy}

\subsection{The operator realization of the twisted extended ribbon algebra}

Let us start by noting that the definition and some properties of ribbon operators were presented in Sec.~\ref{sec:ribbon-operator}. Here we start by recalling the form of the non-zero matrix element of the ribbon operator $F^{(h,g)}(\Gamma)$:
\begin{align}
\bra{\text{fin}}F^{(h,g)}(\Gamma)\ket{\text{i}}= f_A\cdot f_B \cdot f_{AB}\cdot w_h^{\Gamma}(g),
\end{align}
where $w_h^{\Gamma}(g)$ is defined in Eq.(\ref{eq:24}), and the relation between the initial and final states, $\ket{\text{i}},\ket{\text{fin}}$, is explained in Eq.~\eqref{eq:Fmatelem}; the $f_A,f_B,f_{AB}$ are rather complicated phase factors depending only on the degrees of freedom living on ends of $\Gamma$, which we here explicitly define:
\begin{align}
&f_A=\frac{\omega(h_Ah^{-1},h,b_N)\omega(b_N,h,h_Ah^{-1})\omega(e_{N+1}^{-1},b_Nc_{N+1}^{-1},h)}{\omega(c_{N+1},b_Nc_{N+1}^{-1},h)\omega(h_Ah^{-1},h,h_A^{-1})\omega(h_A^{-1},h,h_Ah^{-1})},\notag\\
&f_B=\frac{c_{h_B}(h,h_B)c_{hh_B}(c_1,h)}{c_{hh_B}(h_B,h)c_{c_1}(h,h_B)}\notag\\
&\;\;\;\;\cdot\frac{\omega(h,h^{-1}c_1^{-1}b_0,e_1^{-1})\omega(c_1,h,h^{-1}c_1^{-1}b_0)}{\omega(h_B,h,h_B^{-1}h^{-1})},\notag\\
&f_{AB}=\omega^{-1}(h,h_A^{-1},h_Ah_B).\label{eq:f_factors}
\end{align}
Here the flux in $t_A$ is $h_A=b_N^{-1}c_{N+1}e_{N+1}$ and the flux in $t_B$ is $h_B=b_0^{-1}c_1e_1$ in the initial state $|\{u_i\},\{\tilde g_{ij}\}\rangle$. (see Fig.\ref{fig:FD} for the definitions of $b_0,b_{N}$... degrees of freedom living of the ends of $\Gamma$). Although here we use the specific geometric configuration of the ribbon $\Gamma$ in Fig.\ref{fig:FD}, the definitions of $F^{(h,g)}(\Gamma)$, and $D_{(h,g)}(A)$ ($D_{(h,g)}(B)$) below, can be easily generalized to any geometric configuration of $\Gamma$.

Next, we define the operator $D_{(h,g)}(A)$ ($D_{(h,g)}(B)$) explicitly on end-A (end-B) (also mentioned in the main text), where $h\in GG$, $g\in SG\times GG$. They are defined as an operator in the whole Hilbert space $\mathcal{H}$ (In fact they are also well-defined in $\mathcal{K}(\Gamma)$, because $D_{(h,g)}(A)$,$D_{(h,g)}(B)$ do not change the flux of a 2-simplex inside $\Gamma$), via the matrix elements:
\begin{align}
\bra{\text{f}}D_{(h,g)}(A)\ket{\text{i}}&=\delta_{h_A,h}\cdot c_h(g,b_N)W_6(i)\notag\\
\bra{\text{f}}D_{(h,g)}(B)\ket{\text{i}}&=\delta_{h_B,h}\cdot c_h(g,c_1)W_6(i),
\end{align}
where the phase factor $W_6(i)$ is defined in the main text in Eq.(\ref{eq:76}).

With these definitions, after straightforward but complicated algebra, one can show that,
in both the Hilbert space $\mathcal{H}$ and its subspace $\mathcal{K}(\Gamma)$:
\begin{align}
 D_{(h_2,g_2)}(A)\cdot D_{(h_1,g_1)}(A)&=\delta_{h_1,h_2}\cdot c_{h_1}(g_2,g_1)D_{(h_1,g_2g_1)}(A)\notag\\
 D_{(h_2,g_2)}(B)\cdot D_{(h_1,g_1)}(B)&=\delta_{h_1,h_2}\cdot c_{h_1}(g_2,g_1)D_{(h_1,g_2g_1)}(B).\label{eq:DD_o_a}
\end{align}

In the sub-Hilbert space $\mathcal{K}(\Gamma)$, we have more identities:
\begin{align}
 &F^{(h_1,g_2g_1)}(\Gamma)D_{(h_2h_1,g_2)}(A)\notag\\
=&c_{g_2}(h_2,h_1)c_{h_1}(g_2,g_1)D_{(h_2,g_2)}(A)F^{(h_1,g_1)}(\Gamma),\label{eq:DF_o_a_A}\\
&D_{(h_1h_2,g_2)}(B)F^{(h_1,g_1g_2)}(\Gamma)\notag\\
=&c_{g_2}(h_1,h_2)c_{h_1}(g_1,g_2)F^{(h_1,g_1)}(\Gamma)D_{(h_2,g_2)}(B),\label{eq:DF_o_a_B}
\end{align}
and,
\begin{align}
 &F^{(h_2,g_2)}(\Gamma)F^{(h_1,g_1)}(\Gamma)\notag\\
=&\delta_{g_1,g_2}\cdot c_{g_1}(h_2,h_1)F^{(h_2h_1,g_1)}(\Gamma).\label{eq:FF_o_a}
\end{align}
Eqs.(\ref{eq:DD_o_a},\ref{eq:DF_o_a_A},\ref{eq:DF_o_a_B},\ref{eq:FF_o_a}) are summarized in Eqs.(\ref{eqDDD},\ref{eqFD},\ref{eqDF},\ref{eqFFF}) in the main text.

In addition, it can be easily shown that \emph{$D_{(h ,g)}(A)$, $D_{(h ,g)}(B)$ and $F^{(h ,g)}(\Gamma)$ operators all commute with the global symmetry transformations in $\SG$.}

\subsection{Braiding}
\subsubsection{Braiding between the quasiparticles at end-$A$'s}
Next, we describe the braiding and fusion properties of the quasiparticles created by the $F^{h,g}(\pathgA)$ operators when applied on the ground state $|gs\rangle$. For these purposes, we must consider multiple ribbons. Starting from multiple ribbons $\pathgA_1,\pathgA_2,...\pathgA_N$, we define a sub-Hilbert space $\mathcal{K}(\pathgA_1,\pathgA_2,...\pathgA_N)\subset \mathcal{H}$ to be the one spanned by those states satisfying zero-flux rule everywhere inside $\pathgA_1,\pathgA_2,...\pathgA_N$. Now all the operators $F^{(h,g)}(\pathgA_n)$, $D_{(h,g)}(A_n)$, $D_{(h,g)}(B_n)$, $n=1,2,...N$ are all well-defined in $\mathcal{K}(\pathgA_1,\pathgA_2,...\pathgA_N)$. ($A_n$ and $B_n$ are the two ends of the ribbon-$\pathgA_n$).
\begin{figure*}
\includegraphics[width=0.8\textwidth]{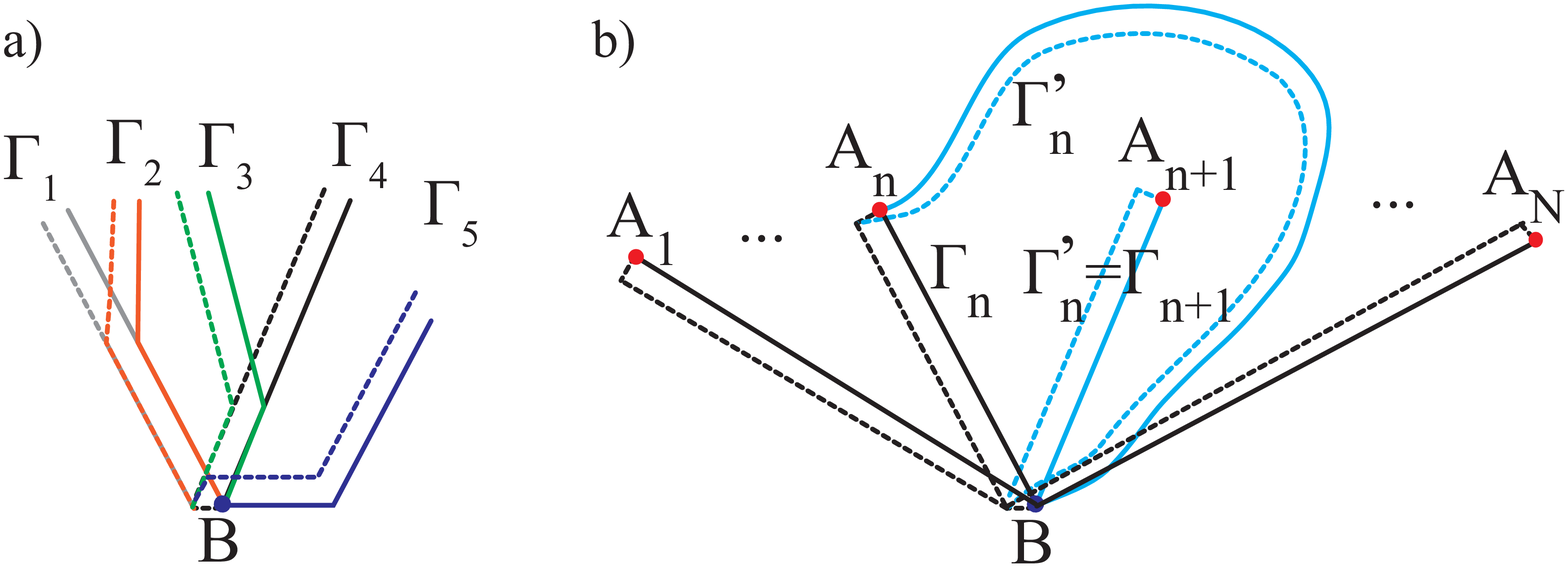}
\caption{Multiple ribbons sharing end-$B$, describing quasiparticles at their end-$A$'s. (a) Example of ordering multiple (five) ribbons, with strings $\pathg_1\ldots\pathg_5$, in a ``counter-clockwise sense''. The ribbons are allowed to have overlaps only over a finite length starting from end-$B$. The figure shows a realization possible on a triangular lattice. (b) The counter-clockwise $180^\circ$ braiding of particles $A_n$ and $A_{n+1}$ in an $N$-particle state, a generalization of Fig.~\ref{fig:ribbons}. The (blue) strings $\pathg$' apply to the braided state.}
\label{fig:multiple_ribbons}
\end{figure*}

In order to make the braiding and fusion algebra of the quasiparticles at both the $A_n$ and $B_n$ ends on the same footing, it is convenient to introduce a slightly modified ribbon operator:
\begin{align}
 \tilde F^{(h,g)}(\pathgA)\equiv F^{(h,g)}(\pathgA) f_{AB}^{-1}, 
\end{align}
where the phase factor $f_{AB}$ is given in Eq.(\ref{eq:f_factors}). $\tilde F^{(h,g)}(\pathgA)$ also commutes with the global symmetry, and satisfies the same operator algebra as $F^{(h,g)}(\pathgA)$ with $D(A),D(B)$ operators (see Eq.(\ref{eq:DF_o_a_A},\ref{eq:DF_o_a_B})). (But the algebra in Eq.(\ref{eq:FF_o_a}) is no longer satisified by $\tilde F^{(h,g)}(\pathgA)$.) We can then construct an excited state using multiple ribbons $\pathgA_1,...\pathgA_N$ that all share the same end-B, but end-$A_n$ are all different from each other. Here by ``sharing the same end-$B$'', we require that the vertices $i_{B_n}$, the triangles $t_{B_n}$, and the edge-$B_n$'s are all the same: $B_1=B_2=...=B_N\equiv B$, $\mbox{edge-}B_1=\mbox{edge-}B_2=...=\mbox{edge-}B_N\equiv\mbox{edge-}B$, Let's only consider the case that $\pathgA_1,...\pathgA_N$ do not overlap/intersect with each other except over a finite length starting from end-$B$, which is enough for our purposes. We can choose an ordering of the ribbons: $\pathgA_n$ is on the counter-clockwise side of $\pathgA_{n+1}$, $\forall n$. (See Fig.\ref{fig:multiple_ribbons}a for a geometric illustration.) An excited state is given by :
\begin{align}
 |\psi_{u_B}^{k_1...k_N}\rangle=\hat P_{u_B}(B)\tilde F^{k_1}(\pathgA_1)\tilde F^{k_2}(\pathgA_2)...\tilde F^{k_N}(\pathgA_N)|gs\rangle,
\end{align}
where we have used $k_1=(h_1,k_1)...k_N=(h_N,g_N)$ to save notation. $\hat P_{u_B}(B)$ is a projection operator that enforces $u_{i_B}=u_B\in SG$ at end-B, which commutes with all the $\tilde F$-operators. $|\psi_{u_B}^{k_1...k_N}\rangle$ only contains quasiparticles at end-$A_n$ and end-$B$, and will be very useful to understand the braiding and fusion algebra.  We denote the sub-Hilbert space spanned by all the states $|\psi_{u_B}^{k_1...k_N}\rangle$ with a fixed $u_B$ as $\mathcal L_{u_B}(A_1,A_2,...A_N,B)$, because one can show that this sub-Hilbert space only depends on the ribbons' ends, but does not depend on the paths of the ribbons. 

We will soon study the braiding and fusion operations of the quasiparticles at the end-$A_n$'s in $\mathcal L_{u_B}(A_1,A_2,...A_N,B)$. These operations only act within the sub-Hilbert space $\mathcal L_{u_B}(A_1,A_2,...A_N,B)$ for a fixed $u_B$, because $u_B$ is always unchanged in them. In fact, we will show that the braiding and fusion algebra (or, the superselection sectors of end-$A_n$ quasiparticles) only involve $D_{(h_1,h_2)}(B)$, with $h_1,h_2\in GG$. 

However, when a general $D_{(h,g)}(B)$ operator ($g=h_g\cdot \tilde g$, $h_g\in GG$ and $\tilde g\in SG$) acts on a state in $\mathcal L_{u_B}(A_1,A_2,...A_N,B)$, it will send the state to a different sub-Hilbert space $\mathcal L_{\tilde g\cdot u_B}(A_1,A_2,...A_N,B)$. In addition, the global symmetry transformation of $\tilde g$ also sends $\mathcal L_{u_B}(A_1,A_2,...A_N,B)$ to $\mathcal L_{{\tilde g}^{-1}\cdot u_B}(A_1,A_2,...A_N,B)$. Therefore if one wants to study the general $D_{(h,g)}(B)$ operators and the global symmetry transformations, one should consider a larger sub-Hilbert space: $\mathcal L(A_1,A_2,...A_N,B)\equiv\oplus_{u_B\in SG}\mathcal L_{u_B}(A_1,A_2,...A_N,B)$. Although we do not have a proof, we believe that $\mathcal L(A_1,A_2,...A_N,B)$ contains all possible excited states at end-$A_n$'s and the end-$B$. (For single ribbon states $|\psi_{u_B}^{k}$, we do have a proof in the main text that they span all possible excited states at the end-$A$ and the end-$B$.)

Note that because $h_{A_n}=\openone$ in $|gs\rangle$, $f_{AB}=1$ due to the canonical form of 3-cocycle. Therefore $|\psi^{k_1...k_N}\rangle$ can be equally created by the original $F^{(h,g)}(\pathgA)$ ribbon operator:
\begin{align}
 |\psi_{u_B}^{k_1...k_N}\rangle= \hat P_{u_B}(B) F^{k_1}(\pathgA_1)  F^{k_2}(\pathgA_2)...  F^{k_N}(\pathgA_N)|gs\rangle,
\end{align}
which is given in Eq.(\ref{eq:48}) in the main text. In fact, to study the braiding and fusion operations of the end-$A_n$ quasiparticles, it does not matter whether $F^{(h,g)}(\pathgA)$ or $\tilde F^{(h,g)}(\pathgA)$ is used, because they give the same algebra. This is why we only use  $F^{(h,g)}(\pathgA)$ operators in the main text for simplicity.

The geometric illustration of the braiding process between an end-$A_n$ and an end-$A_{n+1}$ has been discussed in the main text, but is also shown in Fig.~\ref{fig:multiple_ribbons}b. The following operator identity is crucial to compute the braiding algebra: when $\pathgA_1$ and $\pathgA_2$ share the same end-B but have different end-$A$'s, and $\pathgA_1$ is on the counter-clockwise side of $\pathgA_2$:
\begin{align}
 &\tilde F^{(h_1,g_1)}(\pathgA_2)\tilde F^{(h_2,g_2)}(\pathgA_1)\notag\\
=&c_{h_2}(g_2h_1^{-1},h_1)\tilde F^{(h_2,g_2h_1^{-1})}(\pathgA_1)\tilde F^{(h_1,g_1)}(\pathgA_2)\frac{\omega(h_1,h_2,h_B)}{\omega(h_2,h_1,h_B)}.\label{eq:FF_share_B}
\end{align}
\emph{Here $h_B$ should be understood as an operator that measures the gauge flux in the 2-simplex $t_B$}. The order between the the $F$-term and the $\omega$-term on the right-hand side is therefore important. Eq.(\ref{eq:FF_share_B}) already describes the physical counter-clockwise braiding ($180^{\circ}$) operations $\hat R_{CC}^{n,n+1}$ between quasiparticles at end-$A_n$ and end-$A_{n+1}$ in $\mathcal L(A_1,...,A_N,B)$ completely:

\begin{widetext}
\begin{align}
 &\hat R_{CC}^{n,n+1}|\psi_{u_B}^{k_1,...(h_n,g_n),(h_{n+1},g_{n+1})...,k_N}\rangle\notag\\
=&\hat R_{CC}^{n,n+1}\hat P_{u_B}(B)\tilde F^{k_1}(\pathgA_1)...\tilde F^{(h_{n},g_{n})}(\pathgA_n)\tilde F^{(h_{n+1},g_{n+1})}(\pathgA_{n+1}) ...\tilde F^{k_N}(\pathgA_N)|gs\rangle\notag\\
=&\hat P_{u_B}(B)\tilde F^{k_1}(\pathgA_1)...\tilde F^{(h_{n},g_{n})}(\pathgA_{n+1})\tilde F^{(h_{n+1},g_{n+1})}(\pathgA_{n}) ...\tilde F^{k_N}(\pathgA_N)|gs\rangle\notag\\
=&\frac{\omega(h_n,h_{n+1},h_{n+2}h_{n+3}...h_{N})}{\omega(h_{n+1},h_{n},h_{n+2}h_{n+3}...h_{N})}c_{h_{n+1}}(g_{n+1}h_n^{-1},h_n)\hat P_{u_B}(B)\tilde F^{k_1}(\pathgA_1)...\tilde F^{(h_{n+1},g_{n+1}h_{n}^{-1})}(\pathgA_{n}) \tilde F^{(h_{n},g_{n})}(\pathgA_{n+1})...\tilde F^{k_N}(\pathgA_N)|gs\rangle\notag\\
=&\frac{\omega(h_n,h_{n+1},h_{n+2}h_{n+3}...h_{N})}{\omega(h_{n+1},h_{n},h_{n+2}h_{n+3}...h_{N})}c_{h_{n+1}}(g_{n+1}h_n^{-1},h_n)|\psi_{u_B}^{k_1,...(h_{n+1},g_{n+1}h_{n}^{-1}),(h_{n },g_{n })...,k_N}\rangle\label{eq:Rcc_hat}
\end{align}
\end{widetext}
Here the $h_B$ in Eq.(\ref{eq:FF_share_B}) picks up the accumulated flux in $t_B$: $h_{n+2}h_{n+3}...h_{N}$.

Although everything about braiding can be understood from Eq.(\ref{eq:Rcc_hat}), the $\omega$-factor in it is not convenient. Can we get rid of this phase factor and find the underlying algebraic structure satisfied by $\hat R_{CC}^{n,n+1}$?

When acting on the ground state, $h_B=\openone$, and the $\omega$-term in Eq.(\ref{eq:FF_share_B}) is $1$ due to the canonical form of a 3-cocycle. In this particular case Eq.(\ref{eq:FF_share_B}) becomes Eq.(\ref{eq:51}). The braiding of an excited state with only two ribbons $\pathgA_1,\pathgA_2$ is indeed described by Eq.(\ref{eq:50}). Let's define a formal braiding operator $R_{CC}^{1,2}$ which implements the physical braiding $\hat R_{CC}^{1,2}$ in $|\psi^{k_1k_2}_{u_B}\rangle$:
\begin{align}
 R_{CC}^{1,2}=&R^{qr}D_r^{(1)}(B)\otimes D_q^{(2)}(B)\cdot\sigma\notag\\
=&\sigma\cdot R^{rq}D_r^{(1)}(B)\otimes D_q^{(2)}(B)\notag\\
=&\sigma\cdot \sum_{h_1,h_2\in GG}D^{(1)}_{(h_1,\openone)}(B)\otimes D^{(2)}_{(h_2,h_1)}(B),\label{eq:Rcc_a_12}
\end{align}
where $\sigma$ is the permutation operator: $\sigma|\psi_{u_B}^{k_1k_2}\rangle=|\psi_{u_B}^{k_2k_1}\rangle$, the tensor $R^{(h_1,g_1),(h_2,g_2)}=\delta_{h_1,g_2}\delta_{g_1,\openone}$ is also defined in the main text, and $D_{(h_n,p_n)}^{(n)}(B)$ with $h_n,p_n\in GG$ is defined to be a formal operator that only transforms the $\pathgA_n$ operator $\tilde F^{(h,g)}(\pathgA_{n})$ as if we are in a single ribbon state (see Eq.(\ref{eq:DF_o_a_B})). More precisely, because
\begin{align}
D_{(h_1,p_1)}(B)|\psi_{u_B}^{(\tilde h_1,\tilde g_1)}\rangle=\delta_{h_1,\tilde h_1}\cdot c_{\tilde h_1}(\tilde g_1 p_1^{-1},p_1)|\psi_{u_B}^{(\tilde h_1,\tilde g_1 p_1^{-1})}\rangle,
\end{align}
we have
\begin{align}
&D^{(1)}_{(h_1,p_1)}(B)\otimes D^{(2)}_{(h_2,p_2)}(B) |\psi_{u_B}^{(\tilde h_1,\tilde g_1),(\tilde h_2,\tilde g_2)}\rangle=\delta_{h_1,\tilde h_1}\cdot\notag\\
&\delta_{h_2,\tilde h_2}\cdot c_{\tilde h_1}(\tilde g_1 p_1^{-1},p_1)c_{\tilde h_2}(\tilde g_2 p_2^{-1},p_2)|\psi_{u_B}^{(\tilde h_1,\tilde g_1 p_1^{-1}),(\tilde h_2,\tilde g_2 p_2^{-1})}\rangle,
\end{align}
The key property of these formal tensor product operators is that they commute with any local operator at end-$A_i$, $\forall i$. That is why they are topological operators for the quasiparticles at end-$A_i$'s.

Because of Eq.(\ref{eq:DD_o_a}), the multiplication of the formal operators $D_{(h,p)}^{(n)}(B)$ ($\forall h,p\in GG$) satisfy the following algebra:
\begin{align}
 D_i^{(n)}(B)D_j^{(n)}(B)=\Omega^k_{i j}D_k^{(n)}(B),\label{eq:multiply}
\end{align}
where we used the tensor $\Omega^k_{ij}\equiv\delta_{h_i,h_j}\,\delta_{h_k,h_i}\,\delta_{g_k,g_ig_j}\,c_{h_k}(g_i,g_j)$ defined in Eq.(\ref{eq:38}). 

Eq.(\ref{eq:multiply}) tells us that the quasiparticles created by ribbon $\pathgA_n$ at end-$A_n$ form representation of this algebra. Mathematically, the algebra in Eq.(\ref{eq:multiply}) is called the multiplication in the quasi-quantum double $\mathbf D^{\tilde\omega}(GG)$\cite{Dijkgraaf:1990p7462,Propitius:1993CS_ccy}, where $\tilde \omega\in H^3(GG,U(1))$ is the 3-cocyle induced on $GG$ by the cocycle $\omega\in H^3(SG\times GG,U(1))$ in our model by restricting the elements $x,y,z\in GG$ in $\omega(x,y,z)$. Multiplications in $\mathbf D^{\tilde\omega}(GG)$ is associative. A representation of the multiplication algebra Eq.(\ref{eq:multiply}) is called a representation of the quasi-quantum double $\mathbf D^{\tilde\omega}(GG)$.

Because we will show that the braiding algebra is completely determined by $D_{(h,p)}^{(n)}(B)$ operators with $h,p\in GG$, one knows the braiding properties of a quasiparticle at end-$A_n$ if we know which representation of $\mathbf D^{\tilde\omega}(GG)$ that this particle is in. In fact, the quasiparticle species (or more precisely, its superselection sector) is labeled by an irreducible representation of $\mathbf{D}^{\tilde\omega}(GG)$. And different irreducible representations of $\mathbf D^{\tilde\omega}(GG)$ correspond to different quasiparticle species.

The physical counter-clockwise $360^{\circ}$ braiding for the two particle states is $(\hat R_{CC}^{n,n+1})^2$. Its action in the basis $|\psi^{k_1...k_N}_{u_B}\rangle$ is actually very simple, because the $\omega$-terms in Eq.(\ref{eq:Rcc_hat}) cancel out for $(\hat R_{CC}^{n,n+1})^2$:
\begin{align}
 &(\hat R_{CC}^{n,n+1})^2|\psi_{u_B}^{k_1...k_nk_{n+1}...k_N}\rangle\notag\\
=&D^{(n)}_{(h_n,h_{n+1})}(B)\otimes D^{(n+1)}_{(h_{n+1},h_n)}(B)|\psi_{u_B}^{k_1...k_nk_{n+1}...k_N}\rangle
\end{align}

It is also tempting to formally define the general operator
\begin{align}
 R_{CC}^{n,n+1}=&R^{qr}D_r^{(n)}(B)\otimes D_q^{(n+1)}(B)\cdot\sigma\notag\\
=&\sigma\cdot R^{rq}D_r^{(n)}(B)\otimes D_q^{(n+1)}(B)\notag\\
=&\sigma\cdot \sum_{h_n,h_{n+1}\in GG}D^{(n)}_{(h_n,\openone)}(B)\otimes D^{(n+1)}_{(h_{n+1},h_n)}(B),\label{eq:Rcc_a}
\end{align}
Note that the formal operator $R_{CC}^{n,n+1}$ does not has a hat, which distinguish it from the physical braiding $\hat R_{CC}^{n,n+1}$ in Eq.(\ref{eq:Rcc_hat}). The $\omega$-factor in Eq.(\ref{eq:Rcc_hat}) tells us that, for multiple ribbon states, the physical braiding $\hat R_{CC}^{n,n+1}$ is \emph{not} implemented by $R_{CC}^{n,n+1}$ in the basis $\{|\psi_{u_B}^{k_1,...,k_N}\rangle\}$. But we will show that if we change into certain different basis, $\hat R_{CC}^{n,n+1}$ is still implemented by $R_{CC}^{n,n+1}$!

Let's firstly consider three-ribbon states: $|\psi_{u_B}^{k_1k_2k_3}\rangle$. The physical $\hat R_{CC}^{2,3}$ between the end-$A_2$ and the end-$A_3$ are still implemented by $R_{CC}^{2,3}$. But in this basis, $R_{CC}^{1,2}$ and the physical braiding $\hat R_{CC}^{1,2}$ differ by a phase factor $\frac{\omega(h_1,h_2,h_3)}{\omega(h_2,h_1,h_3)}$, due to the $\omega$-term in Eq.(\ref{eq:FF_share_B}). However, this braiding can still be implemented by $R_{CC}^{1,2}$ if we choose a different basis.

Let's define the state:
\begin{align}
 |\psi_{u_B}^{((k_1,k_2),k_3)}\rangle\equiv \omega^{-1}(h_1,h_2,h_3) |\psi_{u_B}^{k_1k_2k_3}\rangle
\end{align}
For reasons that will become clear in a moment, we also define:
\begin{align}
 |\psi_{u_B}^{(k_1,(k_2,k_3))}\rangle\equiv |\psi_{u_B}^{k_1k_2k_3}\rangle
\end{align}
All states $\{|\psi_{u_B}^{((k_1,k_2),k_3)}\rangle\}$ form another basis of $\mathcal L_{u_B}(A_1,A_2,...A_N,B)$, with differ from the original basis $\{|\psi_{u_B}^{(k_1,(k_2,k_3))}\rangle\}$ only by phase factors. 

It is then clear now that the physical braiding $\hat R_{CC}^{2,3}$ is implemented by $R_{CC}^{2,3}$ in basis $\{|\psi_{u_B}^{(k_1,(k_2,k_3))}\rangle\}$ (but not in the basis $\{|\psi_{u_B}^{((k_1,k_2),k_3)}\rangle\}$), and the physical braiding $\hat R_{CC}^{1,2}$ is implemented by $R_{CC}^{1,2}$ in the basis $\{|\psi_{u_B}^{((k_1,k_2),k_3)}\rangle\}$ (but not in the basis $\{|\psi_{u_B}^{(k_1,(k_2,k_3))}\rangle\}$). Such a basis change is necessary to maintain the same algebraic form of the formal braiding operator $R^{n,n+1}_{CC}$ in Eq.(\ref{eq:Rcc_a}). 

The two different parentheses configurations: $(k_1,(k_2,k_3))$ and $((k_1,k_2),k_3)$ can be viewed as two different ways to ``multiply'' quasiparticles. For reasons that will become clear later, it is better to call these operations as ``co-multiplications''. And different orders of ``co-multiplications'' do not give the same results, which differ by a basis change: mathematically, the ``co-multiplications'' are not associative, but are quasi-associative. It turns out that the ``co-multiplications'' of quasiparticles here have a clear physical meaning: the fusions, which we will discuss shortly.

One can generalize the above observation for 3-ribbon states to multiple-ribbon states, which turns out to be also very useful to represent the fusion algebra. We define:
\begin{align}
  |\psi_{u_B}^{(k_1,(k_2,...(k_{N-2},(k_{N-1},k_N)))...)}\rangle\equiv |\psi_{u_B}^{k_1k_2...k_{N-2}k_{N-1}k_N}\rangle
\end{align}
We now consider an arbitrary parentheses configuration between $k_1,k_2,...,k_N$. For convenience, we use a tree diagram to represent it. For example, $(k_1,(k_2,k_3))$ and $((k_1,k_2),k_3)$ can be represented as figure-(a) and figure-(b) below, while $(k_1,(k_2,...(k_{N-2},(k_{N-1},k_N)))...)$ is represented as figure-(c).
\begin{figure}[H]
 \begin{center}
  \includegraphics[width=0.48\textwidth]{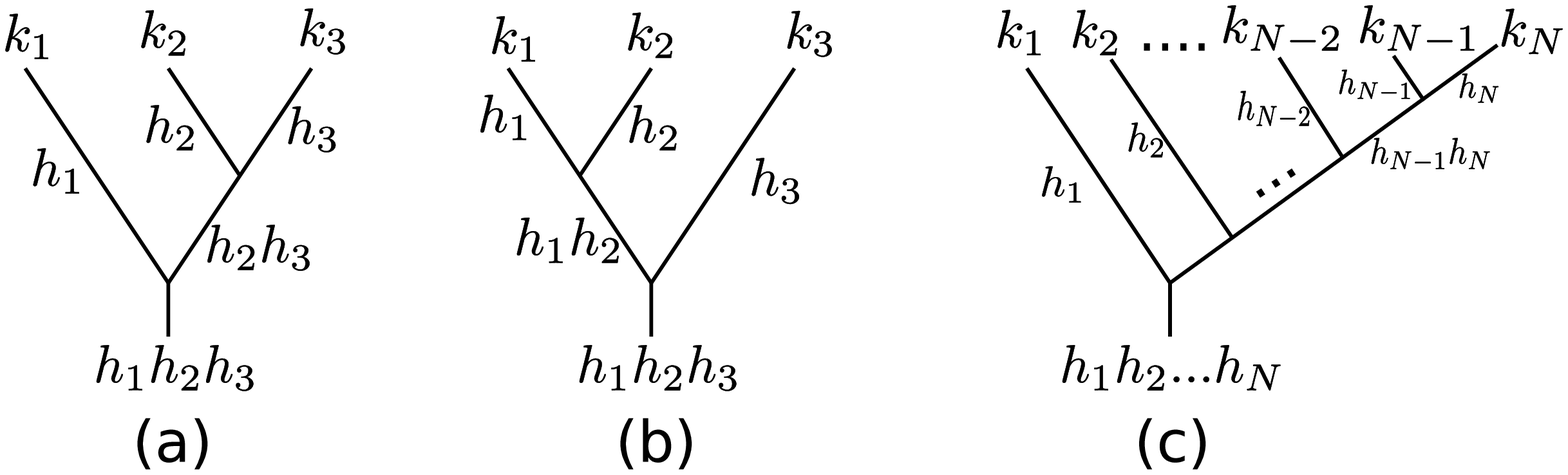}
 \end{center}
\end{figure}
Note that we define a tree diagram not only as a geometric object: First, it contains $k_1=(h_1,g_1),...k_N=(h_N,g_N)$ assigned for the top end points. Second, every edge (line segment) in a tree diagram is also assigned a group element $\in GG$, which is specified as follows. The top edges are assigned as $h_1,h_2,...h_N$, and every lower edge coming out of merging two upper edges is assigned by the product of the group elements in the upper two edges.

We can use a tree diagram to represent any parentheses configuration:
\begin{figure}[H]
 \begin{center}
  \includegraphics[width=0.4\textwidth]{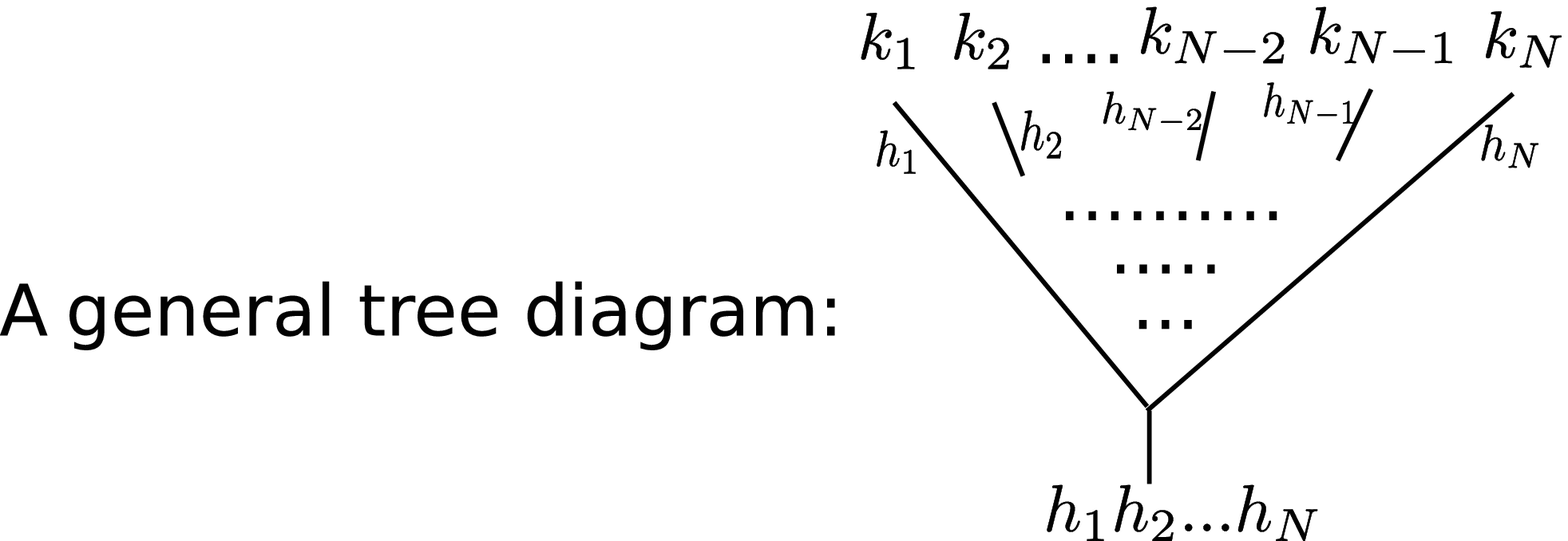}
 \end{center}
\end{figure}
We now define a basis for a fixed tree diagram $\mbox{Tree}_{\alpha}$, where $\alpha$ labels the tree configuration:
\begin{align}
 |\psi_{u_B}(\mbox{Tree}_{\alpha})\rangle\equiv w(\mbox{Tree}_{\alpha})|\psi_{u_B}^{k_1k_2...k_{N-2}k_{N-1}k_N}\rangle,
\end{align}
where $w(\mbox{Tree}_{\alpha})$ is a phase factor which we define below.

Any two tree diagrams with the same set of assigned $k_1=(h_1,g_1),...k_N=(h_N,g_N)$ on the top end points can be deformed into each other by a finite number of so-called F-moves. An F-move is a local deformation of a tree diagram, namely $\mbox{Tree}_{\alpha}$ and $\mbox{Tree}_{\beta}$ are only different locally as shown below:
\begin{figure}[H]
 \begin{center}
  \includegraphics[width=0.45\textwidth]{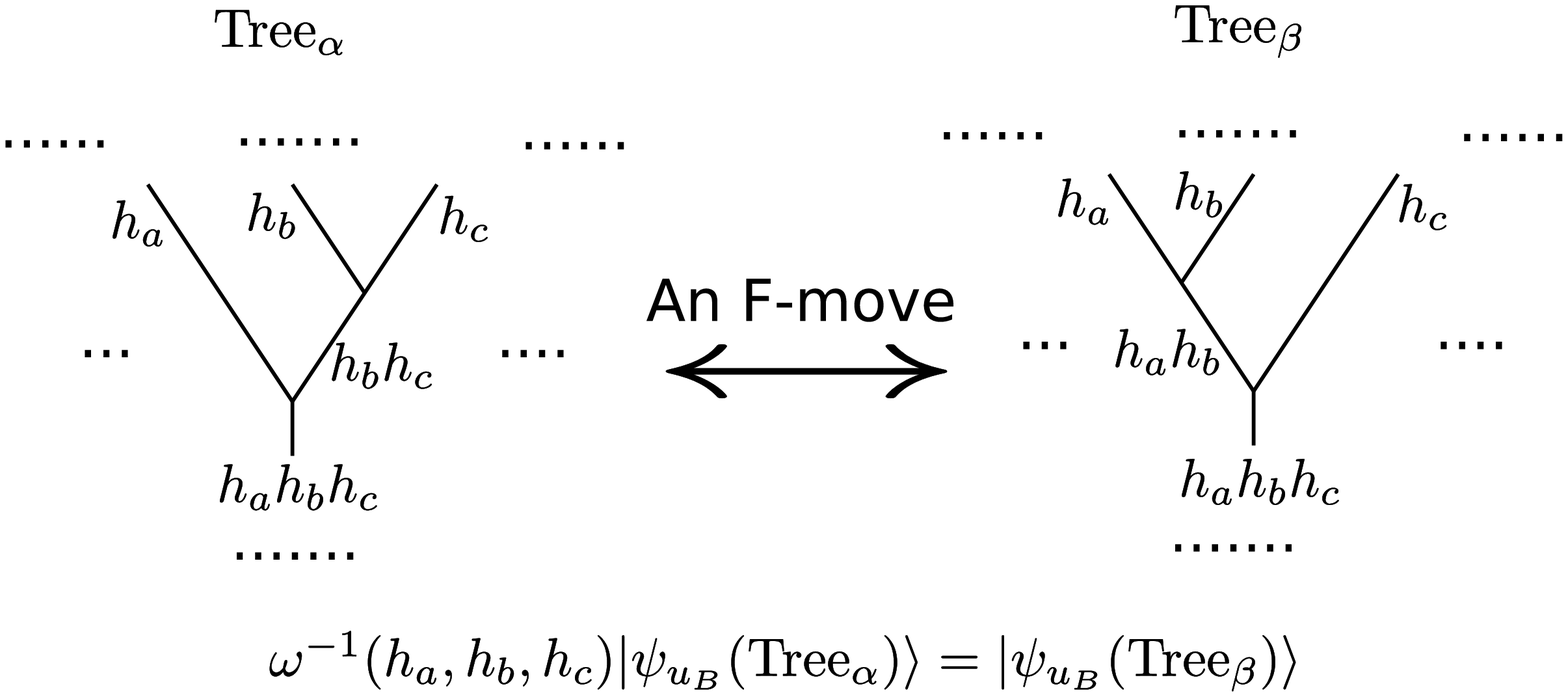}
 \end{center}
\caption{A general F-move.}\label{fig:F-move}
\end{figure}
When a local F-move occurs, we define the state $\omega^{-1}(h_a,h_b,h_c)|\psi_{u_B}(\mbox{Tree}_{\alpha})\rangle=|\psi_{u_B}(\mbox{Tree}_{\beta})\rangle$. In this fashion, starting from the tree diagram of the original basis $(k_1,(k_2,...(k_{N-2},(k_{N-1},k_N)))...)$, we can find the phase factor $w(\mbox{Tree}_{\alpha})$ accumulated during a sequence of F-moves for any tree diagram $\alpha$.

Apparently there are many different possible F-move paths that can connect a given tree diagram with that of $(k_1,(k_2,...(k_{N-2},(k_{N-1},k_N)))...)$. One may wonder whether the accumulated phase factor $w(\mbox{Tree}_{\alpha})$ is the same or not for different F-move paths. It turns out that $w(\mbox{Tree}_{\alpha})$ is independent of which path that one chooses. This is a consequence of the 3-cocycle condition. 3-cocycle condition dictates that the F-moves satisfy a crucial self-consistent condition: the pentagon equation, which in turn indicates that $w(\mbox{Tree}_{\alpha})$ is well-defined --- a consequence of the Mac Lane's coherence theorem\cite{lane1998categories}. We refer interested readers to Ref.\onlinecite{Kitaev:2006p6266} by Kitaev for detailed discussions.

With this definition of $|\psi_{u_B}(\mbox{Tree}_{\alpha})\rangle$, one can show that the physical braiding operation $\hat R_{CC}^{n,n+1}$ between end-$A_n$ and end-$A_{n+1}$ is implemented as $R_{CC}^{n,n+1}$ in any basis in which $k_n$ and $k_{n+1}$ are parenthesized together: $...(k_n,k_n+1)...$. 

Mathematically, the formal operators $R_{CC}^{n,n+1}$ does not satisfy the Yang-Baxter equation, a self-consistent equation for braiding algebra. This dictates that an appropriate changing of basis is required, which is discussed in detail above. With this changing of basis, the formal operators satisfy the so-called quasi-Yang-Baxter equation. These mathematical structures are exactly those in the quasi-quantum double $\mathbf D^{\tilde \omega}(GG)$.

\subsubsection{Braiding between the quasiparticles at end-$B$'s}

\begin{figure*}
  \centering
\includegraphics[width=0.8\textwidth]{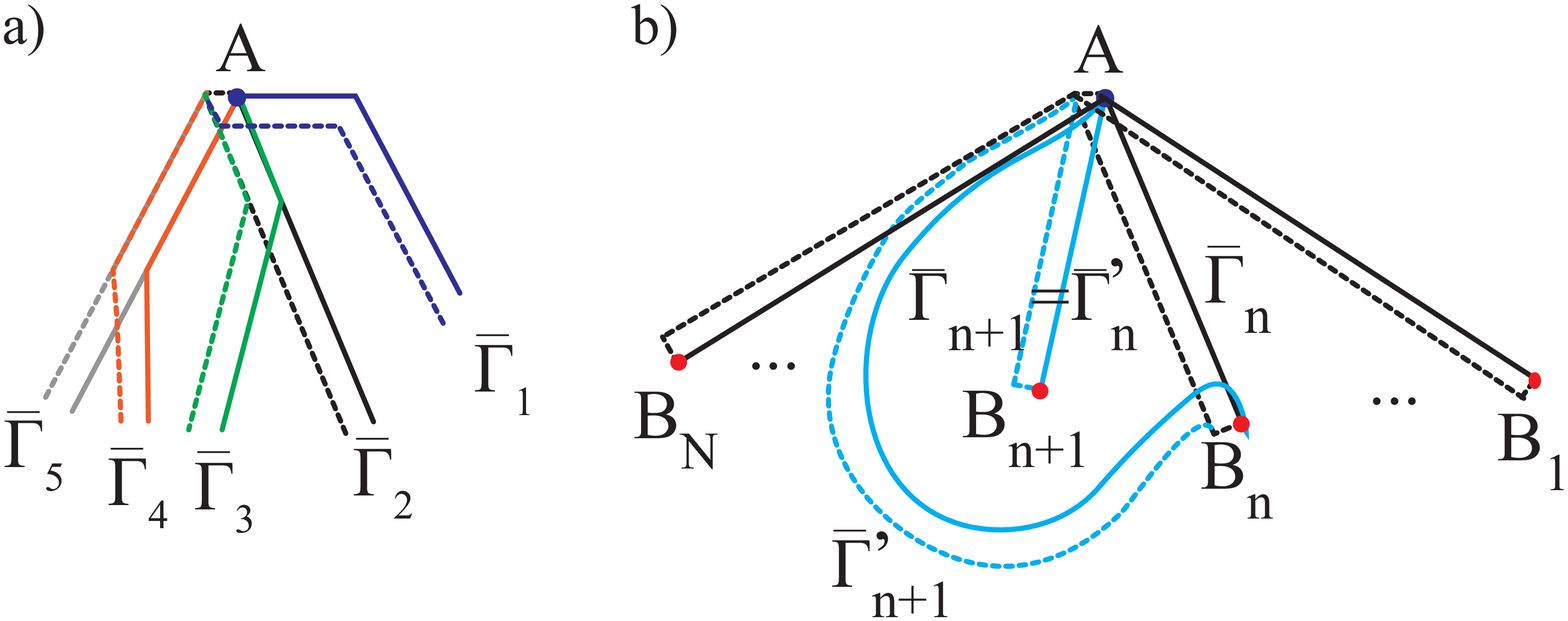}
\caption{Multiple ribbons sharing end-$A$, describing quasiparticles at their end-$B$'s. (a) Example of ordering multiple (five) ribbons, with strings $\pathgB_1\ldots\pathgB_5$, in a ``counter-clockwise sense". The ribbons are allowed to have overlaps only over a finite length starting from end-$A$. The figure shows a realization possible on a triangular lattice. (b) The counter-clockwise $180^\circ$ braiding of particles $B_n$ and $B_{n+1}$ in an $N$-particle state. The (blue) strings ${\pathgB}$' apply to the braided state.}
\label{fig:braid_B_ends}
\end{figure*}

Similarly we can construct an excited state using multiple ribbons $\pathgB_1,...\pathgB_n$ that all share the sa end-A: $A_1=A_2=...=A_N\equiv A$, $\mbox{edge-}A_1=\mbox{edge-}A_2=...=\mbox{edge-}A_N\equiv\mbox{edge-}A$, and that do not overlap/intersect with each other except for a finite length starting from end-A, while end-$B_n$ are all different from each other.  We also choose an ordering: $\pathgB_{i}$ is on counter-clockwise side of $\pathgB_{i+1}$. (See Fig.\ref{fig:braid_B_ends}a for a geometric illustration.) We can create an excited state:
\begin{align}
  |\bar\psi_{u_A}^{k_1...k_n}\rangle=\hat P_{u_A}(A)\tilde F^{k_1}(\pathgB_1)\tilde F^{k_2}(\pathgB_2)...\tilde F^{k_n}(\pathgB_n)|gs\rangle,
\end{align}
which only hosts quasiparticles at end-$B_i$ and end-$A$. Here $\hat P_{u_A}$ is a projector which enforces $u_{i_A}=u_A$ for a fixed element $u_A\in SG$. (Note that it is important to use $\tilde F^{(h,g)}(\pathgB)$ operators, \emph{not} the $F^{(h,g)}(\pathgB)$ operators to construct $|\bar\psi_{u_A}^{k_1...k_N}\rangle$.) We will use $|\bar\psi_{u_A}^{k_1...k_N}\rangle$ to study the braiding properties of the quasiparticles at the end-$B_n$'s. We denote the sub-Hilbert space spanned by $\{|\bar\psi_{u_A}^{k_1...k_N}\rangle\}$ as $\mathcal{L}_{u_A}(B_1,B_2,...,B_N,A)$.

The geometric illustration of the counter-clockwise $180^{\circ}$ braiding of the end-$B_n$ and the end-$B_{n+1}$ is shown in Fig.\ref{fig:braid_B_ends}b. The following operator identity is crucial to understand its underlying algebraic structure. When $\pathgB_1$ and $\pathgB_2$ share the same end-$A$ but have different end-$B$'s, and $\pathgB_1$ is on the counter-clockwise side of  $\pathgB_2$:
\begin{align}
  &\tilde F^{(h_1,g_1)}(\pathgB_2)\tilde F^{(h_2,g_2)}(\pathgB_1)\notag\\
=&c_{h_1}(h_2,g_1h_2^{-1})\tilde F^{(h_2,g_2)}(\pathgB_1)\tilde F^{(h_1,g_1h_2^{-1})}(\pathgB_2)\frac{\omega(h_2,h_1,h_A^{-1})}{\omega(h_1,h_2,h_A^{-1})}.\label{eq:FF_share_A}
\end{align}
Here $h_A$ should be interpreted as an operator measuring the gauge flux in $t_A$. $h_A=\openone$ in $|gs\rangle$ and $h_A=h_1^{-1}h_2^{-1}...h_{N}^{-1}$ in $|\bar\psi_{u_A}^{k_1...k_n}\rangle$.

Eq.(\ref{eq:FF_share_A}) already describes the physical counter-clockwise braiding ($180^{\circ}$) operations $\hat{\bar R}_{CC}^{n,n+1}$ between quasiparticles at end-$B_n$ and end-$B_{n+1}$ in $\mathcal L(B_1,...,B_N,A)$ completely:
\begin{widetext}
\begin{align}
 &\hat{\bar R}_{CC}^{n,n+1}|\bar\psi_{u_A}^{k_1,...(h_n,g_n),(h_{n+1},g_{n+1})...,k_N}\rangle\notag\\
=&\hat{\bar R}_{CC}^{n,n+1}\hat P_{u_A}(A)\tilde F^{k_1}(\pathgB_1)...\tilde F^{(h_{n},g_{n})}(\pathgB_n)\tilde F^{(h_{n+1},g_{n+1})}(\pathgB_{n+1}) ...\tilde F^{k_N}(\pathgB_N)|gs\rangle\notag\\
=&\hat P_{u_A}(A)\tilde F^{k_1}(\pathgB_1)...\tilde F^{(h_{n},g_{n})}(\pathgB_{n+1})\tilde F^{(h_{n+1},g_{n+1})}(\pathgB_{n}) ...\tilde F^{k_N}(\pathgB_N)|gs\rangle\notag\\
=&\frac{\omega(h_{n+1},h_n,h_{n+2}h_{n+3}...h_{N})}{\omega(h_{n},h_{n+1},h_{n+2}h_{n+3}...h_{N})}c_{h_{n}}(h_{n+1},g_{n}h_{n+1}^{-1})\hat P_{u_A}(A)\tilde F^{k_1}(\pathgB_1)...\tilde F^{(h_{n+1},g_{n+1})}(\pathgB_{n}) \tilde F^{(h_{n},g_{n}h_{n+1}^{-1})}(\pathgB_{n+1})...\tilde F^{k_N}(\pathgB_N)|gs\rangle\notag\\
=&\frac{\omega(h_{n+1},h_n,h_{n+2}h_{n+3}...h_{N})}{\omega(h_{n},h_{n+1},h_{n+2}h_{n+3}...h_{N})}c_{h_{n}}(h_{n+1},g_{n}h_{n+1}^{-1})|\bar\psi_{u_A}^{k_1,...(h_{n+1},g_{n+1}),(h_{n },g_{n}h_{n+1}^{-1})...,k_N}\rangle\label{eq:Rbarcc_hat}
\end{align}
\end{widetext}
Here the $h_A^{-1}$ in Eq.(\ref{eq:FF_share_A}) picks up the inverse of the accumulated flux in $t_A$: $h_{n+2}h_{n+3}...h_{N}$.

We can also define the formal braiding operator:
\begin{align}
 \bar R_{CC}^{n,n+1}=&\bar R^{rq} D_r^{(n)}(A)\otimes D_q^{(n+1)}(A)\cdot \sigma\notag\\
=&\sigma \cdot \bar R^{qr} D_r^{(n)}(A)\otimes D_q^{(n+1)}(A)\notag\\
=&\sigma\sum_{h_n,h_{n+1}\in GG}\big[c_{h_n^{-1}}(h_{n+1},h_{n+1}^{-1})c_{h_{n+1}}(h_n,h_n^{-1})\notag\\
&\cdot D_{(h_n,h_{n+1})}^{(n)}(A)\otimes D_{(h_{n+1},\openone)}^{(n+1)}(A)\big]
\end{align}
where $\sigma$ is the permutation operator, and we define the tensor $\bar R^{(h_1,g_1),(h_2,g_2)}$ as:
\begin{align}
\bar R^{(h_1,g_1),(h_2,g_2)}=\delta_{h_1,g_2}\cdot \delta_{g_1,\openone}\cdot c_{h_2^{-1}}(h_1,h_1^{-1})c_{h_1}(h_2,h_2^{-1})
\end{align}
In fact one can show that for an abelian group, $c_{h_1}(h_2,h_2^{-1})=c_{h_1}(h_2^{-1},h_2)$, $\forall h_1,h_2$.
$D_{(h_n,p_n)}^{(n)}(A)$ with $h_n,p_n\in GG$ is a formal operator that transforms the $\pathgB_n$ operator $\tilde F^{(h,g)}(\pathgB_{n})$ as if we are in a single ribbon state (see Eq.(\ref{eq:DF_o_a_A})). More precisely, for instance, because:
\begin{align}
 &D_{(h_1,p_1)}|\bar\psi_{u_A}^{(\tilde h_1,\tilde g_1)}\rangle\notag\\
=&\delta_{h_1^{-1},\tilde h_1}\cdot c_{p_1}^{-1}(\tilde h_1^{-1},\tilde h_1)c_{\tilde h_1}^{-1}(p_1,\tilde g_1)|\bar\psi_{u_A}^{(\tilde h_1,p_1\tilde g_1)}\rangle
\end{align}
we have:
\begin{align}
  &D^{(1)}_{(h_1,p_1)}\otimes D^{(2)}_{(h_2,p_2)}|\bar\psi_{u_A}^{(\tilde h_1,\tilde g_1),(\tilde h_2,\tilde g_2)}\rangle=\delta_{h_1^{-1},\tilde h_1}\cdot \delta_{h_2^{-1},\tilde h_2} \notag\\
&\cdot c_{p_1}^{-1}(\tilde h_1^{-1},\tilde h_1)c_{\tilde h_1}^{-1}(p_1,
\tilde g_1)c_{p_2}^{-1}(\tilde h_2^{-1},\tilde h_2)c_{\tilde h_2}^{-1}(p_2,\tilde g_2)\notag\\
&\cdot |\bar\psi_{u_A}^{(\tilde h_1,p_1\tilde g_1),(\tilde h_2,p_2\tilde g_2)}\rangle
\end{align}
$D_{(h_n,p_n)}^{(n)}(A)$ also satify the multiplication algebra Eq.(\ref{eq:multiply}) in the quasi-quantum double $D^{\tilde\omega}(GG)$.

One can show that for two-ribbon states, the $\omega$-factor in Eq.(\ref{eq:Rbarcc_hat}) is unimportant because it equals one due to the canonical form of a 3-cocyle. In this case one can show that the physical braiding $\hat{\bar R}_{CC}^{1,2}$ is indeed implemented by the formal operator ${\bar R}_{CC}^{1,2}$.

For multiple-ribbon states, the $\omega$-factor in Eq.(\ref{eq:Rbarcc_hat}) becomes important. Similar to the braiding of end-$A$'s, one way to get rid of the $\omega$-factor in Eq.(\ref{eq:Rbarcc_hat}) in the braiding algebra of end-$B$'s is to introduce appropriate basis changes, which we do not dicuss here. After the appropriate basis change, physical braiding $\hat{\bar R}_{CC}^{n,n+1}$ can still be implemented by the formal operators ${\bar R}_{CC}^{n,n+1}$

Finally, the physical counter-clockwise $360^{\circ}$ braiding $(\hat{\bar R}_{CC}^{n,n+1})^2$ also has a simple algebraic form:
\begin{align}
 &(\hat{\bar R}_{CC}^{n,n+1})^2|\bar\psi_{u_A}^{k_1...k_nk_{n+1}...k_N}\rangle\notag\\
=&D^{(n)}_{(h_n^{-1},h_{n+1}^{-1})}(A)\otimes D^{(n+1)}_{(h_{n+1}^{-1},h_n^{-1})}(A)|\bar\psi_{u_A}^{k_1...k_nk_{n+1}...k_N}\rangle
\end{align}

\subsection{Fusion}
We study the fusion of the quasiparticles at the end-$A$'s only. Let consider multiple-ribbon states with ribbons all sharing the same end-$B$, but with different end-$A$'s. 

Firstly let's consider two-ribbon states $|\psi_{u_B}^{k_1k_2}\rangle$. According to the twisted extended ribbon algebra Eq.(\ref{eq:DF_o_a_B}), the $D_{(h,p)}(B)$ operators with both $h,p\in GG$ transform two-ribbon states as:
\begin{align}
 D_{(h,p)}&(B)|\psi_{u_B}^{(\tilde h_1,\tilde g_1),(\tilde h_2,\tilde g_2)}\rangle=\delta_{h,\tilde h_1\tilde h_2}\cdot c_p(h_1,h_2)\notag\\
&\cdot c_{\tilde h_1}(\tilde g_1p^{-1},p)c_{\tilde h_2}(g_2p^{-1},p)|\psi_{u_B}^{(\tilde h_1,\tilde g_1p^{-1}),(\tilde h_2,\tilde g_2p^{-1})}\rangle\notag\\
=&\sum_{h_1\cdot h_2=h,h_1,h_2\in GG}\big[c_g(h_1,h_2)\notag\\
&\cdot D^{(1)}_{(h_1,p)}(B)\otimes D^{(2)}_{(h_2,p)}(B)\big]|\psi_{u_B}^{(\tilde h_1,\tilde g_1),(\tilde h_2,\tilde g_2)}\rangle\notag\\
=&\Lambda_{(h,p)}^{ij}D^{(1)}_i(B)\otimes D^{(2)}_j(B)|\psi_{u_B}^{(\tilde h_1,\tilde g_1),(\tilde h_2,\tilde g_2)}\rangle,
\end{align}
where we have used the formal operators $D^{(n)}_i(B)$ introduced in Eq.(\ref{eq:Rcc_a_12}), and the tensor $\Lambda_{k}^{ij}=\delta_{g_i,g_j}\,\delta_{g_k,g_i}\,\delta_{h_k,h_ih_j}\,c_{g_k}(h_i,h_j)$ defined in Eq.(\ref{eq:38}).

This motivate us to generally define formal operators for multi-ribbon states:
\begin{align}
 D_{r}^{(n)(n+1)}(B)\equiv \Lambda_r^{ij}D_i^{(n)}(B)\otimes D_j^{(n+1)}(B)\label{eq:co-multiply}
\end{align}
Basically, when acting on a multi-ribbon state, $D_{r}^{(n)(n+1)}(B)$ only transforms the $\pathgA_n$ and $\pathgA_{n+1}$ ribbons as if we are in a two-ribbon states. Mathematically, the operation in Eq.(\ref{eq:co-multiply}) is called co-multiplication in the quasi-quantum double $\mathbf {D}^{\tilde\omega}(GG)$, and is often denoted in mathematical literature by:
\begin{align}\label{eq:comultiplicationD}
 \Delta(D_r)\equiv\Lambda_r^{ij}D_i\otimes D_j.
\end{align}

$D_{r}^{(n)(n+1)}(B)$ turns out to be very useful, we will show soon that braiding properties of the fused quasiparticle of $\pathgA_n$ and $\pathgA_{n+1}$ ribbons are completely determined by $D_{r}^{(n)(n+1)}(B)$.

Because $D_{r}^{(n)(n+1)}(B)$ has a physical interpretation of acting $D_{r}(B)$ on the two-ribbon states, $D_{r}^{(n)(n+1)}(B)$ clearly also satify the multiplication algebra Eq.(\ref{eq:multiply}) in the quasi-quantum double $\mathbf {D}^{\tilde\omega}(GG)$. Therefore, if we know the irreducible representations of $\mathbf {D}^{\tilde\omega}(GG)$ (i.e., superselection sectors) for the quasiparticles at end-$A_n$ and end-$A_{n+1}$, the co-multiplication Eq.(\ref{eq:co-multiply}) induces another representaion of $\mathbf{D}^{\tilde\omega}(GG)$, which is generally reducible. One can decompose this induced representation into its irreducible components. Every irreducible component corresponds to one fusion channel. \emph{This procedure defines the fusion rule.}
\begin{figure}
  \centering
\includegraphics[width=0.5\textwidth]{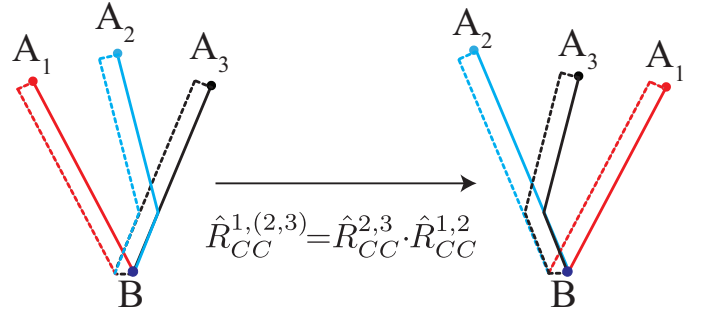}
\caption{Braiding quasiparticle $A_1$ with the fused quasiparticle $A_2A_3$ can be calculated using the shown formula for a two-step process. The resulting braiding operator is determined by the action of two topological operators (Eq.~(\ref{eq:braid_fused})), analogously to the simple case of two-particle braiding; however, in this case one topological operator relates to particle $A_1$, but the other is a ``comultiple'' (Eqs.~(\ref{eq:co-multiply}),(\ref{eq:comultiplicationD})) of topological operators acting on particles $A_2$ and $A_3$.}
\label{fig:fusion_braiding}
\end{figure}

Now let's consider three-ribbon states $|\psi_{u_B}^{k_1k_2k_3}\rangle$, which allows us to study the braiding of a fused quasiparticle with another quasiparticle. Let's imagine that we fuse the end-$A_2$ and the end-$A_3$ quasiparticles first, and braid the fused particle with the end-$A_1$ quasiparticle. We should physically braid both end-$A_2$ and end-$A_3$ with end-$A_1$, which we define as $\hat R_{CC}^{1,(2,3)}\equiv \hat R_{CC}^{2,3}\cdot\hat R_{CC}^{1,2}$ (see Fig.(\ref{fig:fusion_braiding})). To understand the algebraic structure of this procedure, the basis change that we introduced earlier becomes useful now. One can straighforwardly show that:
\begin{align}
 &\hat R_{CC}^{1,(2,3)}|\psi_{u_B}^{(k_1,(k_2,k_3))}\rangle=(\hat R_{CC}^{2,3}\cdot\hat R_{CC}^{1,2})|\psi_{u_B}^{(k_1,(k_2,k_3))}\rangle\notag\\
=&c_{h_1}(h_2,h_3)c_{h_2}(g_2h_1^{-1},h_1)c_{h_2}(g_3h_1^{-1},h_1)\cdot\notag\\
&\;\;\;\;\;\;\;|\psi^{(((h_2,g_2h_1^{-1}),(h_3,g_3h_1^{-1})),(h_1,g_1))}\rangle\notag\\
=&\sigma\cdot c_{h_1}(h_2,h_3)D_{(h_2,h_1)}^{(2)}(B)\otimes D_{(h_3,h_1)}^{(3)}(B)|\psi_{u_B}^{(k_1,(k_2,k_3))}\rangle\\\label{eq:braid_fused}
=&\sigma\cdot R^{rq} D_r^{(1)}(B)\otimes D_{q}^{(2)(3)}(B)|\psi_{u_B}^{(k_1,(k_2,k_3))}\rangle,
\end{align}
where the permutation operator $\sigma$ is defined as $\sigma |\psi_{u_B}^{(k_1,(k_2,k_3))}\rangle=|\psi_{u_B}^{((k_2,k_3),k_1)}\rangle$. Therefore the braiding algebra of the fused particle satisfy the same algebra as in Eq.(\ref{eq:Rcc_a}) after using $D_{q}^{(2)(3)}(B)$ operator.

Similarly, one can imagine to fuse the end-$A_1$ and the end-$A_2$ quasiparticles first, and braid the fused particle with the end-$A_3$ quasiparticle. This physical process is $\hat R_{CC}^{(1,2),3}\equiv \hat R_{CC}^{1,2}\cdot \hat R_{CC}^{2,3}$. One can also show that:
\begin{align}
 &\hat R_{CC}^{(1,2),3}|\psi_{u_B}^{((k_1,k_2),k_3)}\rangle\notag\\
=&\sigma\cdot R^{rq} D_r^{(1)(2)}(B)\otimes D_{q}^{(3)}(B)|\psi_{u_B}^{((k_1,k_2),k_3)}\rangle,
\end{align}
where the permutation operator $\sigma$ is defined as $\sigma |\psi_{u_B}^{((k_1,k_2),k_3))}\rangle=|\psi_{u_B}^{(k_3,(k_1,k_2))}\rangle$.

This discussion can be easily generalized to multiple ribbon states. One can show that the braiding algebra of the fused particle is always represented using the $D_{r}^{(n)(n+1)}(B)$ in a basis where $(k_n,k_{n+1})$ are parenthesized together.

Finally, the fusion algebra is formally represented by the co-multiplication in $\mathbf{D}^{\tilde \omega}(GG)$ in Eq.(\ref{eq:co-multiply}) only in a basis where $(k_n,k_{n+1})$ are parenthesized together. Co-multiplication is not associative but is quasi-associative; namely it becomes associative after the changing of basis: the F-move introduced earlier. Because F-moves satisfy the pentagon equation, one can show that the fusion algebra also satisfy the pentagon equation\cite{Kitaev:2006p6266,Dijkgraaf:1990p7462}, the self-consistent equation for fusion algebra.

\subsection{Summary}
In this section we find the operator realizations of the twisted extended ribbon algebra, and also study the braiding and fusion properties of the topological quasiparticles created by ribbon operators. We find that the topological order in our model is described by the quasi-quantum double $\mathbf{D}^{\tilde \omega}(GG)$, where the cocycle $\tilde \omega\in H^3(GG,U(1))$ is the one naturally induced by the cocycle $\omega\in H^3(SG\times GG,U(1))$ in our model. 

The core mathematical structures of the quasi-quantum double $\mathbf{D}^{\tilde \omega}(GG)$ include the multiplications in Eq.\ref{eq:multiply}, the co-multiplications in Eq.(\ref{eq:co-multiply}) and the changing of basis described by Fig.(\ref{fig:F-move}) (mathematically called associator). The superselection sector of a quasiparticle is determined by the irreducible representation of the multiplication algebra Eq.(\ref{eq:multiply}). The braiding algebra of quasiparticles is determined by the formal braiding operator in Eq.(\ref{eq:Rcc_a}), together with the changing of basis (associator), which satisfy the quasi-Yang Baxter equation. The fusion algebra of quasiparticles is determined by the co-multiplication algebra Eq.(\ref{eq:co-multiply}), together with the changing of basis(associator), which satisfy the pentagon equation. One can further show that the braiding algebra the fusion algebra are compatible: they satify the hexagon equation\cite{Kitaev:2006p6266}.

We have not studied the interplay between the global symmetry $SG$ and the topological order $\mathbf{D}^{\tilde \omega}(GG)$ here. For instance, we have not used $D_{(h,g)}(A)$, $D_{(h,g)}(B)$ operators when $g\notin GG$ except for stating their basic properties in the twisted extended ribbon algebra Eq.(\ref{eq:DD_o_a},\ref{eq:DF_o_a_A},\ref{eq:DF_o_a_B},\ref{eq:FF_o_a}). However, in Sec.\ref{sec:example}, we carefully study the interplay between the global symmetries and the topological orders in some examples. We believe that those studies can be generalized to any phase in our classification.

\section{Particle statistics directly from crossing strings}
\label{sec:part-stat-triv}
\begin{figure*}
  \centering
\includegraphics[width=1\textwidth]{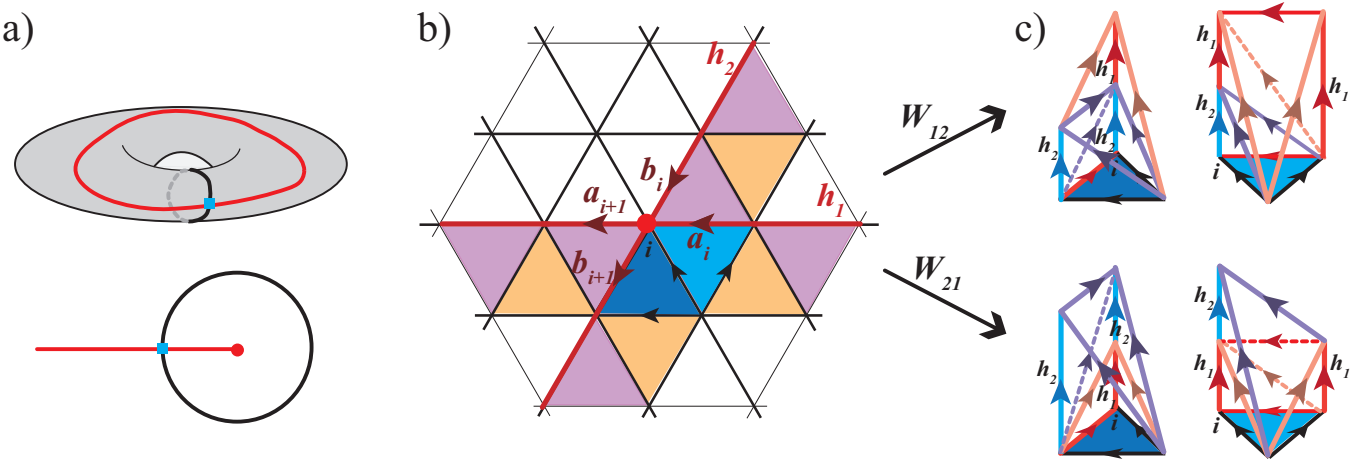}
\caption{Quasiparticle statistics from ribbon commutation. (a) Commutation of two strings (ribbon operators, black and red) at the intersection point (blue square) gives the braiding statistics. (Top) Strings representing tunneling of particle---anti-particle pairs across the system (the surface of torus) can give the self-statistics. (Bottom) Braiding setup independent of system shape. (b) Near the intersection of two ribbon operators $\hat{\wstring}_{h_1}^{\pathg_1}(\gGG_1)$, $\hat{\wstring}_{h_2}^{\pathg_2}(\gGG_2)$, the phase difference occurs due to 3-simplices on top of two blue shaded triangles, detailed in (c) with phases $W_{mn}=\bra{\text{f}}\hat{\wstring}_{h_m}^{\pathg_m}(\gGG_m)\hat{\wstring}_{h_n}^{\pathg_n}(\gGG_n)\ket{\text{i}}$ shown in top ($W_{12}$) and bottom ($W_{21}$) rows.
  Note that elements $b_i$ are defined to be oriented oppositely to our lattice edge orientations, but consistent with definition of ribbon operator phase, Fig.~\ref{fig:5}, hence the appearance of $b^{-1}_{i+1}$ in Eq.~(\ref{eq:28}).
}
\label{fig:stat}
\end{figure*}

Here we provide an alternative, direct approach to braiding statistics, and compare it to that of Section~\ref{sec:braiding-matrix}.

The braiding statistics of two quasiparticles is a topological property that therefore cannot depend on the details of the ribbon operator at its ends. When the system is a torus, the statistical phase of quasiparticles (meaningful for Abelian quasiparticles only) follows from the commutation of operators $\transl_a$ which describe the following process: creating a particle---anti-particle pair, tunneling the particle across the system in direction $a=x,y$, and finally annihilating the pair, as described by the formula~\cite{Oshikawa:2006p5869}
\begin{equation}
  \label{eq:29}
  \transl_x\transl_y=e^{-i 2\theta}\transl_y\transl_x,
\end{equation}
with $\theta$ the (exchange) statistical angle of the quasiparticles, see Fig.~\ref{fig:stat}(a). The ribbon operator $\hat{F}^{(h,g)}(\pathg)$ represents exactly the operation $\transl$ along its $\pathg$. Since the string ends are not involved, we can actually also use $\hat{\wstring}_{h}^{\pathg}(g)$ (see Fig.~\ref{fig:5}).

The angle in Eq.~(\ref{eq:29}) depends only on the commutation relation at the intersection point of the two strings, so even if the system is not a torus, we can consider a braiding operation performed locally\cite{Levin:2012p7190} and realized by an open and closed string that intersect at a single point, Fig.~\ref{fig:stat}(a).

Fig.~\ref{fig:stat} shows the ingredients for calculating the phases $W_{mn}=\bra{\text{f}}\hat{\wstring}_{h_m}^{\pathg_m}(g_m)\hat{\wstring}_{h_n}^{\pathg_n}(g_n)\ket{\text{i}}$, which reveal the statistical angle:
\begin{equation}
  \label{eq:30}
e^{-i 2\theta}=\frac{W_{12}}{W_{21}}.
\end{equation}
The phases $W_{mn}$ differ due to the 3-simplices positioned on top of two lattice triangles which are shared by the two ribbon operators, as well as due to string phases $\phasec_{\pathg_m}^{h_m}$ on two edges, Fig.~\ref{fig:stat}(b).

Using the definition of the 2-cocycle $\ccyc_h$, Eq.~(\ref{eq:14}), we get for the phase factor ratio due to the 3-simplices
\begin{equation}
  \label{eq:28}
\frac{\phasew_{12}}{\phasew_{21}}=\ccyc_{h_1}(h_2,a_i)\,\ccyc_{h_2}(h_1,b^{-1}_{i+1}),
\end{equation}
with $i$ the lattice site of intersection.

Since the definition of statistical angle only makes sense for \textit{Abelian} quasiparticles, we consider only trivial cocycles $\ccyc_h$. This means that we can rewrite the 2-cocycle using a 1-cochain, as in Eq.~(\ref{eq:15}).

In that case, the phase ratio $\phasec_{12}/\phasec_{21}$ is given by Eq.~(\ref{eq:31}), and can easily be simplified, giving
\begin{equation}
  \label{eq:32}
  \frac{\phasec_{12}}{\phasec_{21}}=\frac{\ccye_{h_1}(h_2\cdot a_i)}{\ccye_{h_1}(a_i)}\frac{\ccye_{h_2}(b_{i+1})}{\ccye_{h_2}(h_1^{-1}\cdot b_{i+1})},
\end{equation}
see labels in Fig.~\ref{fig:stat}(a).

Using the property $\ccye_h(\gGG_1^{-1})/\ccye_h(\gGG_2^{-1})=\ccye^{-1}_h(\gGG_1)/ \ccye^{-1}_h(\gGG_2)$, which holds for any 1-cochain describing a trivial canonical 2-cocycle, the total phase ratio $\frac{W_{12}}{W_{21}}= \frac{\phasew_{12}\phasec_{12}}{\phasew_{21}\phasec_{21}}$ finally becomes
\begin{equation}
  \label{eq:33}
e^{-i 2\theta}=\ccye_{h_1}(h_2)\,\ccye_{h_2}(h_1)\frac{\ccye_{h_1}(g_1) \ccye_{h_2}(g'_2)}{\ccye_{h_1}(g'_1) \ccye_{h_2}(g_2)}.
\end{equation}
The group elements appearing here are: $g_1=\prod_j a_j$, $g_2=\prod_k b_k$, i.e. the values for isolated strings, and $g'_1=h_2\prod_j a_j$, $g'_2=h_1\prod_{k}b_k$, i.e. the values after the other string is already applied.

After using the 1-cochain, Eq.~(\ref{eq:15}), on Eq.~(\ref{eq:R2pi}), the result obtained by calculating the braiding matrix by using the algebra of local operators at the string ends almost exactly matches the one here (up to replacing $g_1\rightarrow h_2\cdot g_1$ here). The difference involves a deeper subtle issue we will address now. 
The obtained angle $2\theta$ (either following from Eq.~(\ref{eq:R2pi}) or Eq.~(\ref{eq:33})) should represent $2\pi$ braiding of Abelian quasiparticles in systems with trivial cocycle $\ccyc_h$.
However, for the statistical angle to be physical, one expects that it is invariant under changes of the 3-cocycle by a 3-coboundary, as introduced in Eq.~(\ref{eq:gauge3coc}), because this change does not alter the physical content of the theory. Such transformations of $\ccy$ by a coboundary lead to the following transformation of the 1-cochain $\ccye$ by an arbitrary function $u(x,y)$:
\begin{equation}
  \label{eq:34}
  \ccye_h(\gGG)\rightarrow   \ccye_h(\gGG)\frac{u(h,\gGG)}{u(\gGG,h)}.
\end{equation}
One can see that the factor $\ccye_{h_1}(h_2)\,\ccye_{h_2}(h_1)$ in the result Eq.~(\ref{eq:33}) is invariant, but the rest of the expression is not! (The same is true for result following from Eq.~(\ref{eq:R2pi}).)

The resolution of this puzzle is instructive. Namely, the ribbon operator $F^{(h,g)}(\pathg)$ we used here and in Section \ref{sec:braiding-matrix}, when acting on the ground state, does not necessarily create a quasiparticle pair state with well defined flux $h$ and gauge charge $h_g$. As explicitly shown in the examples of Section~\ref{sec:example}, the gauge charge operator may act non-trivially within the quasiparticle Hilbert space. (This Hilbert space is created by the action of $F^{(h,g)}(\pathg)$ on the ground state, but has to be specified further by discriminating different values of $\SG$ elements $u_C$ at the ends $C=A,B$ of string $\pathg$, Section~\ref{sec:gener-struct-excit}.) This means that physical states, having well-defined gauge charge, can actually be non-trivial linear combinations of states spanned by $F^{(h,g)}(\pathg)$.

Of course, the braiding matrix Eq.~(\ref{eq:R2pi}) contains all information necessary to specify braiding properties of quasiparticles, one only has to pose the right question, which would involve explicit use of physical flux and charge states constructed using the formalism in Section~\ref{sec:example}.

In the present discussion of Abelian quasiparticles, it becomes obvious that the physical quasiparticle states with fixed flux and charge are simply obtained by absorbing the non-invariant factors in Eq.~(\ref{eq:33}). This is achieved easily by using $\bar{F}^{(h,g)}(\pathg)\equiv F^{(h,g)}(\pathg)\ccye_{h}(g)$, which leads to the statistical angle independent on the cocycle within a fixed cocycle equivalence class:
\begin{equation}
  \label{eq:1}
  e^{-i 2\theta}=\ccye_{h_1}(h_2) \ccye_{h_2}(h_1).
\end{equation}
(The same is obtained starting from expression given by Eq.~(\ref{eq:R2pi}).)

Let us now use Eq.~(\ref{eq:1}) on two Abelian topological theories with $Z_2$ order, namely the toric code (TC) and double semion (DS) models (see Appendix~\ref{sec:exampleH_Z1Z2}). These two orders are physically distinguished exactly by the statistics of their quasiparticles.

As demonstrated explicitly in Appendix~\ref{sec:exampleH_Z1Z2}, the TC is recovered by choosing the trivial cocycle $\ccy(g_1,g_2,g_3)=1,\forall g_1,g_2,g_3\in\G$. From these constraints we need to obtain the values of the 1-cochain $\ccye_h(g)$. The constraint $\ccy(g_1,g_2,g_3)=1$ leads to:
\begin{equation}
  \label{eq:78}
  \ccyc_h(g_1,g_2)=1,\quad\text{(TC)},
\end{equation}
which implies that
\begin{equation}
  \label{eq:79}
  \ccye_h(g_1)\,\ccye_h(g_2)=\ccye_h(g_1\cdot g_2),\quad\text{(TC)},
\end{equation}
i.e. the 1-cochain $\ccye$ actually becomes a 1-cocycle. There are two representations of the $Z_2$ group, i.e. in total four solutions for $\ccye$ given by $\ccye_0(g)=1$ or $\ccye_0(g)=(-1)^g$, and $\ccye_1(g)=1$ or $\ccye_1(g)=(-1)^g$.

In the present formalism, a particular ribbon is determined by its flux $\vis$ and one of the particular solutions for the 1-cochain $\ccye$.

The TC has four quasiparticles, $\{\openone,e,m,em\}$, the trivial, charge, flux, and charge-flux bound state, respectively. Assigning $\vis=0$ to $\openone,e$, and $\vis=1$ to $m,em$, we make a choice (not unique) of assigning each of the four solutions to one particular quasiparticle: 1) To $\openone$: $\ccye_h(g)=1$; 2) To $e$: $\ccye_h(g)=(-1)^g$; 3) To $m$: $\ccye_1(g)=(-1)^g$; and 4) To $em$: $\ccye_0(g)=(-1)^g$.

This solution recovers the well-known TC result that $\openone,e,m$ are bosons, $em$ a fermion, and non-trivial mutual statistics is given by $2\theta=\pi$ amongst the $\{e,m,em\}$.

According to the explicit demonstration in Appendix~\ref{sec:exampleH_Z1Z2}, the DS theory is obtained by choosing the second representative 3-cocycle for $\G=Z_2$ given by $\ccy(g_1,g_2,g_3)=-1$ for $g_1=g_2=g_3=1$, and $1$ otherwise. Repeating the analysis from the TC case above, the difference lies in the constraint:
\begin{equation}
  \label{eq:80}
  \ccyc_1(1,1)=-1,\quad\text{(DS)},
\end{equation}
which leads to solutions given by $\ccye_0(g)=1$ or $\ccye_0(g)=(-1)^g$, and $\ccye_1(1)=i$ or $\ccye_1(1)=-i$, while $\ccye_1(0)=1$ always.

The non-trivial excitations in the Abelian topological phase described by DS theory are two semions $s_1$, $s_2$, and their bound state $s_{12}$. Assigning $\vis=0$ to $\openone,s_{12}$, and $\vis=1$ to $s_1,s_2$, we make a choice (not unique) of assigning each of the four solutions to one particular quasiparticle: 1) To $\openone$: $\ccye_1(1)=i$; 2) To $s_{12}$: $\ccye_0(g)=(-1)^g$, $\ccye_1(1)=-i$; 3) To $s_1$: $\ccye_0(g)=(-1)^g$, $\ccye_1(1)=i$; and 4) To $s_2$: $\ccye_1(1)=-i$.

This solution indeed recovers the well-known quasiparticle properties: $\openone,s_{12}$ are bosons; $s_1$, $s_2$ are semions ($2\theta=\pi$); and the non-trivial mutual statistics is $2\theta=\pi$ between $s_{12}$ and either $s_1$ or $s_2$.

\section{Explicit form of models for $Z_2$ topologically ordered phases}
\label{sec:exampleH_Z1Z2}
\begin{figure}
  \centering
\includegraphics[width=0.3\textwidth]{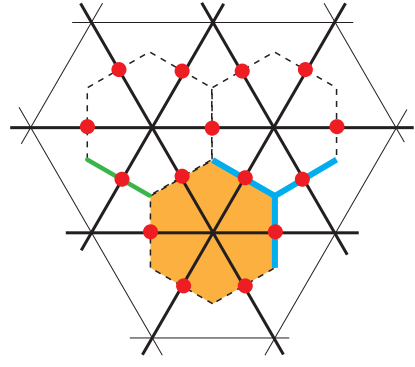}
  \caption{Recovering the two $Z_2$ topological phases from the exactly solvable models: toric code (TC) and double semion theory (DS). The dual lattice is honeycomb (dashed). Red dots mark the positions of group elements $g_{ij}$ assigned to edges $ij$, becoming ``spin-1/2'' states in TC and DS. A honeycomb lattice plaquette $ph$ is shaded. The plaquette operator $\hat{B}_p$ (Fig.~\ref{fig:1}) acts on the six ``spins'' of $ph$. The DS model differs from TC by having an additional phase in its plaquette operator. This phase depends on spins on the six outer legs of $ph$, one of which is marked by green line. These six ``outer'' spins belong to the six tetrahedrons in our $\hat{B}_p$, Fig.~\ref{fig:1}. The $Q_t$ operator in both models just acts on the three spins nearest to a site (spins on blue lines).}
\label{fig:honeycomb}
\end{figure}

Here we consider the well-understood case of $Z_2$ topological order, to demonstrate that our general model explicitly yields models for the two inequivalent phases having such order.
Two well-known models for the two distinct $Z_2$ topological phases are Kitaev's ``toric code'' \cite{Kitaev:2003p6185} (TC) and the ``double semion'' theory\cite{Levin:2005p3468} (DS). We will make a direct comparison to the TC and DS model variants presented in Ref.\onlinecite{Levin:2012p7190}.

For that purpose, we set the symmetry group to be trivial, $\SG=Z_1$, and the gauge group to $\GG=Z_2$. There are only two group elements, which we can label by $\G=\{\openone\equiv+1,a\equiv -1\}$, with $a=a^{-1}$. The group element $g_{ij}=\pm1$ assigned to the $ij$ edge (of the triangular lattice) we will call ``spin'' ($\pm1$=''up/down'') positioned at the mid-point of the edge $ij$. We will now switch to the dual lattice, which is honeycomb, as shown in Fig.~\ref{fig:honeycomb}. The spins are still on the edges of the honeycomb lattice, and the honeycomb plaquette $ph$ contains the same six spins as our original plaquette $p$.

Let us start with the operator $Q_t$, which will be the same in both topologically ordered theories. This operator forces the product of three spins adjacent to a honeycomb lattice site to $+1$, see Fig.~\ref{fig:honeycomb}. It therefore acts as the projector
\begin{equation}
  \label{eq:77}
  Q_v=\frac{1}{2}(1+\prod_{i\in v}\sigma^{(i)}_z)
\end{equation}
on three spins $i$ neighboring the vertex $v$. Comparing directly to the variant of TC and DS models presented in Ref.\onlinecite{Levin:2012p7190}, the first term in $H$, Eq.~(\ref{eq:5}), becomes up to an overall constant just the standard vertex contribution $\sum_v\prod_{i\in v}\sigma^{(i)}_z$ to $H^{TC}$ or $H^{DS}$ in Ref.\onlinecite{Levin:2012p7190}. Further, the product $\prod_{t\in p}Q_t$ appearing in the plaquette terms of $H$ exactly becomes the $\prod_{v\in ph} Q_v$ term in $H^{TC}$ or $H^{DS}$ which projects the flux in the honeycomb plaquette $ph$ to zero (this is the $P_p$ factor in Ref.\onlinecite{Levin:2012p7190}).

At this point, we can discuss the physical interpretation of $Q_v$ and the outcome of this subsection more precisely. In our model, $Q_v=1,\forall v$ gives the zero-flux rule on the lattice. One should recall that if we had been considering the \textit{dual} of the globally symmetric $\SG=Z_2$, $\GG=Z_1$, the zero-flux rule would have been automatically satisfied (see Eq.~(\ref{eq:73})). This fact is equivalent to saying that in the undualized theory the domain walls separating regions of up and down spins have to form closed loops, i.e. ``closed strings'' in the language of string-net models\cite{Levin:2005p3468}. No matter if we start from the dualized $\SG=Z_2$ model or the $\GG=Z_2$ model, the restriction to states satisfying $Q_v=1$ (which is automatic or chosen, respectively) leads to $Z_2$ gauge theory models, and this will be clearly shown below. In fact, Ref. \onlinecite{Levin:2012p7190} clarifies that there are still two physically distinguishable $Z_2$ gauge theories, which represent two different topological orders; both of these were obtained there by dualizing $Z_2$ spin models, automatically enforcing the closed string rule. We will explicitly show that our $\G=\GG=Z_2$ model indeed reduces to the two inequivalent $Z_2$ gauge theories when $Q_v=1$.

One can now ask how do the Kitaev's ``toric code'' \cite{Kitaev:2003p6185} (TC) and the ``double semion'' theory\cite{Levin:2005p3468} (DS) fit in this? Simply, the two different $Z_2$ topological orders are also described by these two different models. (TC and DS can be seen as Ising matter coupled to the two different $Z_2$ gauge fields. \cite{Levin:2012p7190}) The TC and DS models are expected to arise when the restriction to $Q_v=1$ is lifted, i.e. they are ``open string'' models. We will explicitly show below that our $\G=\GG=Z_2$ model indeed gives rise to TC in the absence of that restriction.

Let us move on to the plaquette operator and the Hamiltonian of our model.
Under $\hat{B}^a_{p\equiv i}$, the six group elements $g$ on edges (of triangular lattice) which share the lattice site $i$ are multiplied by $a$. Therefore the six
spins on the honeycomb $ph$ plaquette are ``flipped'', $g=\pm1\rightarrow\mp1$, i.e. acted on by
Pauli matrix $\sigma_x$.
Obviously, $\hat{B}^{\openone}_{p\equiv i}$ does not change the state (no flipped spins).

Next we consider the phase factor of $\hat{B}$. There are exactly two inequivalent classes of 3-cocycles $\ccy$ for the group $\G=Z_2$, as dictated by $H^3(Z_2,U(1))=Z_2$. They are represented by the two distinct choices \cite{Propitius:1995p6856}:
\begin{equation}
  \label{eq:81}
\ccy(-1,-1,-1)=\pm1.  
\end{equation}
In both cases, $\ccy(g_1,g_2,g_3)=1$ for any $g_1,g_2,g_3\in\G$ when they are not all equal to $-1$.

To obtain a model for TC phase, we choose a trivial cocycle, i.e. $\ccy(g_1,g_2,g_3)=1$, $\forall g_1,g_2,g_3\in\G$. The amplitude from Eq.~(\ref{eq:3}) then becomes trivial, $B^{TC}_p=1$. The action of the plaquette operator is
\begin{equation}
  \label{eq:72}
\hat{B}^{TC}_p=\frac{1}{2}(1 + \prod_{l\in ph}\sigma^{(l)}_x),
\end{equation}
with six spins $l$ in the honeycomb plaquette $ph$. This is equivalent to the standard plaquette operator of TC model in Ref.~\onlinecite{Levin:2012p7190}, although in the present case there is a constant term added to the $\sigma_x$ term. This constant shift can be removed by smoothly deforming the $\hat{B}^{TC}_p$ operator without ever closing the gap in the Hamiltonian. (The constant shift is there to make our $\hat{B}^{TC}_p$ a projector, which is its general property.) Our model with $\ccy=1$ is consequently equivalent to the TC model. Under the $Q_v=1$ constraint, obviously we obtain the standard $Z_2$ gauge theory on the honeycomb lattice.

We now turn to the DS model\cite{Levin:2005p3468}.
Compared to the TC model, the only difference is that the honeycomb plaquette operator now has an additional phase factor determined by the value of six spins on ``outer legs'' of the plaquette \cite{Levin:2012p7190}; one of six such legs of the plaquette $ph$ is marked by green line in Fig.~\ref{fig:honeycomb}. This phase in the DS model is assigned by the rule: if exactly two, or exactly six, of the ``outer leg'' spins are in state $-1$, there is a phase factor $-1$ in the plaquette operator. If we again restrict to a ``closed string'' theory, i.e. $Q_v=1$, the lines of $-1$ spins on the honeycomb lattice physically represent the closed domain walls of the dual $Z_2$ spin theory, and therefore there is always an even number of $-1$ legs entering a plaquette $ph$. As mentioned, this closed string restriction of DS model is the second type of $Z_2$ lattice gauge theory.

Let us show that under $Q_v=1$ our model indeed reduces to the $Q_v=1$ restricted DS model from Ref.\onlinecite{Levin:2012p7190}. The six ``outer leg'' spins of the honeycomb lattice are actually lying on the six outer edges of the plaquette $p$ in our model on the triangular lattice (compare Figs.~\ref{fig:honeycomb} and \ref{fig:1}). Although our $\hat{B}_p$ does not change them, they are on the edges of the six tetrahedrons which determine the phase of the operator.

Given the restriction $Q_v=1$, i.e. the zero-flux rule on the triangular lattice, and using $\ccy(-1,-1,-1)=-1$, it is straightforward to show that the phase coming from six tetrahedrons (see $B^{s=1}_p$ in Eq.~(\ref{eq:2})) is equal to $-1$ only in two cases: 1) When the only outer leg spins in state $-1$ are on two legs separated by $120^{\circ}$; 2) When beside the two legs as in case (1), there are two more legs with spin $-1$ separated by $60^{\circ}$.

These rules seem strange, however, this is remedied by a simple change of basis. As explained, for an arbitrary basis state there is an even number of $-1$ outer legs, i.e. segments of dual domain walls, entering the honeycomb plaquette $ph$. These segments of domain walls have to be connected (they do not have ends!) in some way along the six edges of the hexagonal plaquette $ph$. Given the configuration of $-1$ legs entering the $ph$, no matter how we choose to connect them, the number of $ph$ edges we will use for that will have fixed \textit{parity}. It is therefore well-defined to use this number, $\#\text{int.legs}(ph)$, for any basis state. After redefining the basis by $\ket{\text{state of plaquette $p$}}\rightarrow(-1)^{\#\text{int.legs(ph)}}\ket{\text{state of plaquette $p$}}$, our ``strange'' phase factors become exactly the DS model's phase factors described above. Therefore, within the zero-flux manifold of states, our model with $\G=Z_2$ and $\ccy(-1,-1,-1)=-1$ explicitly reduces to the ``closed string'' version of the DS model, which was firstly understood in that form in Ref.\onlinecite{Levin:2012p7190}.

\section{All two-particle states are given by action of ribbon operator, and local operators act projectively on these states}
\label{sec:all-two-particle}

Let us first show that the local operators form a projective representation of the group $\G$ in the Hilbert space $\mathcal{L}(A,B)$ of two excitations at the ends $A,B$ of the ribbon $\pathg$. Let us for concreteness focus on the projected subspace $\mathcal{L}_{u_A}(A,B)$ defined in Eq.~\eqref{eq:58}, and the exact same results follow for $\mathcal{L}_{u_B}(A,B)$.
For brevity, let us also use the double index notation, i.e. $i\equiv(h_i,g_i),j\equiv(h_j,g_j),\ldots$, with $h_i,h_j,\ldots\in\GG$ and $g_i,g_j,\ldots\in\G$. Directly combining the definition of excited states Eq.~\eqref{eq:58} and the operator algebra from Eq.~\eqref{eqFDexpl}, we get:
\begin{align}
  \label{eq:DAexpl}
  D_j(A)\ket{\state^k_{u_A}}&=\delta_{h_j,h_k^{-1}}\,\ccyc^{-1}_{g_j}(h_k,h_k^{-1}) \,\ccyc^{-1}_{h_k}(g_j,g_k)\ket{\state^{(h_k,g_kg_j)}_{\tilde{g}_j\cdot u_A}}\\
  \label{eq:DBexpl}
D_j(B)\ket{\state^k_{u_A}}&=\delta_{h_j,h_k}\,\ccyc_{h_k}(g_j^{-1}g_k,g_j)\ket{\state^{(h_k,g_j^{-1}g_k)}_{u_A}},
\end{align}
where we used the usual factorization $g_j=h'_j\cdot\tilde{g}_j$ with $\tilde{g}_j\in\SG$, $h'_j\in\GG$. Applying the operators twice and using the 2-cocycle identities Eq.~(\ref{2cy6}),~(\ref{2ccy}), we get:
\begin{subequations}  \label{eq:44}
  \begin{align}
    \label{eq:44a}
    D_i(A)D_j(A)\ket{\state^k_{u_A}}&=\ccyc_{h^{-1}_k}(g_i,g_j)\;D_{ij}(A) \ket{\state^k_{u_A}}\\
    \label{eq:44b}
  D_i(B)D_j(B)\ket{\state^k_{u_A}}&=\ccyc_{h_k}(g_i,g_j)\;D_{ij}(B) \ket{\state^k_{u_A}},
\end{align}
\end{subequations}
where we used the obvious shorthand notation $ij=(h_ih_j,g_ig_j)$. The 2-cocycle $c_h$, with $h\in\GG$, therefore determines the projective representation of $\G$.

The projective representation of local operators is actually unitary, as we will next show. In the basis $\ket{\state^k_{u_A}}$ labeled by $k$ ($u_A$ is fixed), the matrix elements of operator $D_{(h,g)}(A)$ from Eq.~(\ref{eq:DAexpl}) are
\begin{equation}
  \label{rep}
\rep(h,g)_{k',k}=\ccyc^{-1}_g(h^{-1},h) \ccyc^{-1}_{h^{-1}}(g,g_{k})\delta_{g_{k'},g g_k}\delta_{h,h^{-1}_k}\delta_{h_k,h_{k'}}.  
\end{equation}

Using the multiplication law, Eq.~(\ref{eq:44a}), it further follows that $\rep(h,g) \rep(h^{-1},g^{-1})=\ccyc_h(g,g^{-1})\rep(\openone,\openone)$, where the identity operator $\rep(\openone,\openone)$ in the basis $\ket{\state^k_{u_A}}$ is represented by the unit matrix. This determines the inverse of the matrix $\rep(h,g)$.

On the other hand, in the orthonormal basis the matrix of adjoint operator is given by the conjugated transpose $M^\dagger_{k',k}=M(h,g)^*_{k,k'}$.

The matrices $M^\dagger$ and $M^{-1}$ differ only by a phase factor containing five 2-cocycle factors. Using the cocycle rules Eqs.~(\ref{2cy6}),~(\ref{2ccy}) and (\ref{2cy1}) it is easy to show that the phase factor is equal to one, and the representation $M$ is unitary. The same can be shown for the representation of $D(B)$.

We note that the projective representation of $\G$ formed by local operators $D(A)$ or $D(B)$, as discussed here, is \textit{not} the algebra defining the quasi-quantum double. The operator algebra within the quasi-quantum double is formed by the ``topological operators'', and depends on cocycle of elements in $\GG$ only. This is discussed in detail in Appendix~\ref{app:extended_ribbon}.

Let us next consider the problem of determining the two-quasiparticle Hilbert space $\mathcal{\tilde L}(A,B)$, where we fix the two excitations at positions $A$ and $B$. We will prove that the Hilbert space $\mathcal{L}(A,B)=\oplus_{u_C}\mathcal{L}_{u_C}(A,B)$ ($C$ is either $A$ or $B$), as introduced after Eq.~\eqref{eq:58}, indeed contains all the states of the two excitations, i.e. it contains $\mathcal{\tilde L}(A,B)$. Our proof follows the one given in Ref.~\onlinecite{Kitaev:2003p6185} for the model generalizing the $Z_2$ toric code to arbitrary finite groups (the ``generalized toric code''); however, our proof has significant changes, rooted in the fact that our model differs significantly from the generalized toric code. (For the present proof, the presence of global symmetries in our model makes the most important difference.)

We note that since clearly the two-quasiparticle Hilbert space $\mathcal{\tilde L}(A,B)$ is a subspace of $\mathcal{K}(\pathg)$, the action of $F(\pathg)$ and $D$ operators appearing below is always well-defined.

To establish our result, we need to define additional useful tensors related to the ones in Sec.~\ref{sec:extend-ribb-algebra}. We recall the double index notation using Latin indices: $i\equiv(h_i,g_i),j\equiv(h_j,g_j),\ldots$, with $h_i,h_j,\ldots\in\GG$ and $g_i,g_j,\ldots\in\G$, while the Kronecker delta function $\delta^i_j\equiv\delta_{h_i,h_j}\delta_{g_i,g_j}$. The antipode tensor $S$ is defined such that the element $S^k_lF^l\otimes D_k$ is the inverse of the element $F^i\otimes D_i$ in the algebra $\mathbb{F}\otimes\mathbb{D}$. This definition results in the following identity and solution for $S$:
\begin{align}
  \label{eq:39}
  &S^k_l\Lambda_p^{lm}\Omega^q_{kn}\delta^n_m=\epsilon_pe^q\\ &S^{(h_1,g_1)}_{(h_2,g_2)}=\delta_{h_1,h_2^{-1}}\,\delta_{g_1,g_2^{-1}}\,\ccyc^{-1}_{g_2}(h_2,h_1)\,\ccyc^{-1}_{h_1}(g_1,g_2).
\end{align}
One can check that the defining identity Eq.~(\ref{eq:39}) holds for the given form of $S$ by using the 2-cocycle identities Eq.~(\ref{2cy6}) and Eq.~(\ref{2cyinv}).

Recall from Sec.~\ref{sec:extend-ribb-algebra} the functions $\epsilon_i\equiv\delta_{h_i,\openone}$ and $e^i\equiv\delta_{g_i,\openone}$, which define the unit and counit of the algebras $\mathbb{F},\mathbb{D}$.
Using these functions we now define $\tau_s\equiv N_{\G}^{-1}e^s$ and
\begin{equation}
  \label{eq:35}
  \hat{C}(A)\equiv\frac{1}{N_{\GG}}\epsilon_iD_i(A),
\end{equation}
where $N_{\GG}=|\GG|$ is the order of the gauge group and $N_{\G}=|\G|$ is the order of the group $\G$. It is important to notice that by its definition, the $\epsilon_i$ constrains only the gauge group element $h_i\equiv\openone$ of the double index $i=(h_i,g_i)$. This means that the operator $\hat{C}(A)$ in Eq.~\eqref{eq:35} projects out any non-zero flux in triangle $t_A$, i.e. it ensures there are no \textit{flux} excitations at that lattice plaquette. However, the action of operator $D(A)$ in $\hat{C}(A)$ is still non-trivial, e.g. since the element $g_i\in\G$ in principle modifies the lattice site element $u_{i_A}$.

We now arrive at the central identity:
\begin{align}
  \label{eq:42} &\tau_s\Omega^s_{mp}S^p_qF^m(\pathg)\hat{C}(A)F^q(\pathg)=\\ \label{eq:centralidentgg}
  &=\frac{1}{N_{\GG}N_{\G}}\delta_{h_r,\openone}\delta_{g_j,\openone}D_j(A)F^r(\pathg)=\\\notag
  &=\frac{1}{N_{\GG}N_{\G}}\openone_{\mathbb{D}_A\otimes \mathbb{F}\otimes \mathbb{D}_B},
\end{align}
where we again used the double summation convention, i.e. $\sum_p=\sum_{h_p\in\GG,g_p\in\G}$, etc. The first equality can be proved by using the explicit definitions of all tensors, the operator algebra, and the 2-cocycle properties. The last line of Eq.~(\ref{eq:42}) is slightly more subtle. Consider an arbitrary matrix element of the operator in Eq.~\eqref{eq:centralidentgg}. The delta functions ensure that the ribbon and local operators do not change any edge or site elements between the initial and final state. Further, their phase factors containing cocycles are also trivial due to our choice of standard cocycle, see Eq.~(\ref{2cy1}). Consider then the matrix element contribution from the ribbon operator: $\delta(u_{i_A}u^{-1}_{i_B},\tilde{g}_r)$, with the obvious factorization $r\equiv(h_r,g_r)=(h_r,h'_r\cdot\tilde{g}_r)$, $h_r,h'_r\in\GG$, $\tilde{g}_r\in\SG$. But there is a summation over $r$, and therefore over $\tilde{g}_r\in\SG$, so the delta function is harmless. The same is true for the delta function coming from the local operator. We therefore see that the quantum amplitude is $1$, and indeed the action equals the identity operator $\openone$ in the algebra $\mathbb{D}_A\otimes \mathbb{F}\otimes \mathbb{D}_B$.

Let us now consider an arbitrary state $\ket{\state}$ in the Hilbert space $\mathcal{\tilde L}(A,B)$. We can define:
\begin{align}
  \label{eq:83}\notag
\ket{\eta^q}&\equiv \hat{C}(A)F^q(\pathg)\ket{\state}\\ G_q&\equiv N_{\GG}N_{\G}\tau_s\Omega^s_{mp}S^p_qF^m(\pathg),
\end{align}
so that
\begin{equation}
  \label{eq:86}
  \ket{\state}=G_q\ket{\eta^q}
\end{equation}
holds due to the identity Eq.~\eqref{eq:42}. The states $\ket{\eta^q}$ actually have \textit{no flux} excitations at $A$ due to the action of $\hat{C}(A)$. Elementary gauge and charge excitations must be in pairs and by construction are symmetric, so $\ket{\eta^q}$ can only describe the ground state $\ket{\gs}$ or a one-particle state that breaks the global symmetry. Since the site elements $u_{i_A},u_{i_B}$ are the local (at $A,B$) degrees of freedom acted on by the global symmetry, we conclude that the $\ket{\eta^q}$ states can be written as linear combinations of the form:
\begin{equation}
  \label{eq:84} \ket{\eta^q}=\sum_{u_1,u_2\in\SG}k^q_{u_1,u_2}\hat{P}_{u_1}(A)\hat{P}_{u_2}(B)\ket{\gs},
\end{equation}
with some coefficients $k^q_{u_1,u_2}$. (These states include the ground state itself.)

Eqs.~(\ref{eq:86}) and (\ref{eq:83}) show that an arbitrary two-particle state is a combination of the form
\begin{equation}
  \label{eq:85} 
\ket{\state}=\sum_{\substack{u_1,u_2\in\SG\\m}}K_{m,u_1,u_2}F^{m}(\pathg)\hat{P}_{u_1}(A)\hat{P}_{u_2}(B)\ket{\gs}.
\end{equation}
By the definition of the ribbon and projector operators, the $\ket{\state}$ is the zero vector unless $u_1\cdot u_2^{-1}=\tilde{g}_m$, where $m\equiv(h_m,g_m)=(h_m,h'_m\cdot\tilde{g}_m)$, $h_m,h'_m\in\GG$, $\tilde{g}_m\in\SG$. One of the three sums over $\SG$ in Eq.~\eqref{eq:85} is therefore superfluous for physical states in $\mathcal{\tilde L}(A,B)$.

We can use $u_1\cdot u_2^{-1}=\tilde{g}_m$ to eliminate either of the $u_{1,2}$ in Eq.~\eqref{eq:85}, which immediately shows that the arbitrary state $\ket{\state}\in\mathcal{\tilde L}(A,B)$ is indeed a linear combination of projected ribbon states from Eq.~(\ref{eq:58}).



\section{Symmetry fractionalization for multiple quasiparticles}
\label{sec:symmfracmulti}

In the examples of Sec.~\ref{sec:constr-local-symm}, we describe the scheme to find the fractionalized symmetry transformations $U_{\sigma}(C)$ ($C=A,B$) for a quasiparticle at location $C$ in the single-ribbon states, see Eq.(\ref{eq:62}) where we assumed Eq.(\ref{eq:64}). In this section, we show that at least when the non-trivial symmetry fractionalization only occurs in a $Z_2$ gauge sector (namely for visons in a gauge sector $Z_2\in GG$), the definition of local operators $U_{\sigma}(C)$ in Eq.(\ref{eq:62}) implements the fractionalized symmetry transformations for multi-particle states created by ribbon operators defined for ribbons $\Gamma_1,\Gamma_2,...,\Gamma_N$ that all share the same end-$B$:
\begin{align}
 &U_{\sigma}(A_1)U_{\sigma}(A_2)...U_{\sigma}(A_N)U_{\sigma}(B)|\psi_{u_B}^{k_1k_2...k_N}\rangle\notag\\
=&U(\sigma)|\psi_{u_B}^{k_1k_2...k_N}\rangle,\;\;\;\;\forall\sigma\in SG\label{eq:general_SF}
\end{align}

In our simple examples in the main text, it is always true that when non-trivial symmetry fractionalization occurs, it always only occur in one $Z_2$ gauge sector. In addition, we belive that all states $\{|\psi_{u_B}^{k_1k_2...k_N}\rangle\}$, $\forall u_B, k_1,k_2,...,k_N$ span a Hilbert space $\mathcal{L}(A_1,A_2,...,A_N,B)$ that contains all possible excitations at those locations. Therefore Eq.(\ref{eq:general_SF}), which we will prove below, indicates that the $U_{\sigma}(C)$ operators are the general fractionalized symmetry transformations in our examples.

It is straightforward to show that (both $GG$ and $SG$ are Abelian):
\begin{align}
 U(\sigma)|\psi_{u_B}^{k_1k_2...k_N}\rangle=|\psi_{\sigma^{-1}u_B}^{k_1k_2...k_N}\rangle
\end{align}
while
\begin{widetext}
\begin{align}
 &D_{(h_1^{-1},\sigma^{-1})}(A_1)D_{(h_2^{-1},\sigma^{-1})}(A_2)...D_{(h_N^{-1},\sigma^{-1})}(A_N)D_{(h_1h_2...h_N,\sigma^{-1})}(B)|\psi_{u_B}^{k_1k_2...k_N}\rangle\notag\\
=&\epsilon_{h_1,\sigma^{-1},u_{A_1}u_B^{-1}}\epsilon_{h_2,\sigma^{-1},u_{A_2}u_B^{-1}}...\epsilon_{h_N,\sigma^{-1},u_{A_N}u_B^{-1}}\cdot \xi(g,h_1,h_2,...h_N)\cdot |\psi_{\sigma^{-1} u_B}^{k_1k_2...k_N}\rangle\notag\\
=&\epsilon_{h_1,\sigma^{-1},u_{A_1}}\epsilon_{h_2,\sigma^{-1},u_{A_2}}...\epsilon_{h_N,\sigma^{-1},u_{A_N}}\epsilon_{h_1h_2...h_N,\sigma^{-1},u_B^{-1}}\cdot \xi(g,h_1,h_2,...h_N)\cdot |\psi_{\sigma^{-1} u_B}^{k_1k_2...k_N}\rangle\notag\\
=&\epsilon_{h_1^{-1},\sigma^{-1},u_{A_1}}^{-1}\epsilon_{h_2^{-1},\sigma^{-1},u_{A_2}}^{-1}...\epsilon_{h_N^{-1},\sigma^{-1},u_{A_N}}^{-1}\epsilon_{h_1h_2...h_N,\sigma^{-1},u_B}^{-1}\cdot \xi(g,h_1,h_2,...h_N)\cdot |\psi_{\sigma^{-1} u_B}^{k_1k_2...k_N}\rangle,
\end{align}
\end{widetext}
where we used the assumption: $\epsilon_{h,\sigma^{-1},h'}=1$, $\forall h,h'\in\GG$ and $\sigma\in\SG$, and the basic properties of the tensor $\epsilon_{x,y,z}$. The extra phase $\xi(g,h_1,h_2,...h_N)$ is defined as:
\begin{widetext}
\begin{align}
 \xi(g,h_1,h_2,...h_N)&=c_{\sigma^{-1}}^{-1}(h_1^{-1},h_1)c_{\sigma^{-1}}^{-1}(h_2^{-1},h_2)...c_{\sigma^{-1}}^{-1}(h_N^{-1},h_N)\cdot c_{\sigma^{-1}}(h_1,h_2h_3...h_N)c_{\sigma^{-1}}(h_2,h_3...h_N)....c_{\sigma^{-1}}(h_{N-1},h_N)
\end{align}
\end{widetext}
In our examples, non-trivial symmetry fractionalization occurs for a single $Z_2$ subgroup $Z_2^{SF}\in GG$ and $GG=Z_2^{SF}\times GG'$. Let us denote a group element $h\in GG$ as $h=(h_{SF},h')$, with $h_{SF}\in Z_2^{SF}$ and $h'\in GG'$. This indicates that $c_{\sigma^{-1}}(h_a,h_b)\neq 1$ can occur only when both $h_{a,SF}$ and $h_{b,SF}$ have non-trivial components in $Z_2^{SF}$, i.e., $h_{a,SF}=h_{b,SF}=1$, and generally $c_{\sigma^{-1}}(h_a,h_b)=c_{\sigma^{-1}}(h_{a,SF},h_{b,SF})$. Let's denote $c_{\sigma^{-1}}(h_{a,SF}=1,h_{b,SF}=1)=\eta$ (we use the $\{0,1\}$ notation for $Z_2^{SF}$.). Under these conditions, it is easy to show that $\xi(g,h_1,h_2,...h_N)=\eta^{-N_{SF}/2}$, where $N_{SF}$ is the total number of visons (at the end-$A_i$ and end-$B$) in the $Z_2^{SF}$ sector for the state $|\psi_{u_B}^{k_1k_2...k_N}\rangle$. Therefore, using $U_{\sigma}(C)$ defined in Eq.(\ref{eq:62}) for an excitation with a gauge flux $h$,
\begin{align}
 U_{\sigma}(C)=\sqrt{c_{\sigma^{-1}}(h,h^{-1})}\cdot \epsilon_{h,\sigma^{-1},u_C}\cdot D_{(h,\sigma^{-1})}(C),
\end{align}
it follows that Eq.(\ref{eq:general_SF}) is indeed satisfied.





\bibliographystyle{apsrev}
\bibliography{seto}
\end{document}